\documentclass[12pt]{article}
\usepackage{fullpage,amsmath,amssymb,mathtools,natbib,hyperref}
\usepackage{titling,enumitem}
\usepackage[skip=-2pt]{caption} 
\usepackage{datetime,bigints}
\usepackage[dvipsnames]{xcolor}
\usepackage[most]{tcolorbox}
\usdate
\usepackage{multirow,algorithm} 
\usepackage{float}
\usepackage{algorithm,bbm}
\usepackage[]{algpseudocode}
\usepackage{arydshln}
\usepackage[section]{placeins}
\usepackage[titletoc,toc,title]{appendix}
\usepackage{fancyvrb} 
\usepackage{listings}
\usepackage{bm}
\usepackage{xcolor,colortbl} 

\xdefinecolor{gray}{rgb}{0.8,0.8,0.8}
\xdefinecolor{blue}{RGB}{58,95,205}
\DeclareCaptionFormat{listing}{\rule{\dimexpr\textwidth+17pt\relax}{0.4pt}\vskip1pt#1#2#3}
\definecolor{gray}{gray}{0.85}
\definecolor{LightCyan}{rgb}{0.88,1,1}

\newcolumntype{a}{>{\columncolor{gray}}c}
\newcolumntype{b}{>{\columncolor{white}}c}

\usepackage{xr,caption}

\newcommand{\argmin}{\text{argmin}}

\newcommand{\sgn}{s\textsc{gn}}
\newcommand{\gn}{\textsc{gn}}
\newcommand{\gd}{\textsc{gd}}

\usepackage{thmtools} 
\usepackage{amsthm}
{
      \theoremstyle{plain}
      
      \newtheorem{theorem}{Theorem}
      \newtheorem{example}{Example}
      \newtheorem{proposition}{Proposition}
      \newtheorem{lemma}{Lemma}
      
      \newtheorem{assumption}{Assumption}
      
  }

\renewcommand{\arraystretch}{1.5}

\def\lQ{\scalebox{-1}[1]{''}}

\makeatletter
\renewenvironment{abstract}{%
    \if@twocolumn
      \section*{\abstractname}%
    \else 
      \begin{center}%
        {\bfseries \normalsize\abstractname\vspace{\z@}}
      \end{center} \vspace{-0.5cm}%
      \quotation
    \fi}
    {\if@twocolumn\else\endquotation\fi}
\makeatother

\begin{document}

  \title{Noisy, Non-Smooth, Non-Convex\\
  Estimation of Moment Condition Models}
  \author{\Large Jean-Jacques Forneron\thanks{Department of Economics, Boston University, 270 Bay State Road, Boston, MA 02215 USA.\newline Email: \href{mailto:jjmf@bu.edu}{jjmf@bu.edu}, Website: \href{http://jjforneron.com}{http://jjforneron.com}.   \newline Part of this research was conducted while I was visiting Yale University, I am grateful for their hospitality. I would like to thank David Lagakos and Marc Rysman for their inquiries about non-linear estimation which motivated this research, as well as Xiaohong Chen, Jasmin Fliegner, Christian Gouri\'eroux,  Hiro Kaido, Yuichi Kitamura, Dennis Kristensen, Simon Lee, Jessie Li, Serena Ng, Zhongjun Qu, Christoph Rothe, Daniel Wilhelm, and Liang Zhong for helpful comments and suggestions, as well as participants at the CEME Conference at Duke, NY Camp Econometrics, CEMMAP Econometric workshop, CIREQ Econometrics Conference, North American Summer Meeting of the Econometric Society, the Bonn-Mannheim, BC/BU, Columbia, LSE, NY FED, UCL, U. of Toronto, Yale, and Warwick econometric workshops.}
  } \date{\today}
  \maketitle 

  \begin{abstract}  
    A practical challenge for structural estimation is the requirement to accurately minimize a sample objective function which is often non-smooth, non-convex, or both.  This paper proposes a simple algorithm designed to find accurate solutions without performing an exhaustive search. It augments each iteration from a new Gauss-Newton algorithm with a grid search step. A finite sample analysis derives its optimization and statistical properties simultaneously using only econometric assumptions. After a finite number of iterations, the algorithm automatically transitions from global to fast local convergence, producing accurate estimates with high probability. Simulated examples and an empirical application illustrate the results.
  \end{abstract}
  

  \baselineskip=18.0pt
  \thispagestyle{empty}
  \setcounter{page}{0}
  
\newpage

\section{Introduction}
Generalized Method of Moments (GMM) and Simulated Method of Moment (SMM) estimations are routinely used to evaluate structural economic models on empirical data. Textbook examples include estimating models of dynamic discrete choice \citep{hotz1993}, market equilibrium \citep{Berry1995}, general equilibrium \citep{christiano2005}, and export participation \citep{eaton2011}. In practice,  estimates are computed using numerical optimization methods. The focus of this paper is on a common problem faced by practitioners: optimization for GMM, SMM and related methods tend to be very difficult, and while econometric theory often takes optimization for granted, practitioners do need to successfully minimize the GMM objective function to get valid empirical results. 

Three issues are common to structural estimations. Oftentimes i. the objective function is non-convex: fast convex optimizers can return local optima which are not consistent estimates, ii. because of sampling uncertainty and simulation noise: optimization can either succeed or fail with some probability (even if the population problem is well posed), and iii. the sample objective function can be non-smooth or discontinuous: fast derivative-based methods become unstable. Issue i. depends on both the model and choice of moments. Also, convexity is not invariant to non-linear reparameterizations of the coefficients. Issues ii. and iii. are finite sample by nature; iii. is common in SMM estimations of discrete choice models. 

This paper tackles these estimation challenges in two steps. First, it proposes a new Gauss-Newton algorithm that combines non-smooth moments with smoothed Jacobian estimates. The algorithm is shown to converge quickly, in an area that is nearly as large as in the smooth population problem. The resulting estimates are asymptotically unbiased under mild conditions on the bandwidth. Existing methods that rely on smoothing produce estimates that can exhibit asymptotic bias unless the bandwidth vanishes quickly enough; faster than the optimal rate for optimization. Such undersmoothing is not required here, making the estimates more robust to the choice of bandwidth. The local convergence analysis relies on standard local identification conditions.

Second, the paper shows that augmenting each Gauss-Newton iteration with a grid-search step results in a globally convergent algorithm. After a finite number of iterations, the combined steps preserve the fast local convergence rate, albeit with a slightly slower rate. This contrasts with commonly used global optimizers that converge slowly. The global convergence results leverage standard global identification conditions.

The idea here is that the local step alone converges quickly to the target estimates, but only from a local initial value. The global step alone is globally convergent from any initial value, but slow. When using them concurrently, from an arbitrary initial value, the global step can improve upon the local one for a certain number of iterations -- determined by the global identification condition and the choice of sequence used in the global step. Afterwards, convergence is driven by the local step. This makes convergence more robust without sacrificing too much performance. Here, the algorithm adaptively transitions from global to local optimization without user input. The fast convergence occurs without smoothing the sample moments in either the local or global step. This is achieved by exploiting the specific formulation of the GMM objective function. 

This paper adopts a non-asymptotic (finite sample) approach to study the optimization and statistical properties of the algorithm simultaneously. This allows to better understand the effects that tuning parameters can have on both the estimation and the estimates. For instance, smoothing should facilitate optimization but could also degrade the properties of the estimator. Here, the derivations show how both are affected by the choice of smoothing parameters. The finite-sample approach is also conceptually attractive because it gives results that are valid across repeated samples. Textbook optimization results often consider fixed objective functions \citep{nocedal-wright:06,bertsekas2016}. However, in empirical work, each dataset is associated with its own, unique, objective function. In that sense, guarantees over repeated samples provide a sense of robustness against sampling uncertainty. 

The first extension of the main results considers momentum, also known as the Polyak heavy-ball, which is often used to accelerate the convergence of stochastic gradient descent algorithms. Here, it can be used to reduce the effect of sampling noise on optimization while maintaining the rate of convergence. In practice, it smoothes the optimization path, optimal values for the tuning parameters are tabulated. The second extension proposes a computationally attractive Monte Carlo quasi-Newton implementation of the algorithm which makes it applicable to models where smoothing is not analytically tractable.  
 
Simulated and empirical examples illustrate the results. A dynamic discrete choice model illustrates the algorithm's advantages  over a benchmark which minimizes a smoothed objective function. From a distant starting value, the proposed algorithm performs well whilst the benchmark systematically fails to find consistent estimates. When initialized at the true value, the benchmark is consistent but sensitive to the choice of smoothing parameter. In contrast, the algorithm is fairly robust to the smoothing parameter. This is in line with theory. A thorough comparison with several global optimizers highlights the good performance of the proposed algorithm - both in terms of numerical accuracy and computation time. Similar to this paper, several of these optimizers are hybrid: they combine local and global iterations. However, they often rely on heuristic rules and lack theoretical guarantees for convergence, unlike the proposed method. This is reflected in the relative performance of these methods. Then, a simple Aiyagari model illustrates the results on a small but fairly challenging estimation problem where commonly used optimizers fail to converge. The algorithm performs well for a range of tuning parameters. Finally, the empirical application considers a model of joint retirement decision from \citet{honore2018}. Their estimation of $30$ coefficients takes more than 5 hours whereas the algorithm introduced in this paper finds accurate estimates within 11 minutes.

Lastly, note that the results cover smooth GMM estimations as a special case. The idea to combine a local and global search step to design a fast converging optimizer naturally extends to non-GMM settings such as smooth maximum likelihood and non-linear least-squares estimations. A key difference is that the local optimization results here rely on approximate local linearity of moments whereas in these settings derivations would require local strong convexity and a convex optimizer. Convexity requires further assumptions beyond the first derivatives and can be more restrictive than the assumptions used here. This is illustrated for a simple quantile estimation below. Because these estimations would involve a different set of assumptions and derivations, they are not covered in this paper.

\paragraph{Outline of the paper.} After a brief literature review, Section \ref{sec:algos} briefly describes the setting and introduces the algorithm with a simple illustration. Section \ref{sec:main} provides the main assumptions, then derives local and global convergence results in finite samples. Section \ref{sec:extend} provides two extensions of the algorithm and derives their properties. Section \ref{sec:MC_Emp} provides simulated and empirical applications to illustrate the properties and performance of the method compared to a number of popular global optimizers. The Appendices provide proofs as well as additional simulation and empirical results.

\section*{Overview of the Problem and Literature} 

Optimization is a defining feature of non-linear models evaluated by M-estimation:
\begin{align*}\hat\theta_n = \text{argmin}_{\theta \in \Theta} Q_n(\theta), 
\end{align*}
where $Q_n$ is the sample objective function, $\theta$ are parameters of interest belonging to a compact, finite-dimensional set $\Theta$. While some estimators have closed-form solution, this is the case of OLS regressions, non-linear models generally require numerical methods to find $\hat\theta_n$. Statistical software provide empirical researchers with a wide array of optimizers. 

When choosing among these methods, two properties are important to consider: convergence guarantees, typically over a class of functions, and convergence rates within that class. The former matters since a failed optimization yields inconsistent estimates, the latter is critical when $Q_n$ is costly to evaluate. The following expands on these two properties.\footnote{Since most optimization results consider a fixed objective $Q_n$, sampling uncertainty is not discussed yet.} The goal here is to find, after $b \geq 1$ iterations, a $\theta_b$ such that: 
\begin{align*} 
  \|\theta_b-\hat\theta_n\| \leq \text{err}, 
\end{align*} 
at a desired tolerance level $\text{err} >0$. Suppose $Q_n$ has a certain set of features, one could ask: how large does $b$ need to be to guarantee the error $\|\theta_b-\hat\theta_n\|$ is at most $\text{err}$? This is the complexity of the optimization problem \citep{Nemirovsky1983}. If $Q_n$ is smooth and strongly convex,\footnote{$Q_n$ is strongly convex if its Hessian $H_n$ is continuous and $0 < \underline{\lambda} \leq \lambda_{\min}[H_n(\theta)] \leq \lambda_{\max}[H_n(\theta)] \leq \overline{\lambda} < \infty$.} then $b \geq O( |\log(\text{err})| )$ so that to divide $\text{err}$ by $10$, an additional $O(\log[10])$ iterations are required. Gradient descent, and (quasi)-Newton methods can achieve this rate of convergence. However, when $Q_n$ is non-convex, they may only converge to a local optimum. If $Q_n$ is $r$-times continuously differentiable but otherwise arbitrary, then $b \geq O( \text{err}^{-d_\theta/r} )$ iterations are required, where $d_\theta = \text{dim}(\theta)$. For continuous problems, $r=1$, this rate is achieved by a grid search. Here, the curse of dimensionality is apparent: to divide $\text{err}$ by $10$, the number of iterations needs to be multiplied by $10^{d_\theta} = 10,000$ for $d_\theta=4$.\footnote{To put this into context, the average GMM estimation published between 2016 and 2018 in the American Economic Review involved $d_\theta = 10$, the median $d_\theta = 6$ \citep{jjlz22}.} \citet[p14]{Nesterov2018} and \citet[p915]{andrews1997} illustrate that as $d_\theta$ increases, optimization quickly is practically infeasible under this rate of convergence. Computer software provides a number of global optimizers, however, as \citet[Sec1]{griewank1981} points out, those that do not cover the set $\Theta$ are ``heuristic,'' in the sense that global convergence is not guaranteed.\footnote{A sequence $(\theta_k)_{k \geq 0}$ will be said to cover $\Theta$ if it is \textit{dense}, i.e. $\sup_{\theta \in \Theta} (\inf_{ 0 \leq \ell \leq k}\|\theta-\theta_\ell\|) \to 0$, as $k \to \infty$. These types of methods are referred to as \textit{space-covering} in \citet[p13]{griewank1981}.} Several popular methods are used as benchmark for the method introduced in this paper in Section \ref{sec:MC_Emp}, and their known theoretical properties are reviewed there.

Convergence guarantees are a very desirable property since they ensure that a given method will perform well in a variety of settings. A common denominator of GMM estimations are the conditions required for consistency and asymptotic normality. This paper solely relies on these conditions to derive convergence results so that the algorithm proposed below performs well in a range of empirical settings and datasets. 

Since many GMM problems are non-convex, a number of authors have proposed strategies to speed-up convergence. Building on \citet{robinson1988}, \citet{andrews1997} proposed a stopping-rule to produce a consistent first-step estimate, with $\text{err} = o_p(1)$, which is followed by a few Newton-Raphson iterations. The main challenge is in computing the consistent estimate which still requires a global search. \citet{chernozhukov2003} introduce a quasi-Bayesian framework where MCMC sampling replaces optimization. The random-walk Metropolis-Hastings algorithm converges under weak conditions. However, convergence rates, which measure performance, are mostly derived for log-concave distributions.\footnote{See e.g. \citet{mengersen1996}, \citet{brooks1998}, \citet{belloni2009}.}

A common strategy to handle non-smooth sample moments is to use smoothing. The complexity bounds above then apply to the smoothed optimization problem. \citet[pp1000-1001]{mcfadden1989} and \citet{bruins2018} suggest to replace an indicator function by a smooth approximation in simulation-based estimation of discrete choice models. Several papers consider smoothing in quantile regressions \citep{kaplan2017,fernandes2021,he2021}. A poor choice of smoothing parameter can have a significant impact on the estimates and their statistical properties \citep[see][Table 1, p127]{kaplan2017}. Another approach is to apply subgradient methods to the unsmoothed quantile loss function \citep{lee2022}. Other applications of smoothing for econometric estimation include \citet{fermanian2004}, \citet{altissimo2009}, \citet{kristensen2012}.  For these, it is possible to apply numerical bias correction to reduce smoothing bias as in \citet{kristensen2017}; this is asymptotically valid under further regularity conditions \citep[e.g.][Sec3.4.1]{bruins2018}. For indirect inference, \citet{frazier2019} consider an alternative simulation method for discrete outcomes to get smooth simulated moments. This applies to a specific class of models and moments. \citet{ackerberg2009}, \citet{sauer2021} propose to use importance sampling to smooth the objective function; however, a poor choice of sampling distribution inflates simulation noise \citep{ackerberg2009}. \citet{hong2015} study the asymptotic properties of M-estimates defined by a first-order condition computed by finite-differences; optimization properties are not derived. Finite-differences yield non-smooth derivative estimates when the objective is itself non-smooth, this makes the properties of the associated derivative-based optimizer unclear.  
Here, the local and global search step involve unsmoothed moments making the estimates more robust to the choice of bandwidth. This is illustrated in the next Section. Despite using unsmoothed moments, convergence results are comparable to the smooth population problem, in finite samples. This contrasts with generic non-smooth optimizers, including subgradient methods, which are generally slower than smooth ones. 



\section{Noisy, Non-Smooth, Non-Convex Estimation} \label{sec:algos}
\subsection{The Smoothed Gauss-Newton Algorithm} \label{sec:thealgo}
Let $\overline{g}_n(\theta) = \frac{1}{n} \sum_{i=1}^n g(\theta;x_i)$ be the sample moments, where $x_i$ are iid random variables and $\theta$ are the parameters of interest. The goal is to find an estimate of $\theta^\dagger \in \Theta$ solving the moment condition $g(\theta^\dagger):=\mathbb{E}[g(\theta^\dagger;x_i)]=0$. In practice, this entails finding an approximate minimizer $\hat\theta_n \in \Theta$ of the sample objective function $Q_n(\theta) = \|\overline{g}_n(\theta)\|_{W_n}^2 = \overline{g}_n(\theta)^\prime W_n \overline{g}_n(\theta)$ satisfying:
\begin{align}
  \|\overline{g}_n(\hat\theta_n)\|_{W_n}^2 \leq \inf_{\theta \in \Theta}  \|\overline{g}_n(\theta)\|_{W_n}^2 + o_p(n^{-1}), \label{eq:approxi_min}
\end{align}
where $(W_n)_{n \geq 1}$ is a sequence of symmetric and strictly positive definite weighting matrices with limit $W >0$. To simplify the analysis, $W_n$ is assumed not to depend on $\theta$.\footnote{This excludes continuously-updated and allows for two-step estimations.} Algorithm \ref{algo:sgn} is designed to find such an approximate minimizer. It relies on the Jacobian $G_{n,\varepsilon}$ of smoothed moments $\overline{g}_{n,\varepsilon}$ using a bandwidth $\varepsilon$, computed by convolution smoothing. Let $\phi$ be the standard Gaussian density, the smoothed moments and Jacobian are:
\begin{align*}
  \overline{g}_{n,\varepsilon}(\theta) &:= \mathbb{E}_{Z}[\overline{g}_n(\theta + \varepsilon Z)] =  \int \overline{g}_n(\theta + \varepsilon Z)\phi(Z)dZ,\\  G_{n,\varepsilon}(\theta) &:= \partial_\theta \overline{g}_{n,\varepsilon}(\theta) = \frac{1}{\varepsilon} \int \overline{g}_{n}(\theta + \varepsilon Z) Z^\prime \phi(Z)dZ
\end{align*} 

\begin{algorithm}[h] 
  \caption{Smoothed Gauss-Newton (\sgn)} \label{algo:sgn} 
        \begin{algorithmic}
          \State 1) \textbf{Inputs} (a) a learning rate $\gamma \in (0,1)$, (b) a smoothing parameter $\varepsilon >0$, and (c) a sequence $(\theta^b)_{b \geq 0}$ covering the parameter space $\Theta$. 
        \State 2) \textbf{Iterations} 
        \State set $b=0$, $\theta_0 = \theta^0$
        \Repeat
              \State compute $\theta_{b+1} = \theta_b - \gamma \Big[ G_{n,\varepsilon}(\theta_b)^\prime W_n G_{n,\varepsilon}(\theta_b) \Big]^{-1} G_{n,\varepsilon}(\theta_b)^\prime W_n \overline{g}_n(\theta_b)$ \Comment{Local Step}
              \State if $\|\overline{g}_n(\theta^{b+1})\|_{W_n} < \|\overline{g}_n(\theta_{b+1})\|_{W_n}$, set $\theta_{b+1} = \theta^{b+1}$ \Comment{Global Step}
              \State increment $b := b+1$
        \Until{a stopping critera is met}
        \State 3) \textbf{Output} (a) estimates: $\tilde\theta_n = \argmin_{0 \leq j \leq b} \|\overline{g}_n(\theta_{j})\|_{W_n}$,\\ (b) asymptotic variance: $\tilde{V}_n = ( \tilde{G}_{n,\varepsilon}^\prime W_n \tilde{G}_{n,\varepsilon} )^{-1} ( \tilde{G}_{n,\varepsilon}^\prime W_n \Sigma_n W_n \tilde{G}_{n,\varepsilon} ) ( \tilde{G}_{n,\varepsilon}^\prime W_n  \tilde{G}_{n,\varepsilon} )^{-1}$,\\ 
        where $\tilde{G}_{n,\varepsilon} = G_{n,\varepsilon}(\tilde{\theta}_n)$, and $\Sigma_n$ is an estimate of $\Sigma = \lim_{n\to\infty} n \text{var}[\overline{g}_n(\theta^\dagger)]$
        \end{algorithmic}
\end{algorithm}

The computation of $G_{n,\varepsilon}$ is discussed in the next subsection and Section \ref{sec:extend}, Algorithm \ref{algo:sqn}. Algorithm \ref{algo:sgn} alternates between a local and a global search step. The main idea is that, from the current guess $\theta_{b}$, the local step proposes a new candidate $\theta_{b+1}$. When the moments are non-smooth, the smoothed Jacobian $G_{n,\varepsilon}$ is used to approximate the unfeasible Gauss-Newton local iteration. The global step proposes a competing candidate $\theta^{b+1}$, if the fit is strictly better, $\theta_{b+1}$ is discarded and $\theta^{b+1}$ becomes the current iterate. Otherwise, the value from the local step is kept. 
Note that, away from the solution, the local step might produce values $\theta_{b+1}$ outside of $\Theta$. One possibility is to set $\|\overline{g}_n(\theta_{b+1})\|_{W_n} = +\infty$, then the global step always binds. Another standard approach is to project $\theta_{b+1}$ inside $\Theta$, that is to set each coefficient that is outside its bounds slightly below/above its respective upper/lower bound. Appendix \ref{sec:Code} provides R code for Algorithm \ref{algo:sgn} with the extensions from Section \ref{sec:extend}.

Algorithm \ref{algo:sgn} returns the best iterate $\tilde{\theta}_n$, which is a valid estimator if at least one of the $\theta_j$ is close to minimizing $Q_n$. For inference, standard errors for the $j$-th coefficient of $\theta$ can be readily computed as $\text{se}(\tilde{\theta}_{n,j}) = \sqrt{\tilde{V}_{n,jj}/n}$; $\tilde{V}_{n,jj}$ denotes the $j$-th diagonal entry of $\tilde{V}_{n}$. The variance $\Sigma$ can be estimated in a standard fashion, e.g. $\Sigma_n = 1/n \sum_{i=1}^n g(\tilde{\theta}_n;x_i) g(\tilde{\theta}_n;x_i)^\prime - \overline{g}_n(\tilde{\theta}_n)\overline{g}_n(\tilde{\theta}_n)^\prime$ for an iid sample. Note that Gauss-Newton iterations alone can be globally convergent without convexity under rank conditions \citep{jjlz22}. Here, the global step ensures that convergence will hold over a broader class of problems. A common approach to finding a global solution is the multi-start strategy which runs a local optimizer many times from different starting values. The number of starting values required for global convergence depends on 1) the choice of algorithm and 2) the objective function; so it is unknown in practice. This paper simply adds a grid search step to get global convergence in a single run of the Algorithm. As discussed earlier, covering the set $\Theta$ is required to guarantee convergence. Running both steps at each iteration yields faster convergence than using them in succession as in the two-step approach of \citet{robinson1988}, \citet{andrews1997}. Differences with these methods are discussed in more detail in Appendix \ref{apx:comp}. 

Potential applications of Algorithm \ref{algo:sgn} include quantile regressions and SMM estimations of static or dynamic discrete choice models. Some smooth problems are made non-smooth by numerical approximations. For instance, if an income process is discretized to solve for the optimal policy, the approximated model is discrete and the moments are potentially non-smooth. Continuous models with discrete changes can also exhibit non-smoothness. In finance, this is the case of jumps-diffusion models estimated by SMM \citep{carrasco2000}.

\subsection{Choice of Tuning Parameters} \label{sec:tuning}
The algorithm has three tuning parameters: the learning rate $\gamma$, the bandwidth $\varepsilon$, and the covering sequence $(\theta^b)_{b \geq 0}$. The following provides specific recommendations for each of them.

\paragraph{Learning rate.} The Gauss-Newton normalization $[ G_{n,\varepsilon}(\theta_b)^\prime W_n G_{n,\varepsilon}(\theta_b) ]^{-1}$ ensures local convergence for any choice of $\gamma \in (0,1)$. In contrast, gradient descent methods require $\gamma > 0$ to be sufficiently small. The main tradeoff in choosing $\gamma$ is that a larger learning rate leads to faster local convergence but over a smaller local area around the solution. In the applications, a smaller value $\gamma=0.1$ balances speed and the area of convergence quite well.  


\paragraph{Bandwidth.} For optimization, setting $\varepsilon = c \cdot n^{-1/4}$ is rate optimal. The constant $c$ can be chosen to minimize a mean squared prediction error as follows. Construct a grid $\theta^1,\dots,\theta^B$, this can be the same as the covering sequence described below. Then for each $j = 1,\dots,B$ find $\theta^j_1,\dots,\theta^j_K$ the K-nearest neighbors to $\theta^j$ from the same grid and compute:
\[ \text{MSE}_{j}^\ell(\varepsilon) = \| \overline{g}_n(\theta^j) - \overline{g}_n(\theta^j_\ell) - G_{n,\varepsilon}(\theta^j_\ell)(\theta^j - \theta^j_\ell) \|^2. \]
Compute the average $\text{MSE}(\varepsilon)$ of $\text{MSE}_{j}^\ell(\varepsilon)$ over $j$ and $\ell$. This measures how well $G_{n,\varepsilon}$ can predict $\overline{g}_n(\theta^j)$ from $K$ nearby directions $\theta^j_\ell$. The moments and Jacobians can be computed once, ideally in parallel, before evaluating the $\text{MSE}$s. Take $\varepsilon_1,\dots,\varepsilon_m = c_1 \cdot n^{-1/4},\dots,c_m \cdot n^{-1/4}$ and pick the $\varepsilon$ among these which minimizes the $\text{MSE}$ above. For regressions, it can be useful to normalize variables to have mean $0$ and variance $1$ so that the bandwidth is scale-free. The applications use $B = 250$, $K=10$, set $c_1,\dots,c_m$ to $8 = 2^{3},2^2,\dots,2^{-4},2^{-5} = 0.003$. 

The choice of a MSE criterion is related to the role of $G_{n,\varepsilon}$ in the local step. It is used to approximate $\overline{g}_n(\theta)$ around $\theta = \theta_{b-1}$, the local step minimizes this linear surrogate. The MSE criterion measures the accuracy of the linear approximation.

\paragraph{Covering Sequence.} the choice of sequence $(\theta^b)_{b \geq 0}$ matters much for the performance of Algorithm \ref{algo:sgn}. The main property required is that the so-called dispersion of the sequence:
\[ D(\theta^1,\dots,\theta^b) = \sup_{\theta \in \Theta} \left( \inf_{j=1,\dots,b} \|\theta-\theta^j\| \right) \]
be as small as possible for each $b$. The dispersion measures how well the sequence $(\theta^1,\theta^2,\dots)$ covers the parameter space $\Theta$, i.e. how quickly, as $b$ increases, the distance between any $\theta$ and the closest $\theta^j$ decreases. Intuitively, the goal is to have one of $\{\theta^1,\dots,\theta^b\}$ that is as close as possible to $\hat{\theta}_n$, wherever it may be within $\Theta$, at each iteration $b$. This is only possible if the parameter space is bounded, so it is necessary to set bounds for the parameters. Bounds are always required for global non-convex optimization. Many parameters are naturally bounded: the discount rate is between $0$ and $1$, the risk-aversion coefficient is typically assumed to be between $0$ and $40$. For less obvious cases, it can be useful to estimate a simpler or reduced-form model to get a gross approximation of the parameter's magnitude. A good example is \citet{salanie2019} who approximate non-linear demand coefficients using a Taylor expansion and then use two-stage least-squares to approximate the coefficients. 

Given a set of bounds for each coefficient, write the parameters space as $\Theta = [\underline{\theta}_1,\overline{\theta}_1] \times \dots \times [\underline{\theta}_{d_\theta},\overline{\theta}_{d_\theta}]$. In the special case where $\Theta = [0,1] \times \dots \times [0,1]$, there are sequences for which the dispersion $D(\theta^1,\dots,\theta^b)$ is close to its theoretical lower bound, these are called \textit{low-dispersion sequences}. This is discussed in more detail in the next Section. The Sobol and Halton sequences are two popular low dispersion sequences \citep{judd1998,Lemieux2009}. Both are readily available in R, Julia, Matlab, and Python. Finally, to construct the covering sequence for $\Theta$, take one of the Sobol or Halton sequence, denoted $\vartheta^b$, and adjust each of its coordinates $\theta^b_j = \underline{\theta}_j + [\overline{\theta}_j - \underline{\theta}_j] \vartheta_j^b$ to match the desired bounds. 

Several factors can indicate that the user-specified $\Theta$ is too small to contain the solution $\hat{\theta}_n$. 1) If the local step exits $\Theta$ in some direction, and the global step cannot improve the fit inside $\Theta$, then the bounds could be widened in that direction. 2) If, after many iterations, $Q_n(\tilde{\theta}_n)$ is significantly non-zero, the bounds may be too narrow and could be enlarged. It may also be that the model is misspecified, a case that is not covered in the result.  

\subsection{Additional Implementation Details} \label{sec:implement}

\paragraph{Computing the Jacobian.} First, the following Lemma applies a change of variable argument similar to \citet{powell1989} or \citet{hazan2016} to derive some useful identities.
\begin{lemma}[Useful Identities] \label{lem:identities} For any $\theta\in \Theta$ and $\varepsilon >0$, we have:
  \begin{itemize} \setlength\itemsep{0em}
      \item[i.] $\overline{g}_{n,\varepsilon}(\theta) = \int_Z \overline{g}_n(\theta + \varepsilon Z)\phi (Z)dZ = \varepsilon^{-1} \int_u \overline{g}_n(u)\phi (\frac{u-\theta}{\varepsilon})du$,
      \item[ii.] $G_{n,\varepsilon}(\theta) = -\frac{1}{\varepsilon}\int_Z \overline{g}_n(\theta + \varepsilon Z)\phi^\prime (Z)dZ = -\varepsilon^{-2}\int_u \overline{g}_n(u) \phi^\prime (\frac{u-\theta}{\varepsilon})du$,
      \item[iii.] $G_{\varepsilon}(\theta) = -\frac{1}{\varepsilon}\int_Z g(\theta + \varepsilon Z)\phi^\prime (Z)dZ = \int_Z G(\theta + \varepsilon Z)\phi(Z)dZ$.
  \end{itemize}
\end{lemma}

While in theory the smoothing can be applied to any estimation problem, $\overline{g}_{n,\varepsilon}$ and $G_{n,\varepsilon}$ typically do not have closed-form. This is the case for the applications in Section \ref{sec:MC_Emp}. Nevertheless, an unbiased Monte Carlo estimator of $G_{n,\varepsilon}$ can be computed as follows:
\begin{align*}
  \hat{G}_{n,\varepsilon}(\theta) = \frac{1}{\varepsilon L} \sum_{\ell = 0}^{L-1} \big[\overline{g}_{n,\varepsilon}(\theta + \varepsilon Z_\ell) - \overline{g}_{n,\varepsilon}(\theta)\big] Z_\ell^\prime,
\end{align*}
using random draws $Z_\ell \overset{iid}{\sim} \mathcal{N}(0,I_{d_\theta})$ with $L \geq 1$. The mean-zero adjustment $\overline{g}_{n,\varepsilon}(\theta)$ yields better finite-$L$ properties. Other choices of densities $\phi$ could be used as long as they are: 1) continuously differentiable, and 2) satisfy $\int Z \phi(Z)dZ = 0$, $\int \|Z\|^2 \phi(Z) dZ < \infty$, $\int \|\partial_Z \phi(Z)\| dZ < \infty$. The Monte Carlo approximation $\hat{G}_{n,\varepsilon}$ does require $\phi$ to be a proper distribution, which excludes higher-order kernels.

A similar estimator was proposed in \citet{polyak1987}, \citet{polyak1990} for minimizing smooth and globally convex objectives by stochastic gradient descent (s\textsc{gd}), and was later studied in \citet{nesterov2017}. Convergence for gradient descent is slower than Gauss-Newton. \citet{chen2015} proposed a similar Jacobian for sieves. In an extension of the main results, a more computationally attractive quasi-Newton implementation of $\hat{G}_{n,\varepsilon}$ is introduced. Its non-asymptotic local convergence properties are derived. All simulated and empirical examples in the paper are based on this implementation. 
When the sample Jacobian $G_n$ exists, and satisfies appropriate regularity conditions, it can be used in the Local Step of Algorithm \ref{algo:sgn} instead of $G_{n,\varepsilon}$.

\paragraph{Regularization.} 
Away from the solution, the Jacobian $G_{n,\varepsilon}$ could be nearly singular. This is a concern since inverting an ill-conditioned matrix can be inaccurate and unstable. In this case, it is recommend to invert $G_{n,\varepsilon}(\theta_b)^\prime W_n G_{n,\varepsilon}(\theta_b) + \lambda I_d$ instead, with $\lambda \geq 0$, in the local step to stabilize it. This is know as the Levenberg–Marquardt (LM) algorithm. It can be particularly useful when $G_{n,\varepsilon}$ is evaluated by Monte Carlo and/or quasi-Newton to reduce sensitivity of the inverse to the approximation error. Section \ref{sec:sqn} gives specific recommendations for setting the penalty $\lambda_b$ at each iteration it that setting.

\paragraph{Stopping Criteria.} Standard convergence criteria for global optimization include that: a maximum number of iterations has been reached, or $Q_n$ has not declined more than some tolerance level (e.g. $\text{tol} = 10^{-6}$) for a certain number of iterations (e.g. $100$). When these occur, the optimization terminates. In the applications, Algorithm \ref{algo:sgn} was terminated when a maximal number of iterations was reached, to keep the computation time fixed.

\citet{andrews1997} proposes an asymptotically valid data-dependent stopping criterion for GMM specifically. The following sketches the idea in the context of Algorithm \ref{algo:sgn}. For simplicity, suppose $W_n$ is the optimal weighting matrix and standard regularity conditions hold. Take $c_{p,n}$ to be the $1-\alpha$ quantile of a $\chi^2_{p-d_\theta}$ distribution, where $p$ is the number of estimating equations, $p-d_\theta$ is the number of overidentifying restrictions. \citet[Table I, p921]{andrews1997} suggests $\alpha = 5\%$ and tabulates values. One can also use the quantiles of a $\chi^2_{p}$ distribution, which is less conservative. Suppose at iteration $k \geq 1$ that $Q_n(\theta_k) \leq c_{p,n}/n$. Then, under regularity conditions, $\theta_k$ is a consistent estimator with $\| \theta_k - \theta^\dagger \| \leq O_p(n^{-1/2})$ and $\| \theta_k - \hat{\theta}_n \| \leq O_p(n^{-1/2})$. From there, run the algorithm for an additional $j \geq 1$ iterations. The algorithm's properties roughly imply that $\|\theta_{b} - \hat{\theta}_n\| \leq (1-\overline{\gamma})^j \| \theta_k - \hat{\theta}_n \|$ for $b=k+j$ and a $\overline{\gamma}$ positive that is smaller than $\gamma$. Then applying one of the two stopping criteria above gives a valid estimator, using $j_{\max} \gg \log(n)$ additional iterations. 



\subsection{Illustation: a pen and pencil example} The following illustrates several aspects of Algorithm \ref{algo:sgn}, including differences between the local step and smoothing the objective function as well as convergence issues under non-convexity. Consider estimating the $t$-th quantile of $x_i \overset{iid}{\sim} F$, $t\in (0,1)$. The population and sample moments are $F(\theta)-t=0$, and:
\[\overline{g}_n(\theta) = \frac{1}{n} \sum_{i=1}^n \left[\mathbbm{1}_{x_i \leq \theta} -t\right] = 0, \]
or more compactly $F_n(\theta)-t=0$; where $F_n$ is the empirical CDF of $x_i$. For any choice of bandwidth $\varepsilon$, the smoothed moment condition and its Jacobian are:\footnote{This is derived using $\mathbb{E}_{Z}(\mathbbm{1}_{x_i \leq \theta + \varepsilon Z}) = \Phi([\theta-x_i]/\varepsilon)$.}
\[\overline{g}_{n,\varepsilon}(\theta) = \frac{1}{n} \left[\sum_{i=1}^n \Phi\left(\frac{\theta-x_i}{\varepsilon}\right)-t\right],\quad  G_{n,\varepsilon}(\theta) = \frac{1}{n\varepsilon} \sum_{i=1}^n \phi\left( \frac{\theta-x_i}{\varepsilon} \right), \]
where $\Phi$ and $\phi$ are the standard Gaussian CDF and density. Notice that $G_{n,\varepsilon}(\theta)=f_{n,\varepsilon}$ is the kernel density estimator of the density $f$.  In this example, the local step in Algorithm \ref{algo:sgn} involves the non-smooth CDF $F_n$ and the smoothed density $f_{n,\varepsilon}$:
\[ \theta_{b+1} = \theta_b - \gamma \frac{ F_n(\theta_b)-t}{ f_{n,\varepsilon}(\theta_b) }. \]
The main appeal is that an exact solution $\hat\theta_n$ with $F_n(\hat\theta_n)-t=0$ is a fixed-point of these iterations, regardless of the bandwidth $\varepsilon$. This implies that \textit{if the Algorithm is convergent}, then the solution is robust to the choice of bandwidth. 

Applying Gauss-Newton to the smoothed moments further substitutes $\overline{g}_n$ for $\overline{g}_{n,\varepsilon}(\theta) = F_{n,\varepsilon}(\theta) - t$, $F_{n,\varepsilon}(\theta) = \int_{-\infty}^\theta f_{n,\varepsilon}(u)du$ is the smoothed CDF. This results in the iterations:
\[ \theta_{b+1} = \theta_b - \gamma \frac{ F_{n,\varepsilon}(\theta_b)-t}{ f_{n,\varepsilon}(\theta_b) }. \]
Here, a fixed-point $\hat\theta_{n,\varepsilon}$ must satisfy $F_{n,\varepsilon}(\hat\theta_{n,\varepsilon})-t=0$. This is the $t$-th quantile of $F_{n,\varepsilon}$ which satisfies $\hat\theta_{n,\varepsilon} = \hat\theta_n + O_p(\varepsilon^2)$. The smoothing bias, captured by $\varepsilon^2$, is only negligible when $\varepsilon = o(n^{-1/4})$ which is more restrictive.

Although estimating a quantile is simple, the GMM formulation can be non-convex. For simplicity take $x_i \sim \mathcal{N}(0,1)$ and, to focus on non-convexity specifically, consider the smooth population loss $Q(\theta) = [ \Phi(\theta) - t ]^2$ with $t = 0.7$. Numerical calculations show that the Hessian $\partial^2_{\theta} Q(\theta)$ is positive for $\theta \in [-0.69,1.15]$, there $Q$ is locally convex, but is negative for $\theta \in [-3,-0.7]$ and $\theta \in [1.16,3]$, there $Q$ is locally concave. As a result, an off-the-shelf BFGS optimizer fails to minimize the population $Q$ from $\theta = -3$, it returns $\theta = 5.02$ instead of $\theta^\dagger = 0.52$. The sample problem is also non-convex. 

Even in this basic example, local optimization does not necessarily converge from an arbitrary value. To resolve this: the global step will produce, if the covering sequence has low dispersion, a value $\theta^b$ that is sufficiently close to $\theta^\dagger$ from which the local step is convergent. From that iteration onwards, the local step is convergent. Hence, the global step ensures the optimizer converges from an arbitrary starting value.

\begin{figure}[ht] \centering \caption{Illustration: estimating a sample quantile} \label{fig:quantile}
  \includegraphics[scale=0.55]{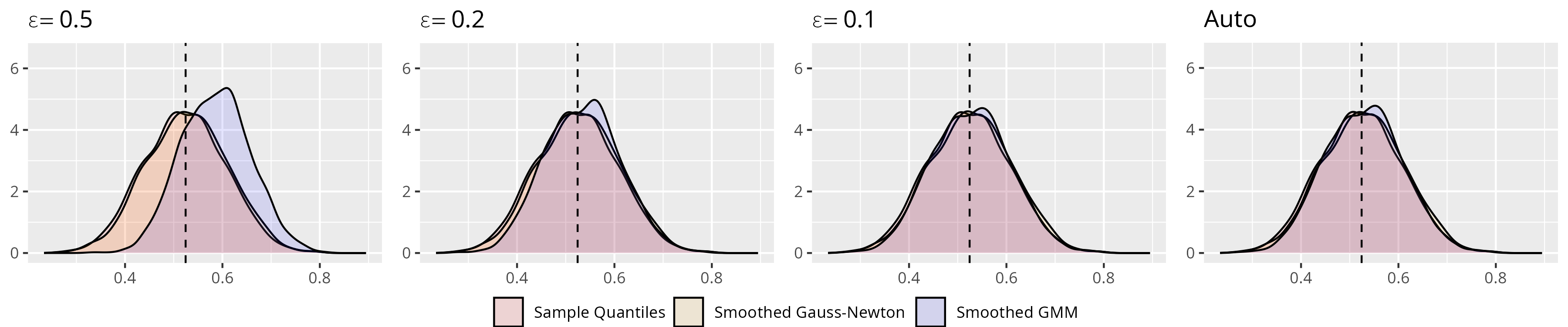}\\
  {\footnotesize \textbf{Legend:} Distribution of estimators for: Sample quantile = R built-in quantile estimator, \sgn = Algorithm \ref{algo:sgn}, Smoothed GMM = minimizing GMM loss with smooth moments. $x_i \sim \mathcal{N}(0,1)$, $n = 250$, quantile level $t= 0.7$. Auto: MSE minimizing bandwidth.   } 
\end{figure}
\begin{table}[ht]
  \centering \caption{Illustration: estimating a sample quantile}  \label{tab:quantile} \setlength\tabcolsep{4.0pt}
  \renewcommand{\arraystretch}{0.935} 
  {\small \begin{tabular}{r|c|ccc|ccc|ccc|ccc} 
    \hline \hline
  & & \multicolumn{3}{c|}{$\varepsilon=0.5$} & \multicolumn{3}{c|}{$\varepsilon=0.2$} & \multicolumn{3}{c|}{$\varepsilon=0.1$} & \multicolumn{3}{c}{Auto} \\ 
  & $\theta^\dagger$ & $\hat\theta_n$ & \sgn & s\textsc{gmm} &  $\hat\theta_n$ & \sgn & s\textsc{gmm} & $\hat\theta_n$ & \sgn & s\textsc{gmm} & $\hat\theta_n$ & \sgn & s\textsc{gmm} \\ 
    \hline
    avg & 0.524 & 0.523 & 0.529 & 0.586 & 0.523 & 0.528 & 0.535 & 0.523 & 0.528 & 0.527 & 0.523 & 0.528 & 0.529 \\ 
    std & - & 0.083 & 0.083 & 0.071 & 0.083 & 0.084 & 0.077 & 0.083 & 0.084 & 0.080 & 0.083 & 0.084 & 0.079 \\ 
    size$_1$ & - & 0.030 & 0.029 & 0.045 & 0.050 & 0.051 & 0.032 & 0.063 & 0.062 & 0.051 & 0.057 & 0.058 & 0.041 \\ 
    size$_2$ & - & 0.039 & 0.040 & 0.053 & 0.039 & 0.038 & 0.025 & 0.039 & 0.039 & 0.032 & 0.039 & 0.038 & 0.029 \\ 
     \hline \hline
  \end{tabular}}\\
  {\footnotesize \textbf{Legend:} \sgn = Algorithm \ref{algo:sgn}, s\textsc{gmm} = GMM with smoothed moments. Auto: MSE minimizing $\varepsilon$. avg, std = average and standard deviation of estimates. size$_1$ = rejection rate for testing $\theta = \theta^\dagger$ at 5\% significance level using standard errors from Algorithm \ref{algo:sgn}. size$_2$ = same as size$_{1}$ but with analytical standard errors, $se(\hat{\theta}_n) = \sqrt{t(1-t)/n}/f_{n,\varepsilon}(\hat\theta_n)$, and $\varepsilon = 1.06 \hat{\sigma}_x n^{-1/5}$. }
\end{table}

Figure \ref{fig:quantile} illustrates the difference between Algorithm \ref{algo:sgn} and smoothing the GMM objective. It compares estimates based on R's \textit{quantile} function\footnote{The sample quantile function in R returns an estimate obtained by a linear interpolation between order statistics, see e.g. \citet{hyndman1996} for details. As such, it does not exactly solve the moment condition $\overline{g}_n(\theta)=0$, and its finite sample properties can differ slightly from the GMM estimator.} with both methods. Estimates from Algorithm \ref{algo:sgn} and the sample quantiles are nearly identical, for all $\varepsilon$. The blue curve corresponds to $\hat\theta_{n,\varepsilon}$, its bias is visible when $\varepsilon = 0.5,0.2$. Since $n^{-1/4} = 0.25$, the choice $\varepsilon = 0.1$ represents the undersmoothing regime. The bandwidth minimizing the MSE criterion is also reported under Auto, it generally close to the smaller value, i.e. $\varepsilon \simeq 0.13$. Table \ref{tab:quantile} further illustrates the bias of the estimates and its implications for inference. Inferences using $\hat\theta_{n,\varepsilon}$ are size distorted compared to sample quantiles and Algorithm \ref{algo:sgn}. The size of the test also depends on the choice of bandwidth. The table also reports the size when standard errors are computed as in Algorithm \ref{algo:sgn} (size$_1$) or, as a benchmark, using the analytical expression and Silverman's rule of thumb to evaluate the unknown density term (size$_2$).

In this example, smoothing replaces the indicator function with the same $\Phi(\cdot/\varepsilon)$ for all $i$. This is generally not the case. Consider a quantile regression with moments $\overline{g}_n(\theta) = \frac{1}{n} \sum_{i=1}^n [\mathbbm{1}( y_i - x_i^\prime \theta )-t]w_i$,  $w_i$ are the instruments. The smoothed moments are $\overline{g}_{n,\varepsilon}(\theta) = \frac{1}{n} \sum_{i=1}^n [\Phi( \frac{y_i - x_i^\prime \theta}{\varepsilon\|x_i\|_2} )-t]w_i$. Here, the bandwidth is effectively $\varepsilon \|x_i\|_2$ which is individual specific. Similarly, the smoothed Jacobian is $G_{n,\varepsilon}(\theta) = \frac{-1}{n \varepsilon} \sum_{i=1}^n \frac{x_i^\prime}{\|x_i\|_2}\phi( \frac{y_i - x_i^\prime \theta}{\varepsilon\|x_i\|_2} )w_i$ which, compared to smoothing the indicator function directly, has the additional $\|x_i\|_2$ term in the denominator. Notice that all moments of $x_i/\|x_i\|_2$ are finite even when $x_i$ has infinite moments. 

\section{Main Results} \label{sec:main}
This section gives the main assumptions and derives the main results. The finite sample convergence results are derived. The role of the tuning parameters is discussed and the role of covering sequences $(\theta^b)_{b \geq 0}$ on the global convergence properties is explored in more details. 

\subsection{Assumptions}

\begin{assumption} \label{ass:gn} Suppose: 
  i. $\theta^\dagger \in \text{int}(\Theta)$, $\Theta \subset \mathbb{R}^{d_\theta}$ is compact and convex,
  ii. $g(\cdot) = \mathbb{E}[g(\cdot;x_i)]$ is continuously differentiable on $\Theta$,
  iii. the Jacobian $G(\theta^\dagger) := \partial_\theta g(\theta^\dagger)$ has full rank, 
  iv. there exists $L_G \geq 0$ such that for any $(\theta_1,\theta_2)\in \Theta^2$, $\|G(\theta_1)-G(\theta_2)\| \leq L_G \|\theta_1-\theta_2\|$,
  v. for all $\eta >0$, there exists $\delta(\eta) >0$ such that $\inf_{\|\theta-\theta^\dagger\| \geq \eta}\|g(\theta)\|_W \geq \delta(\eta),$ where $\delta(\cdot)$ is a continuous and weakly increasing function, vi. there exists $0<\underline{\lambda}_W \leq \overline{\lambda}_W < \infty$ such that $\underline{\lambda}_W \leq \lambda_{\min}(W) \leq \lambda_{\max}(W) \leq \overline{\lambda}_W$.
\end{assumption}

Assumption \ref{ass:gn} are the conditions required for the population quantities. Conditions ii, iii, vi imply $Q(\theta) = \|g(\theta)\|^2_W$ behaves like $\|\theta-\theta^\dagger\|^2$ around the solution $\theta^\dagger$. Conditions ii, iv hold in the particular case where $g(\cdot)$ is twice continuously differentiable with bounded second derivative. Condition iii and v are textbook local and global identification assumptions. These two conditions are particularly important for the results. In practice, it is important to check if the two conditions hold since they determine whether the optimization is well-posed and affect the speed of convergence. Often, this is difficult to verify analytically; \citet{forneron2024} provides a numerical procedure to detect whether these conditions fail.

\begin{assumption} \label{ass:sample_gn} Suppose: 
  i. for all $\theta \in \Theta$, $\mathbb{E}(\|g(\theta;x_i)\|^2) <\infty$,
  ii. for some $L_g \geq 0$, $\psi \in (0,1]$, and any $\delta>0$, $[\mathbb{E}(\sup_{\|\theta_1-\theta_2\|\leq \delta}\|g(\theta_1;x_i)-g(\theta_2;x_i)\|^2)]^{1/2} \leq L_g \delta^\psi$, iii. $W_n \overset{p}{\to} W$, there exists $0<\underline{\lambda}_W \leq \overline{\lambda}_W < \infty$ such that $\underline{\lambda}_W \leq \lambda_{\min}(W_n) \leq \lambda_{\max}(W_n) \leq \overline{\lambda}_W$, with probability 1.
\end{assumption}
Assumption \ref{ass:sample_gn} gives additional conditions for the sample moments. The assumptions allow for a variety of estimation problems, including simulated method of moments estimations with non-smooth moments. As an example, consider a simulated method of moments estimation of a static discrete choice model using:
\[ g(\theta;x_i) = \frac{1}{S} \sum_{s=1}^S z_i ( y_i - \mathbbm{1} \{ w_i^\prime \theta + u_i^s \geq 0 \} ), \]
where $x_i = (y_i,z_i,w_i,u_i^1,\dots,u_i^S)$, $y_i$ is a binary outcome, $z_i$ are instruments, $w_i$ regressors, $u_i^s$ are simulation draws and $S \geq 1$ the number of simulated samples. For a given value of $\theta$, $\mathbbm{1} \{ w_i^\prime \theta + u_i^s \geq 0 \}$ is a simulated discrete outcome. For this model, Assumption \ref{ass:sample_gn} i holds if $z_i$ has finite variance. Condition ii holds with $\psi = 1/2$ under additional moment conditions for $w_i,z_i$ and if $u_i^s$ has a continuous density. The GMM quantile regression discussed above also falls within this class. Other examples include moments that satisfy what is known as a stochastic Lipschitz condition: 
\[ \| g(\theta_1;x_i) - g(\theta_2;x_i) \| \leq L(x_i)\|\theta_1 - \theta_2\|, \text{ for any } \theta_1,\theta_2 \in \Theta, \]
where $L(x_i)$ has finite variance. This holds in the special case where $g(x_i;\theta)$ is smooth: i.e. continuously differentiable and its derivative is bounded by $L$, i.e. $\|\partial_\theta g(x_i;\theta)\| \leq L(x_i)$ for all $\theta$. For these models, condition i holds since $\Theta$ is bounded, and ii holds with $\psi = 1$. Assumption \ref{ass:sample_gn} i, ii is related to the so-called stochastic equicontinuity property. \citet[Sec4]{pakes1989}, \citet{andrews1994}, \citet{newey-mcfadden-handbook} list additional examples, including models which feature censoring, sample selection, simulated multinomial and dynamic discrete choice, and further discuss the so-called $L_2$-smoothness condition ii.


Under Assumptions \ref{ass:gn}-\ref{ass:sample_gn}, it can be shown that the approximate minimizer $\hat\theta_n$ is $\sqrt{n}$-consistent and asymptotically Gaussian:
\[ \sqrt{n}(\hat\theta_n - \theta^\dagger) \overset{d}{\to} \mathcal{N}(0,V), \]
where $V$ is the asymptotic variance of $\hat\theta_n$, see e.g. \citet[Th3.4]{newey-mcfadden-handbook}.

\subsection{Convergence Results}
Given Assumption \ref{ass:gn}, \ref{ass:sample_gn} above, the following first provides large sample convergence properties of Algorithm \ref{algo:sgn}. Then, several finite sample results leading to large sample ones are stated; they indicate which rate is optimal for the smoothing parameter $\varepsilon$. Results for the smooth population objective are given in Appendix \ref{sec:popu}, as a baseline without sampling uncertainty and non-smoothness. The results below are stated in terms of an approximate minimizer:
\begin{align}
  \hat\theta_n = \theta^\dagger - (G^\prime W_n G)^{-1} G W_n \overline{g}_n(\theta^\dagger), \label{eq:theta_hat}
\end{align}
where $G = G(\theta^\dagger)$, it can be shown to satisfy (\ref{eq:approxi_min}).

\begin{theorem} \label{th:sgn_global00} Suppose Assumptions \ref{ass:gn}-\ref{ass:sample_gn} hold and $D(\theta^0,\dots,\theta^b) \to 0$ as $b \to \infty$. Take $\gamma \in (0,1)$, there exists $k \geq 0$ such that as $n\to \infty$, $\varepsilon \to 0$, and $\sqrt{n}\varepsilon \to \infty$:
\begin{align} \sqrt{n}\|\theta_{k+j}- \hat\theta_n\| = o_p(1), \label{eq:asymequiv} \end{align}
for any $j \geq 1$ such that $j/\log(n) \to \infty$. For the same choice of $k$ and $j$:
\begin{align} \sqrt{n}(\theta_{k+j} - \theta^\dagger) \overset{d}{\to} \mathcal{N}(0,V), \label{eq:asymnorm} \end{align}
  where $V = (G^\prime W G)^{-1} G^\prime W \Sigma W G  (G^\prime W G)^{-1}$, $G = G(\theta^\dagger)$, $\Sigma = \text{var}[g(x_i;\theta^\dagger)]$.
\end{theorem}

Theorem \ref{th:sgn_global00} (\ref{eq:asymequiv}) implies that after $b=k+j$ iterations, the iterate $\theta_b$ from Algorithm \ref{algo:sgn} is an approximate minimizer of the GMM objective and (\ref{eq:asymnorm}) implies that it is a $\sqrt{n}$-asymtotically normal estimator of the parameters of interest $\theta^\dagger$. As in the last step of Algorithm \ref{algo:sgn}, given an appropriate estimator $\Sigma_n$ of $\Sigma$, the smoothed Jacobian $G_{n,\varepsilon}(\tilde{\theta}_n)$ can be used to compute standard errors and make inferences on the true $\theta^\dagger$. 

There are two phases in the convergence which contribute to the required number of iterations $b = k+j$. 
First, the Algorithm performs a global search. For a well-chosen covering sequence, $D(\theta^0,\dots,\theta^b) \to 0$ implies that there exists $k$ such that the global step necessarily produces a $\theta^\ell$, $0 \leq \ell \leq k$, that is close enough to $\hat{\theta}_n$ for 1) the local Gauss-Newton step to be convergent, and 2) there are no value away from $\hat{\theta}_n$ that can improve the fit -- i.e. if a $\theta^b$ lowers the objective value in the global step, with $b \geq k$, then it is also locally convergent. This implies that the Algorithm remains local after at most $k$ iterations. Importantly, 1) hinges on the local identification condition, and 2) on the global identification condition  (respectively Assumptions \ref{ass:gn} iii. and v.). 

After those $k$ iterations, Algorithm \ref{algo:sgn} automatically transitions into a fast local convergence phase. Only an additional $j$ iteration, increasing slowly with $n$, are required to produce asymptotically valid estimates that satisfy (\ref{eq:asymequiv}) and (\ref{eq:asymnorm}). The main bottleneck of the Algorithm is typically the finite number of iterations $k$, which depends on the GMM objective and the choice of covering sequence. The latter is now discussed in more detail before presenting several finite sample local convergence results leading to the main Theorem above. 

\paragraph{Choice of covering sequence.} What is particularly attractive in this result is the automatic transition from $k$ global steps to the $j$ local ones without user input. Because $k$ only depends on the GMM objective and $(\theta^\ell)_{\ell \geq 0}$, the Algorithm adapts to the complexity of the optimization problem in a single run. Notice that this property comes from using the local and global steps concurrently. Methods that use them successively, as in e.g. \citet{robinson1988}, \citet{andrews1997} do not have this property as explained in Appendix \ref{apx:comp}. 


To get good convergence results, the covering sequence should reach a neighborhood of $\hat{\theta}_n$ using as few iterations $k$ as possible. This is measured by dispersion $D$ considered in the choice of covering sequence above. For each $k \geq 1$, covering number arguments imply that there exists $(\theta^0,\dots,\theta^k)$ which minimize the dispersion with $D(\theta^0,\dots,\theta^k) = O(k^{-1/d_\theta})$. The problem is that, for $k+1$, the optimal set $(\theta^0,\dots,\theta^k,\theta^{k+1})$ does not use the previous $(\theta^0,\dots,\theta^k)$. The optimal cover for a given $k$ does not necessarily define a proper sequence over $k$. Importantly, the ordering of the points in the sequence matters to ensure the discrepancy $D(\theta^0,\dots,\theta^k)$ is as small as possible for each $k \geq 1$. 

In the special case where $\Theta = [0,1]^{d_\theta}$, \citet[Theorem 3]{niederreiter1983} showed that low-discrepancy sequences, such as the recommended Sobol or Halton sequence, satisfy:
\[ D_{k-1} \leq O( \sqrt{d_\theta} \log(k)k^{-1/d_\theta}), \]
which is close to the theoretical lower bound $O(k^{-1/d_\theta})$ at each $k$. In comparison, his result implies that uniform random draws $\theta^j \overset{iid}{\sim} \mathcal{U}_{\Theta}$ only achieve $D_{k-1} = O_p(k^{-1/[2d_\theta]})$, which is slower. Differences between random and low dispersion sequences are illustrated in Figure \ref{fig:cover}, Appendix \ref{apx:cover}. Finally, although $\Theta$ needs to be bounded to have $k < \infty$, it need not be finite-dimensional. For instance, \citet[Th1]{gallant1987} prove compactness of the Hermite polynomial density space. This implies existence of a finite cover, a sufficient condition for $k < \infty$. Finite covers also exist for the class of smooth functions on bounded convex sets \citep[Th2.7.1]{VanderVaart1996}. Constructing a covering sequence with low-dispersion in these settings could be challenging, however. Additionally, \citet{chen2015} consider smoothing of the Jacobian in an non-parametric setting, as required for the local step. Extensions to the infinite-dimensional setting are not considered here.

\paragraph{Local Convergence.} The following considers the convergence properties of local step only:
\begin{align} \theta_{b+1} = \theta_b - \gamma [G_b^\prime W_n G_b]^{-1} G_b^\prime W_n \overline{g}_n(\theta_n), \quad b=0,1,\dots, \label{eq:step1} \end{align}
where $G_b = G_{n,\varepsilon}(\theta_b)$. As noted earlier, the Gauss-Newton updates rely on a smoothed Jacobian to approximate an infeasible derivative when the problem is non-smooth. Local convergence requires $G_{n,\varepsilon}(\hat{\theta}_n)^\prime W_n \overline{g}_n(\hat{\theta}_n)$ to be as small as possible, in order to behave like a first-order condition for the minimization problem. To this end, Lemma \ref{lem:smooth_foc} below measures the stability of (\ref{eq:step1}) at $\theta=\hat\theta_n$.

\begin{lemma} \label{lem:smooth_foc} 
Suppose Assumptions \ref{ass:gn}-\ref{ass:sample_gn} hold, then for any $\varepsilon >0$ and $c_n \geq 1$:
\[ \|G_{n,\varepsilon}(\hat\theta_n)^\prime W_n \overline{g}_n(\hat\theta_n)\| \leq C_{1} (c_nn^{-1/2})^{1+\psi} \left( 1 + \frac{c_n n^{-1/2}}{\varepsilon} + \frac{\varepsilon}{(c_nn^{-1/2})^\psi} \right) := \Gamma_{n,\varepsilon}, \]
with probability $1-(1+C)/c_n$ for some universal constant $C$, and $C_1$ which only depends on $p=\text{dim}(\overline{g}_n)$, $\Theta$, $\Sigma = \text{var}[g(\theta^\dagger;x_i)]$, and the constants in Assumptions \ref{ass:gn}-\ref{ass:sample_gn}.
\end{lemma}

The term $\Gamma_{n,\varepsilon}$ measures how close $\hat\theta_n$ is to being a fixed-point of (\ref{eq:step1}).\footnote{In just-identified models for which an exact solution $\overline{g}_n(\hat\theta_n)=0$ exists, it is possible to set $\Gamma_{n,\varepsilon}=0$ by using the exact minimizer.} Here $c_n \geq 1$ controls the probability level for which the bound holds over repeated samples, the bound is informative for $c_n \geq 1+C$. It will be useful for understanding the relation between the probability of a successful optimization and the choice of tuning parameters. The bandwidth $\varepsilon$ appears in $\Gamma_{n,\varepsilon}$ in both a numerator and a denominator. The optimal bandwidth for minimizing $\Gamma_{n,\varepsilon}$ is $\varepsilon \asymp (c_n n^{-1/2})^{(1+\psi)/2}$, i.e. $\varepsilon = O(n^{-1/2})$ for the continuous case ($\psi=1$), and $\varepsilon = O(n^{-3/8})$ for the discontinuous one ($\psi=1/2$). Using smoothed instead of sample moments in Algorithm \ref{algo:sgn} would require to substitute $\overline{g}_n(\hat\theta_n)$ for $\overline{g}_{n,\varepsilon}(\hat\theta_n)$ in Lemma \ref{lem:smooth_foc} above. Then, two additional terms would appear in $\Gamma_{n,\varepsilon}$, respectively of order $\varepsilon^2$ and $\sqrt{\varepsilon}n^{-1/2}$. In particular, $\varepsilon^2$ would add smoothing bias to the estimates. 

In the following, $\sigma_{\min}$ denotes the smallest singular value, i.e. $\sigma_{\min}[G(\theta)] = \sqrt{ \lambda_{\min}[G(\theta)^\prime G(\theta)] }$ where $\lambda_{\min}$ returns the smallest singular value of the square matrix $G(\theta)^\prime G(\theta)$.

\begin{proposition} \label{prop:local_conv0} Suppose Assumptions \ref{ass:gn}-\ref{ass:sample_gn} hold. Take $\gamma \in (0,1)$ and $\overline{\gamma} \in (0,\gamma)$, $c_n \geq 1$. Let $0 < \underline{\sigma} < \sigma_{\min}[G(\theta^\dagger)]$, $R_G>0$ be such that $\|\theta-\theta^\dagger\| \leq R_G \Rightarrow \underline{\sigma} \leq \sigma_{\min}[G(\theta)]$, and $\kappa_W = \overline{\lambda}/\underline{\lambda} \geq 1$. There are constants $C_a,C_{\sigma},M_{1,Z},C_2$ and:
\[ R_{n,G} = R_G - 2 C_{a} c_n n^{-1/2}, \quad R_{n,\varepsilon} = \left( \frac{\gamma- \overline{\gamma}}{\gamma} - M_{1,Z}\varepsilon \right) \frac{\underline{\sigma}_{n,\varepsilon}}{\sqrt{\kappa_W}L_G}. \]  
such that, with probability $1-(1+C)/c_n$, for all $\|\theta_b-\hat{\theta}_n\| \leq \min(R_{n,G},R_{n,\varepsilon})$:
\begin{align}
  \|\theta_{b+1}-\hat\theta_n\| \leq (1-\overline{\gamma})\|\theta_b - \hat\theta_n\| + \gamma \Delta_{n,\varepsilon}(\|\theta_b-\hat\theta_n\|), \label{eq:contract_ns}
\end{align}
where: $\Delta_{n,\varepsilon}(\|\theta_b-\hat\theta_n\|) \leq \frac{C_2}{\underline{\sigma}_{n,\varepsilon}^2} \left( \Gamma_{n,\varepsilon} +  \frac{(c_n n^{-1/2})^2}{\varepsilon}\|\theta_b-\hat\theta_n\|^\psi + \frac{c_n n^{-1/2}}{\varepsilon}\|\theta_b-\hat\theta_n\| \right)$\\ and $\underline{\sigma}_{n,\varepsilon} = \underline{\sigma} - C_\sigma[ \frac{c_n n^{-1/2}}{\varepsilon} + \varepsilon]$.
\end{proposition}

Proposition \ref{prop:local_conv0} gives the local convergence properties of a single iteration of the local step. Recall that the local step relies on $G_{n,\varepsilon}$ to linearly approximate the moments around $\theta_b$. Since the moments are non-smooth, the approximation does not become more accurate as $\theta$ approaches $\theta_b$. The approximation does become more accurate as the sample size $n$ increases and the bandwidth $\varepsilon$ decreases. With this in mind, the result implies that, with high probability, the local step produces a $\theta_{b+1}$ which reduces the distance to $\hat{\theta}_n$ by a factor $(1-\overline{\gamma})$ up to an approximation error, measured by $\Delta_{n,\varepsilon}$ which involves non-smoothness of the moments and smoothing bias in the Jacobian. Note that, unlike e.g. \citet{dennis1996} in the smooth case, the derivations do not rely on local convexity arguments which would define a different local neighborhood. For instance, the quantile problem $[\Phi(\theta)-t]^2$ is globally non-convex. Nevertheless, for the quantile example there always exists a $\underline{\sigma}$ such that $R_G = \Theta$ when it is bounded since the density $\phi$ is strictly positive everywhere.

The local convergence neighborhood is of size $\min(R_{n,G},R_{n,\varepsilon})$ which depends on several factors: features of the moments $R_G$, $\underline{\sigma}$, $L_G$, $\kappa_W$, and tuning parameters $\varepsilon,\gamma$.\footnote{The constants $C_a,C_{\sigma},M_{1,Z},C_2$ are made more explicit in Proposition \ref{prop:local_conv} in the Appendix.} $R_G$ measures the area where the singular values of $G$ are a least $\underline{\sigma}$, a measure of local identification. $L_G$ measures the non-linearity of the moments; $L_G=0$ implies $g(\cdot)$ is linear in $\theta$. $\kappa_W$ bounds the condition number of $W_n$; setting $W_n = I_d$ yields $\kappa_W=1$. A larger $L_G,\kappa_W$ and/or a smaller $R_G,\underline{\sigma}$ makes the area of convergence smaller. Very non-linear models (large $L_G$) or less sharply identified (small $R_G,\underline{\sigma}$) are harder to estimate since the area of local convergence is reduced. Using the optimal weighting matrix is more efficient from a statistical perspective. However, if it is poorly conditioned (large $\kappa_W$) then estimation is more difficult. This motivates less efficient choices such as $W = (\Sigma + \lambda)^{-1}$ where $\Sigma = \text{var}[g(x_t;\theta^\dagger)]$ and $\lambda > 0$ is a regularization parameter which brings $\kappa_W$ closer to $1$.

The bandwidth $\varepsilon$ also affects the area of convergence via $R_{n,\varepsilon}$. If $\varepsilon \to 0$ and $\sqrt{n}\varepsilon \to 0$, then $\min(R_{n,G},R_{n,\varepsilon}) = R - o(1)$, where $R = \min(R_G,[\gamma-\overline{\gamma}]\underline{\sigma}/[\gamma L_G \sqrt{\kappa_W}])$ is the area of local convergence in the smooth population problem (see Lemma \ref{lem:cv_local}). A larger learning rate $\gamma$ allows for faster convergence, measured by $\overline{\gamma} < \gamma$, but over a smaller neighborhood. For smooth problem, it is possible to have $\gamma \nearrow 1$ as $\theta_b$ approaches $\hat{\theta}_n$ to make convergence faster. Here, a larger $\gamma$ increases the effect of $\Delta_{n,\varepsilon}$ which can make local optimization less stable. Hence, the recommendation to use a smaller $\gamma$.

The optimal choice of bandwidth, for maximizing the area of local convergence $R_{n,\varepsilon}$, is $\varepsilon = O( n^{-1/4} )$. This choice of bandwidth is larger than for minimizing $\Gamma_{n,\varepsilon}$ above. Lastly, a smaller value $\underline{\sigma}$ makes Algorithm \ref{algo:sgn} more sensitive to the choice of $\varepsilon$ given its role in $\underline{\sigma}_{n,\varepsilon}$. The optimal choice for maximizing $\underline{\sigma}_{n,\varepsilon}$ is also $\varepsilon = O( n^{-1/4} )$. 

\begin{theorem} \label{th:sgn_local0} Assume, without loss of generality, that $R_{n,\varepsilon} \leq R_{n,G}$. Take $\gamma \in (0,1)$, $\overline{\gamma} \in (0,\gamma)$. Suppose Assumptions \ref{ass:gn}-\ref{ass:sample_gn} hold, for $\varepsilon$, $c_nn^{-1/2}$ small enough, with probability $1-(1+C)/c_n$, uniformly in $\|\theta_0-\hat\theta_n\| \leq R_{n,\varepsilon}$, for all $b \geq 0$:
  \begin{equation}
  \begin{aligned} \|\theta_{b}-\hat\theta_n\| \leq &\Big(1-\overline{\gamma} \Big)^b \|\theta_0-\hat\theta_n\| +  \frac{\gamma}{\overline{\gamma}}\frac{C_2}{\underline{\sigma}_{n,\varepsilon}^2} \Big(\Gamma_{n,\varepsilon}  + C_{n,\varepsilon}(\psi) \Big), \label{eq:th1_main} 
  \end{aligned}
  \end{equation}
  where $C_{n,\varepsilon}(\psi) = 0$ for $\psi = 1$ and $\sqrt{n}C_{n,\varepsilon}(\psi) = o(1)$ if $\varepsilon = o(1)$ and $\sqrt{n}\varepsilon \to +\infty$. For the same $c_n = O(1)$,  
    $b \geq \frac{\log(\Gamma_{n,\varepsilon})-\log(\|\theta_0-\hat\theta_n\|)}{\log(1-\overline{\gamma}/2)}$ implies $\sqrt{n}\|\theta_{b}-\hat\theta_n\| = o_p(1).$ 
\end{theorem}

While Proposition \ref{prop:local_conv0} derived properties for a given iteration of the local step, Theorem \ref{th:sgn_local0} describes the full optimization path $(\theta_0,\theta_1,\dots)$ from a local starting value $\theta_0$. The final step of Algorithm \ref{algo:sgn} returns $\tilde \theta_n$ such that $\|\overline{g}_n(\tilde\theta_n)\|_{W_n} = \min_{j=0,\dots,b} \|\overline{g}_n(\theta_j)\|_{W_n} \leq \|\overline{g}_n(\theta_b)\|_{W_n}$. 



The Theorem states that only $b = O(|\log[\Gamma_{n,\varepsilon}]|) = O(\log[n])$ iterations are needed to achieve $\|\theta_b-\hat\theta_n\| = O_p(\Gamma_{n,\varepsilon}) = o_p(n^{-1/2})$ if $\varepsilon \to 0$ and $\sqrt{n}\varepsilon \to 0$. The term $\Gamma_{n,\varepsilon}$ captures, among other things, the smoothing bias in the estimates. Using an optimal rate $\varepsilon = O(n^{-1/4})$ yields $\Gamma_{n,\varepsilon} = O_p(n^{-3/4})$ for both the continuous ($\psi=1$) and discontinuous case ($\psi=1/2$). This is negligible compared to sampling uncertainty, as required to prove (\ref{eq:asymnorm}) in Theorem \ref{th:sgn_global00}. 

In general, the local step alone does not guarantee convergence. Theorem \ref{th:sgn_global}, Appendix \ref{apx:proofs_main}, extends Theorem \ref{th:sgn_local0} to a global convergence result when using both global and local steps.

\section{Extensions of the Main Results} \label{sec:extend}
This section provides two extensions for the local step in Algorithm \ref{algo:sgn} and Proposition \ref{prop:local_conv0}. The first extension adds momentum to (\ref{eq:step1}), a.k.a. the Polyak heavy-ball \citep{polyak1964}. It is commonly used in stochastic gradient descent to accelerate convergence. Here it can be used to either accelerate convergence or maintain the rate of convergence while reducing the effect of sampling noise on optimization. The second extension builds a quasi-Newton approximation of $G_{n,\varepsilon}$ which is useful when the smoothed Jacobian does not have closed-form.
\subsection{Momentum: Acceleration or Noise Reduction} \label{sec:accelerate}
The first extension considers a modification of the local search step:
\begin{align}
  \theta_{b+1} = \theta_b - \gamma (G_b^\prime W_n G_b)^{-1}G_b^\prime W_n \overline{g}_n(\theta_b) + \alpha(\theta_b-\theta_{b-1}), \tag{\ref{eq:step1}'}
\end{align}
where $\theta_{-1} = \theta_0$, $G_b = G_{n,\varepsilon}(\theta_b)$ and $\alpha \in [0,1)$ is the momentum parameter. Two new quantities affect convergence, the companion matrix $A(\gamma,\alpha)$ and the effective rate $\gamma(\alpha)$:
  \[ A(\gamma,\alpha) = \left( \begin{array}{cc} 1-\gamma + \alpha & -\alpha\\ 1 & 0\end{array} \right),\quad 1-\gamma(\alpha) = \sigma_{\max}[A(\gamma,\alpha)],\]
For any $\gamma \in (0,1)$, there exists $\alpha \in (0,1]$ such that $\gamma(\alpha) > \gamma$, and a $\alpha^\star$ which maximizes $\gamma(\alpha)$. Table \ref{tab:alpha_star} provides a selection of combinations $(\gamma,\alpha^\star)$. The last row measures $\gamma/\overline{\gamma(\alpha)} < 1$ which implies that setting $\overline{\gamma}/\gamma > 1$ is now possible: momentum can affect the convergence rate / sensitivity to sampling noise tradeoff in Proposition \ref{prop:local_conv0}.

\begin{table}[ht] \centering \caption{Values of $\gamma$ and optimal choice of $\alpha$} \label{tab:alpha_star}
  \setlength\tabcolsep{4.5pt}
  \renewcommand{\arraystretch}{0.935} 
  {\small \begin{tabular}{r|cccccccc}
    \hline \hline
    $\gamma$ & 0.01 & 0.05 & 0.1 & 0.2 & 0.3 & 0.4 & 0.6 & 0.8\\ \hline
    $\alpha^\star$ & 0.81 & 0.60 & 0.47 & 0.31 & 0.21 & 0.14 & 0.05 & 0.01\\
    $\gamma(\alpha^\star)$ & 0.10 & 0.22 & 0.32 & 0.45 & 0.54 & 0.63 & 0.77 & 0.89\\ 
    $\gamma/\gamma(\alpha^\star)$ & 0.10 & 0.22 & 0.32 & 0.45 & 0.55 & 0.63 & 0.78 & 0.90\\
    \hline \hline
  \end{tabular}}
\end{table}

Notice that $\theta_{b+1} - \alpha(\theta_b-\theta_{b-1}) = \theta_b - \gamma (G_b^\prime W_n G_b)^{-1}G_b^\prime W_n \overline{g}_n(\theta_b)$ is the same as (\ref{eq:step1}) in Proposition \ref{prop:local_conv0}, so convergence of (\ref{eq:step1}') follows from the same derivations. The results are given for $\boldsymbol{\theta_b} = (\theta_b^\prime,\theta_{b-1}^\prime)^\prime$ to denote $\theta_b$ and its lagged value $\theta_{b-1}$; similarly $\boldsymbol{\hat\theta_n} = (\hat\theta_n^\prime,\hat\theta_n^\prime)^\prime$.

\begin{proposition} \label{prop:local_conv2}
  Suppose Assumptions \ref{ass:gn}-\ref{ass:sample_gn} hold. Take $\gamma \in (0,1), \alpha \in [0,1)$ and $\overline{\gamma} \in (0,\gamma(\alpha))$. Take $c_n \geq 1$ and $R_G$ as in Lemma \ref{lem:cv_local}. Uniformly in $\theta_{b} \in \Theta$ such that $\|\theta_b - \theta^\dagger\| \leq R_{n,G}:=R_G - C_a c_n n^{-1/2}$:
  \begin{align}
  \|\boldsymbol{\theta_{b+1}} - \boldsymbol{\hat\theta_n} \| \leq &\left(1-\gamma(\alpha) + \gamma \underline{\sigma}_{n,\varepsilon}^{-1} \sqrt{\kappa_W} L_G\|\theta_b - \hat\theta_n\| + M_{1,Z}\varepsilon \right) \|\boldsymbol{\theta_{b}} - \boldsymbol{\hat\theta_n} \| \notag \\  &+ \gamma \Delta_{n,\varepsilon}(\|\theta_b-\hat\theta_n\|) 
  \end{align}
  with probability $1-(1+C)/c_n$. The remainder $\Delta_{n,\varepsilon}$ is the same as in Proposition \ref{prop:local_conv}.
  Furthermore, if $\theta_b$ is such that:
  \begin{align}
    \|\theta_b-\hat\theta_n\| \leq \left( \frac{\gamma(\alpha)- \overline{\gamma}}{\gamma} - M_{1,Z}\varepsilon \right) \frac{\underline{\sigma}_{n,\varepsilon}}{\sqrt{\kappa_W}L_G} := R_{n,\varepsilon}(\alpha), \label{eq:radius2}
  \end{align}
  then, with probability $1-(1+C)/c_n$:
  \begin{align}
    \|\boldsymbol{\theta_{b+1}} - \boldsymbol{\hat\theta_n} \| \leq \left(1-\overline{\gamma} \right) \|\boldsymbol{\theta_{b}} - \boldsymbol{\hat\theta_n} \| + \gamma \Delta_{n,\varepsilon}(\|\theta_b-\hat\theta_n\|). \label{eq:contract_ns2}
  \end{align}

\end{proposition}
The convergence is now stated in terms of $\boldsymbol{\theta_{b+1}}-\boldsymbol{\hat\theta_n}$, similar to a first-order vector autoregression.  The main advantage of using momentum is that it can allow to maintain the rate of convergence but using a smaller learning rate $\gamma$. This can potentially result is a larger area of local convergence, without sacrificing speed of local convergence. 

\subsection{A Monte Carlo quasi-Newton Implementation} \label{sec:sqn}

In many empirical applications, the smoothed Jacobian $G_{n,\varepsilon}$ is not available in closed form. The following proposes a computationally attractive approximation.

\begin{algorithm}[h] 
  \caption{quasi-Newton approximation $\hat{G}_b$ of $G_{n,\varepsilon}(\theta_b)$} \label{algo:sqn} 
        \begin{algorithmic}
        \State 1) \textbf{Input:} $L \geq d_\theta$,
        \State 2) \textbf{Moment Update:}
        \If{$b=0$}  \Comment{Initialization}
        \State draw $Z_{-\ell} \sim \mathcal{N}(0,I_{d_\theta})$, $\ell = 0,\dots,L-1$
        \State compute $Y_{-\ell} = \frac{1}{\varepsilon}[\overline{g}_n(\theta_{0} + \varepsilon Z_{-\ell})-\overline{g}_n(\theta_{0})]$
        \Else  \Comment{Update}
          \State draw $Z_{b} \sim \mathcal{N}(0,I_{d_\theta})$
          \State compute $Y_b = \frac{1}{\varepsilon}[\overline{g}_n(\theta_{b} + \varepsilon Z_{b})-\overline{g}_n(\theta_{b})]$
        \EndIf
        \State 3) \textbf{Least-Squares Approximation:}
        \State de-mean $\tilde{Z}_{b-\ell} = Z_{b-\ell} - \sum_{\ell=0}^{L-1} Z_{b-\ell}/L$
        \State compute $\hat G_L(\theta_b) = \sum_{\ell=0}^{L-1} Y_{b-\ell} \tilde{Z}_{b-\ell}^\prime \left( \sum_{\ell=0}^{L-1} \tilde{Z}_{b-\ell} \tilde{Z}_{b-\ell}^\prime  \right)^{-1}$
        \end{algorithmic}
\end{algorithm}
Algorithm \ref{algo:sqn} involves an additional tuning parameter: $L \geq d_\theta$. A larger value of $L$ reduces the Monte Carlo error but involves more lagged values $\theta_{b-\ell}$, $\ell = 0,\dots,L-1$, which slows convergence. The algorithm only adds one moment evaluation per iteration, rather than $L \gg 1$ for a direct Monte Carlo implementation. The theoretical results require $L \to \infty$, but the divergence can be slow. Updating several directions at each $b$ could speed-up convergence as it improves the approximation quality.\footnote{Updating several directions in parallel, e.g. $2$ or $4$, can improve the finite-sample optimization properties because, for the same choice of $L$, $\hat{G}_b$ depends on fewer lagged values, e.g. $L/2$ or $L/4$.} For simplicity, the results are only derived for the sample mean estimator: $\hat G_L(\theta_b) = \frac{1}{L} \sum_{\ell=0}^{L-1} \frac{1}{\varepsilon} [ \overline{g}_n(\theta_{b-\ell} + \varepsilon Z_{b-\ell}) - \overline{g}_n(\theta_{b-\ell}) ]Z_{b-\ell}^\prime$.
The following extends Proposition \ref{prop:local_conv0} when $\hat{G}_{L}(\theta_b)$ is used in (\ref{eq:step1}). It is assumed that the Algorithms runs for at most $1 \leq b_{\max} < \infty$ iterations (a stopping criterion).
\begin{proposition} \label{prop:local_conv3}
  Suppose Assumptions \ref{ass:gn}-\ref{ass:sample_gn} hold. There exists $0 < \hat{R}_{G} \leq R_G$ such that uniformly in $\mathcal{E}_b := (\max_{-L+1 \leq \ell \leq 0} \|\theta_{b-\ell} - \hat\theta_n\|) \leq \hat{R}_{G}$, we have with probability $1-(5+C)/c_n$:
  \begin{align}
    \|\theta_{b+1}-\hat\theta_n\| &\leq (1-\gamma + \gamma \hat{\underline{\sigma}}_{n,\varepsilon}^{-1}\sqrt{\kappa_W}L_G[ \|\theta_b -\hat\theta_n\| + \mu_{d_\theta} \mathcal{E}_b ] )\|\theta_b - \hat\theta_n\| \notag\\& \quad + \frac{\gamma}{\hat{\underline{\sigma}}_{n,\varepsilon}^2} \hat{\Delta}_{n,\varepsilon}(\|\theta_b-\hat\theta_n\|,\mathcal{E}_b), \label{eq:Eb}
  \end{align}
  where $\mu_{d_\theta} = \mathbb{E}(\|ZZ^\prime - I_{d_\theta}\|)$, $\hat{\underline{\sigma}}_{n,\varepsilon} = \underline{\sigma}/2 - C_{\sigma,2}(c_nn^{-1/2}\varepsilon^{\psi-1} + \varepsilon + L^{-1/2})\delta_n^{3/2}$, and $\delta_n = \log(c_n) + \log(b_{\max}+L+1)$. The remainder has the form:
  \begin{align*}
    \hat{\Delta}_{n,\varepsilon}(\|\theta_b-\hat\theta_n\|,\mathcal{E}_b) = C_3 \left( \hat{\Gamma}_{n,\varepsilon} + \delta_n^{3/2}[c_nn^{-1/2}\varepsilon^{\psi-1} + \varepsilon + L^{-1/2}]\mathcal{E}_b + c_nn^{-1/2}\|\theta_b-\hat\theta_n\|^\psi  \right),
\end{align*}
with $\hat{\Gamma}_{n,\varepsilon} = c_nn^{-1/2} \left[ (c_nn^{-1/2})^\psi + \delta_n^{3/2}(c_nn^{-1/2}\varepsilon^{\psi-1} + \varepsilon + L^{-1/2}) \right].$
Suppose that:\\ $\mathcal{E}_b \leq \frac{\overline{\gamma}-\gamma}{\gamma} \frac{\hat{\underline{\sigma}}_{n,\varepsilon}}{L_G \sqrt{\kappa_W}(1+\mu_{d_\theta})} := \hat{R}_{n,\varepsilon}$,
then with probability $1-(5+C)/c_n$, we have:
\[ \|\theta_{b+1}-\hat\theta_n\| \leq (1-\overline{\gamma})\|\theta_b-\hat\theta_n\| + \frac{\gamma}{\hat{\underline{\sigma}}_{n,\varepsilon}^2} \hat{\Delta}_{n,\varepsilon}(\|\theta_b-\hat\theta_n\|,\mathcal{E}_b). \]
\end{proposition}
Compared to Proposition \ref{prop:local_conv0}, the number of iterations $b_{\max}$ now plays a role in the results. This is because $\hat{G}_{L}(\theta_b)$ is computed by Monte Carlo and the worst-case approximation error can be made arbitrarily large over infinitely many iterations. The dependence is only logarithmic because $\hat G_L(\theta_b) \simeq \frac{1}{L} \sum_{\ell=0}^{L-1} \partial_\theta g(\theta_{b-\ell})Z_{b-\ell} Z_{b-\ell}^\prime,$
where $Z_{b-\ell} Z_{b-\ell}^\prime$ are iid Wishart distributed. Relying on properties of this distribution, the proof derives exponential tail bounds for the singular values of $\hat{G}_b$. This allows $b_{\max}$ to increase linearly or polynomially in $n$ so that the Monte Carlo error term $\delta_n$ only diverges logarithmically.

The probability bound is now $1-(5+C)/c_n$ to account for the Monte Carlo error. Comparing $\Delta_{n,\varepsilon}$ with $\hat{\Delta}_{n,\varepsilon}$ in Propositions \ref{prop:local_conv0}, \ref{prop:local_conv3} highlights the importance of the lags, measured by $\mathcal{E}_b$, in the convergence. The term $\hat{\Gamma}_{n,\varepsilon}$ depends on $(nL)^{-1/2}$ so that $L \to \infty$ is required. This divergence can, in theory, be arbitrarily slow. In practice, setting $L \geq \max(25,1.5 \times d_\theta)$ typically yields good results. 

The proof of the Proposition highlights several factors which can affect convergence when using $\hat{G}_L$. Algorithm \ref{algo:sqn} relies on Monte Carlo draws $Z_{b-\ell}$ and past updates $Y_{b-\ell}$, two sources of approximation. If the matrix $G_{n,\varepsilon}^\prime W_n G_{n,\varepsilon}$ is ill-conditioned, because $\underline{\sigma}$ is small, the inverse $(\hat{G}_L(\theta_b)^\prime W_n \hat{G}_L(\theta_b))^{-1}$ can be sensitive to Monte Carlo noise. This is measured by $\hat{\sigma}_{n,\varepsilon}$ in the Proposition. Also, if $\theta_b$ changes rapidly, $\theta_{b-\ell} - \theta_b$ is large and past values $Y_{b-\ell}$ give a poor approximation of the current $G_{n,\varepsilon}(\theta_b)$ when the underlying moments $\overline{g}_{n}$ are very non-linear. 

\paragraph{Regularization.} To counter these two issues, the Levenberg-Marquardt (LM) algorithm which inverts $(\hat{G}_L(\theta_b)^\prime W_n \hat{G}_L(\theta_b) + \lambda_b I_d)$ instead of $(\hat{G}_L(\theta_b)^\prime W_n \hat{G}_L(\theta_b))$ in the local step can be particularly useful. Here, $\lambda_b$ penalizes for the quality of the approximation $\hat{G}_L(\theta_b) - G_{n,\varepsilon}(\theta_b)$. The applications use $\lambda_b = \tilde{\lambda} \textsc{rmse}_b \times \|W \hat{G}_{L}(\theta_b)\|$ where $\textsc{rmse}_b$ is the root mean squared error in the regression of $Y_{b-\ell}$ on $\tilde{Z}_{b-\ell}$ and $\tilde{\lambda} = 10^{-3}$. This is implemented using R in Appendix \ref{sec:Code}. When the moments are linear, $\textsc{rmse}_b = 0$  and $\lambda_b = 0$  for all $b$. When the moments are more non-linear, $\textsc{rmse}_b$ and $\lambda_b$ are larger. This shrinks the size of the update $\theta_{b} - \theta_{b-1}$. Increased stability comes at the cost of slower convergence.  In simulated examples below, this also helps stabilize the local step when $\varepsilon$ is chosen to be very small. 

 
\section{Simulated and Empirical Examples} \label{sec:MC_Emp}
\subsection{Dynamic Discrete Choice} \label{sec:ddc}

The first simulated example adapts an example from \citet{bruins2018}. The model is a simple panel dynamic discrete choice model. The data generating process (dgp) is given by:
\begin{align*}
  y_{it} = \mathbbm{1}\{ x_{it}^\prime \beta + u_{it} > 0 \}, \quad 
  u_{it} =  e_{it} + \rho e_{it-1}, \quad e_{it} \overset{iid}{\sim} \mathcal{N}(0,1),
\end{align*}
where $x_{it}$ are iid, strictly exogenous regressors, $i=1,\dots,n$ and $t=1,\dots,T$. The parameter of interest is $\theta = (\beta,\rho)$. The model is estimated by matching OLS estimates from a linear regression of $y_{it}$ on $x_{it}$ and $y_{it-1}$ between sample and simulated datasets. Because of the indicator function $\mathbbm{1}$, the objective function is discontinuous in $\theta$. The simulations use $n=250$, $T=10$, $\text{dim}(\beta) = 14$, $\beta^\dagger_{1}=\dots=\beta^\dagger_{5} = 1/\sqrt{5}$, $\beta^\dagger_{6}=\dots=\beta^\dagger_{14}=0$, $\rho^\dagger = 0.7$. The distant starting value is $\theta_0 = (0,\dots,0)$.

\begin{figure}[ht] \caption{Dynamic Discrete Choice: Distribution of Objection Function} \label{fig:obj_ddc}
  \centering
    \includegraphics[scale=0.55]{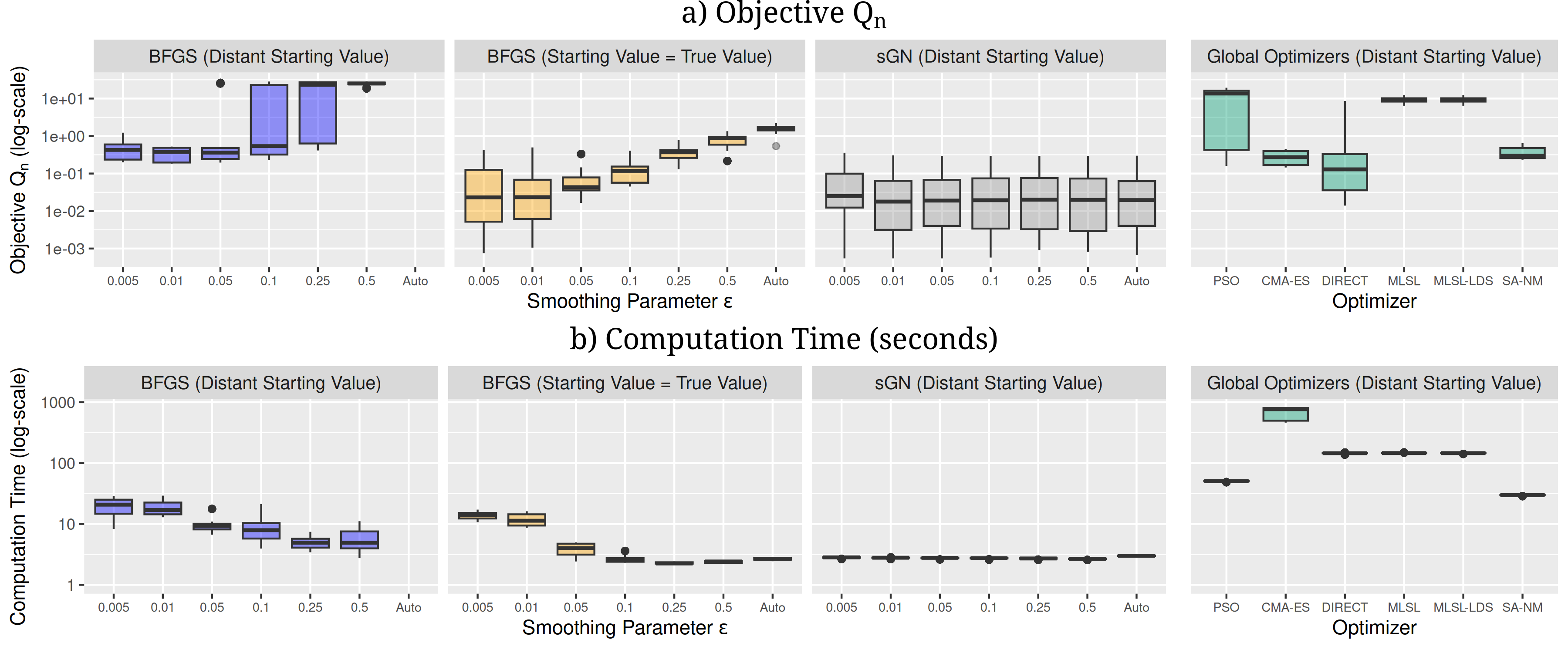}\\
  {\footnotesize \textbf{Legend:} $200$ Monte Carlo replications. Objective $Q_n$ = minimized objective value, Computation Time = running times in seconds. Auto: MSE minimizing $\varepsilon$ from Section \ref{sec:algos}. s\textsc{gn} = Algorithm \ref{algo:sgn} run for $250$ iterations, with $\theta^0 = (0,\dots,0)$. \textsc{bfgs} algorithm applied to objective computed with smoothed moments as in \citet{bruins2018}, initialized at 1) distant starting value $\theta_0 = (0,\dots,0)$, or 2) true value $\theta_0 = \theta^\dagger$. Global optimizers (\textsc{pso}, \textsc{cma-es}, \textsc{direct}, \textsc{mlsl}, \textsc{mlsl-lds}, \textsc{sa-nm}) applied to unsmoothed objective, initialized at $\theta_0 = (0,\dots,0)$, with default tuning parameters. }
\end{figure}

Figure \ref{fig:obj_ddc} compares Algorithm \ref{algo:sgn} with a local BFGS algorithm with smoothed moments, replacing $\mathbbm{1}\{\cdot > 0\}$ with $\Phi(\cdot / \varepsilon)$ in the dgp, where $\Phi$ is the CDF of the standard Gaussian. It also reports results for the unsmoothed problem using a battery of global optimizers: Particle Swarm Optimization \citep[\textsc{pso},][]{clerc2010}, Covariance matrix adaptation evolution strategy \citep[\textsc{cma-es},][]{hansen2003}, Dividing Rectangles \citep[\textsc{direct},][]{jones1993}, Multi-Level Single-Linkage with random numbers and low-dispersion sequence \citep[\textsc{mlsl}, \textsc{mlsl-lds},][]{rinnooy1987}, and Simulated Annealing followed by Nelder-Mead (\textsc{sa-nm}), the latter is commonly used in Economics. These methods apply to continuous and discontinuous problems. \textsc{direct}, Simulated Annealing and \textsc{mlsl} have global convergence guarantees for continuous problems, given appropriate tuning \citep[e.g.][]{belisle1992,locatelli1998}. Similar to Algorithm \ref{algo:sgn}, \textsc{mlsl} and \textsc{mlsl-lds} are hybrid methods. It is a type of multi-start algorithm which combines random draws or the Sobol sequence, respectively, with a local search algorithm. At each iteration, one or several local searches are run until convergence. See Appendix \ref{apx:MultStart} for convergence properties of multi-start Algorithms relative to Algorithm \ref{algo:sgn}. \textsc{cma-es} has convergence guarantees for a class of scale-invariant functions.\footnote{$Q$ is scale-invariant if $Q(\theta^\dagger + \Delta_1) \leq Q(\theta^\dagger + \Delta_2) \Leftrightarrow Q(\theta^\dagger + \rho \Delta_1) \leq Q(\theta^\dagger + \rho \Delta_2)$ for any $\rho >0$. See \citet{toure2023} for the results and their Assumptions F1, F2 for the conditions on $Q$.} \textsc{pso} is heuristic, i.e. does not have formal convergence guarantees.

The top panel compares the final value of the objective $Q_n$ among different methods and tuning parameters. BFGS only converges when initialized near the true value, the computational cost increases when the starting value is further away and when the bandwidth becomes smaller. Global optimizers do not perform as well despite a heavy computational cost (bottom panel). Algorithm \ref{algo:sgn} is run using the distant starting value and the Sobol sequence. It is fairly robust to the choice of bandwidth. The MSE minimizing bandwidth (Auto) gives satisfying result, the time reported in the bottom panel includes the bandwidth selection step. Although \textsc{mlsl-lds} also relies on the Sobol sequence with a local search, its performance is much worse than Algorithm \ref{algo:sgn}. 

  \begin{table}[ht] \caption{Dynamic Discrete Choice: Bias and Mean Absolute Error} \label{tab:par_ddc}
    \centering \setlength\tabcolsep{2.5pt}
    \renewcommand{\arraystretch}{0.9} 
    {\small \begin{tabular}{l|ccccccc||ccccccc}
      \hline \hline
      & \multicolumn{7}{c||}{Coefficient $\beta_1$} & \multicolumn{7}{c}{Coefficient $\rho$}\\
      \hline
      $\varepsilon$ & Auto & 0.5 & 0.25 & 0.1 & 0.05 & 0.01 & 0.005 & Auto & 0.5 & 0.25 & 0.1 & 0.05 & 0.01 & 0.005 \\ 
      \hline
      & \multicolumn{14}{c}{Estimation Bias: $\mathbb{E}(\tilde{\theta}_n) - \theta^\dagger$} \\
      \hline \rowcolor{gray}
      s\textsc{gn} & 0.01 & 0.01 & 0.01 & 0.01 & 0.01 & 0.01 & -0.01 & 0.00 & 0.00 & -0.00 & 0.00 & 0.00 & 0.00 & -0.02 \\ 
      \textsc{bfgs}$_1$ & - & 22.41 & 14.40 & 8.24 & 1.90 & 0.49 & 0.51 & - & 0.30 & 0.30 & 0.30 & 0.30 & 0.30 & 0.29 \\ 
      \textsc{bfgs}$_2$ & 0.13 & 0.02 & -0.00 & 0.01 & 0.01 & 0.05 & 0.01 & -0.02 & -0.07 & -0.06 & -0.02 & -0.01 & 0.03 & -0.01 \\ \hline
      & \multicolumn{14}{c}{Mean Absolute Estimation Error: $\mathbb{E}( | \tilde{\theta}_n - \hat{\theta}_n|) /\text{sd}(\hat{\theta}_n)$} \\
      \hline \rowcolor{gray}
      s\textsc{gn} & -0.03 & 0.23 & 0.23 & 0.26 & 0.25 & 0.33 & 0.66 & -0.10 & 0.20 & 0.19 & 0.23 & 0.20 & 0.24 & 0.53 \\ 
      \textsc{bfgs}$_1$ & - & 610.79 & 392.21 & 224.43 & 51.37 & 12.96 & 13.62 & - & 6.19 & 6.19 & 6.19 & 6.16 & 6.14 & 6.07\\ 
      \textsc{bfgs}$_2$ & 3.01 & 0.37 & 0.52 & 0.34 & 0.28 & 1.39 & 0.87 & -0.50 & 1.70 & 1.35 & 0.64 & 0.42 & 0.82 & 0.69 \\  
       \hline \hline
    \end{tabular}}\\
    {\footnotesize \textbf{Legend:} $200$ Monte Carlo replications. Auto: MSE minimizing $\varepsilon$ from Section \ref{sec:algos}. s\textsc{gn} = Algorithm \ref{algo:sgn} run for $250$ iterations. \textsc{bfgs}$_{1,2}$: \textsc{bfgs} algorithm applied to objective computed with smoothed moments as in \citet{bruins2018}, \textsc{bfgs}$_1$ initialized at $\theta_0 = (0,\dots,0)$, \textsc{bfgs}$_2$ initialized at true $\theta_0 = \theta^\dagger$.}
    \end{table}

    Table \ref{tab:par_ddc} compares estimates between BFGS with a smoothed indicator and Algorithm \ref{algo:sgn}. The table reports the estimates' $\tilde{\theta}_n$ bias as well as the mean absolute difference between $\tilde{\theta}_n$ and $\hat{\theta}_n$, found using Algorithm \ref{algo:sgn} initialized at $\theta^\dagger$, run for $1250$ iterations with a small bandwidth. The mean absolute error is normalized by the standard deviation of the estimates to be on the same scale as a critical value for a t-test, e.g. $1.96$. Algorithm \ref{algo:sgn} is generally close to $\hat{\theta}_n$ whereas smoothed GMM is more sensitive to the choice of bandwidth. For the smallest $\varepsilon = 0.005$, the Levenberg-Marquardt algorithm stabilizes the local step significantly but slows convergence so that $250$ iterations do not systematically produce a valid estimator. 


\subsection{Aiyagari Model} \label{sec:Ayia}
The second simulated example compares the estimation of a simple heterogeneous agent model. These models are key for evaluating the distributional effects of macroeconomic policies. They are, however, very difficult to estimate. As a result, many report calibrated, rather than estimated, results. The following illustrates the properties of Algorithm \ref{algo:sgn} for an SMM estimation of a textbook Aiyagari model. The parameters of interest are $\theta=(\beta,\gamma,\mu,\rho,\sigma)$ which are the discount factor, risk-aversion, average log-income, persistence of log-income, and the standard deviation of log-income shocks. 
This type of estimation is challenging because the model is discretized, approximately solved by value function iterations, and the $5$ parameters are estimated using $10$ non-smooth moments (quantiles). The model and solution method are described in more detail in Appendix \ref{apx:Aiyagari}. 


\begin{table}[ht] \caption{Aiyagari Model: Estimates and Computation Times}
  \label{tab:Aiyagari_tab}
  \centering \setlength\tabcolsep{2.5pt}
  \renewcommand{\arraystretch}{0.9} 
  {\small 
  \begin{tabular}{l|c|ccccc|c}
    \hline \hline
    & $Q_n$ & $\beta$ & $\gamma$ & $\mu$ & $\rho$ & $\sigma$ & Time  \\ \hline
    True Value & - & 0.970 & 3.000 & 1.872 & 0.700 & 0.200 & (hr:mn)\\ \hline
    \multicolumn{8}{c}{s\textsc{gn}}\\
    \hline
    $\varepsilon = 0.5$ & 0.617 & 0.969 & 2.942 & 1.860 & 0.711 & 0.201 & 00:34 \\ 
    $\varepsilon = 0.2$ & 0.582 & 0.968 & 3.030 & 1.860 & 0.712 & 0.201 & 00:34 \\ 
    $\varepsilon = 0.1$ & 0.570 & 0.968 & 3.029 & 1.860 & 0.712 & 0.201 & 00:34 \\ 
    $\varepsilon = 0.05$ & 0.568 & 0.968 & 3.034 & 1.860 & 0.712 & 0.201 & 00:34\\ 
    $\varepsilon = 0.01$ & 0.587 & 0.968 & 3.023 & 1.860 & 0.712 & 0.201 & 00:34 \\ \hline
    \multicolumn{8}{c}{Global Optimizers}\\
    \hline
    \textsc{nm} & 20696 & 0.991 & 19.353 & 1.861 & 0.689 & 0.202 & 00:02\\ 
    \textsc{sa} & 25296 & 0.979 & 42.166 & 2.262 & 0.720 & 0.049  & 01:06 \\ 
    \textsc{sa-nm} & 620 & 0.781 & 20.034 & 3.310 & 0.930 & 0.044 & 01:09\\ 
    \textsc{pso} & 1.370 & 0.974 & 2.379 & 1.829 & 0.702 & 0.207 & 04:42\\ 
    \textsc{cma-es} & 21.479 & 0.942 & 4.927 & 1.961 & 0.741 & 0.183 & 02:27 \\ 
    \textsc{direct} & 58.262 & 0.950 & 4.689 & 2.156 & 0.776 & 0.150 & 04:26\\ 
    \textsc{msls} & 22.740 & 0.971 & 2.721 & 1.742 & 0.680 & 0.229 & 06:08\\ 
    \textsc{msls-lds} & 47.216 & 0.987 & 0.225 & 1.684 & 0.668 & 0.239 & 05:51\\  
     \hline \hline
  \end{tabular}
  }\\
{\footnotesize \textbf{Legend:} $Q_n$ = minimized objective value, $\beta,\gamma,\mu,\rho,\sigma$ = coefficients at the True Value, and for the minimized objective value. Time = running time in hours:minutes (hr:mm). s\textsc{gn} = Algorithm \ref{algo:sgn} with $\gamma = 0.1$, $\alpha = 0.47$, $B = 250$ iterations. Global optimizers (\textsc{pso}, \textsc{cma-es}, \textsc{direct}, \textsc{mlsl}, \textsc{mlsl-lds}, \textsc{sa-nm}) applied with default tuning parameters. }
\end{table}

Table \ref{tab:Aiyagari_tab} compares the minimized objective values, parameter estimates, and computation time for one simulated sample with a range of bandwidths $\varepsilon \in \{0.5,0.2,0.1,0.05,0.01\}$, $\varepsilon = 0.1$ is the MSE minimizing bandwidth.\footnote{ The LM algorithm is applied to improve stability of the local step over the range of tuning parameters. }  and the global optimizers considered in the previous example.\footnote{The global optimizers have a maximum of $50000$ function evaluations, to keep runtime below $6$ hours.} The table suggests that $\gamma$ is the hardest parameter to estimate, as several methods find reasonable estimates for the others. \textsc{pso} finds accurate estimates, except for the coefficient risk-aversion $\gamma$. Figure \ref{fig:Aiyagari1} and Table \ref{tab:Aiyagari_tab2}, Appendix \ref{apx:Aiyagari} illustrate the optimization paths for Algorithm \ref{algo:sgn} and considers additional values for the tuning paramaters.

\subsection{Interdependent Durations} \label{sec:Durations}
The empirical example replicates the estimation of the joint duration model of employment by \citet{honore2018}. They model the joint decision of optimal retirement age for married couples. In their setting, husbands and wives engage in a Nash bargaining game to determine when each will retire. Unlike single-agent duration models, their specification does not have a closed-form likelihood so they consider simulation-based estimation. Because the simulated outcomes are discrete, the resulting objective function is discontinuous and fairly difficult to minimize. They use a repeated succession of 5 optimizers.\footnote{Their procedure: ``The following loop of procedures was used until a loop produced a change in the parameter estimate of less than $10^{-5}$. [\dots] (a) particle swarm [\dots], (b) Powell’s conjugate direction method, (c) downhill simplex using Matlab’s fminsearch routine (d) pattern search using Matlab’s built-in routine, (e) particle swarm focusing on the jump-parameters [\dots].'' See \citet[p1330]{honore2018}.} The following compares their estimation from an accurate initial guess with Algorithm \ref{algo:sgn} initialized using a randomized Sobol sequence.
\begin{table}[ht] \caption{Interdependent Duration Estimates: \citet{honore2018} and s\textsc{gn}}
  \label{tab:est_hp_short}
  \centering \setlength\tabcolsep{2.5pt}
  \renewcommand{\arraystretch}{0.9} 
  {\small \begin{tabular}{l|cc|aa||cc|aa} \hline \hline
  & \multicolumn{4}{c||}{Coefficients for Wives} & \multicolumn{4}{c}{Coefficients for Husbands}\\ \hline
  & \multicolumn{2}{c|}{\footnotesize Honor\'e \& de Paula} & \multicolumn{2}{c||}{s\textsc{gn}} & \multicolumn{2}{c|}{\footnotesize Honor\'e \& de Paula} & \multicolumn{2}{c}{s\textsc{gn}}\\ \hline
  \multirow{2}{*}{\footnotesize $\delta$} & 
  $1.052$ & $1.064$ & $1.065$ & $1.066 $ & 1.052 & 1.064 & $1.065$ & $1.066$ \\ 
  & {\footnotesize { $(0.039) $}} & {\footnotesize { $(0.042) $}} & 
  {\footnotesize { $(0.039) $}} & {\footnotesize { $ (  0.037) $}}
  & {\footnotesize {$ (  0.039) $}} & {\footnotesize {$ (  0.042) $}} & {\footnotesize {$ (  0.039) $}} & {\footnotesize {$ (  0.037) $}} \\ \hline
  \multirow{2}{*}{\footnotesize $\theta_1$} & 
  $1.244$ & $1.244$ & $1.238$ & $1.224$ & $1.169$ & $1.218$ & $1.179 $ & $1.188$ \\ 
  & {\footnotesize { $(0.054) $}} & {\footnotesize { $(0.054) $}} & 
  {\footnotesize { $(0.055) $}} & {\footnotesize { $ (  0.050) $}}& 
  {\footnotesize { $(0.043) $}} & {\footnotesize { $ (  0.058) $}}& {\footnotesize {$ (  0.043) $}} & {\footnotesize {$ (  0.040) $}}\\ \hline
  \multirow{2}{*}{\footnotesize $\ge$ 62 yrs-old} & 
  $10.640$ & $13.446$ & $11.194$ & $11.602 $ & $31.532$ & $39.824$ & $32.817$ & $34.268$\\ 
  & {\footnotesize { $(5.916) $}} & {\footnotesize { $(5.694) $}} & 
  {\footnotesize { $(7.818) $}} & {\footnotesize { $ (  5.692) $}}& {\footnotesize { $(11.356) $}} & {\footnotesize { $ (11.372) $}}& 
  {\footnotesize {$ (  8.131) $}} & {\footnotesize {$ (  7.672) $}}\\ \hline
  \multirow{2}{*}{\footnotesize $\ge$ 65 yrs-old} & 
  $10.036$ & $12.326$ & $10.613$ & $11.721 $ & $25.696$ & $29.254$ & $26.352$ & $26.000$\\ 
 & {\footnotesize { $ ( 11.555) $}} & {\footnotesize { $ ( 7.495) $}} & 
 {\footnotesize { $ ( 10.067) $}} & {\footnotesize { $ ( 10.897) $}}& {\footnotesize { $ ( 9.497) $}} & {\footnotesize { $ ( 11.229) $}}& {\footnotesize {$ ( 13.215) $}} & {\footnotesize {$ ( 14.289) $}}\\ \hline
  {\footnotesize {\vdots}} & {\footnotesize {\vdots}} & {\footnotesize {\vdots}} & {\footnotesize {\vdots}} & {\footnotesize {\vdots}} & {\footnotesize {\vdots}} & {\footnotesize {\vdots}} & {\footnotesize {\vdots}} & {\footnotesize {\vdots}} \\\hline
  {\footnotesize Starting Obj. Value} & $93.70$ & $89.77$ & $2.10^4$ & $5.10^4$ & {\footnotesize { -}} & {\footnotesize { -}} & {\footnotesize { -}} & {\footnotesize { -}} \\ 
  {\footnotesize Final Obj. Value } & $0.470$ & $0.758$ & $0.306$ & $0.364$ & {\footnotesize { -}} & {\footnotesize { -}} & {\footnotesize { -}} & {\footnotesize { -}} \\    \hline
  {\footnotesize Number of Coef. } & $12$ & $30$ & $12$ & $30$ &  {\footnotesize { -}} & {\footnotesize { -}} & {\footnotesize { -}} & {\footnotesize { -}} \\ 
  {\footnotesize Number of Obs.} & $     1227 $ & $     1227 $ & $     1227 $ & $     1227 $ &  {\footnotesize { -}} & {\footnotesize { -}} & {\footnotesize { -}} & {\footnotesize { -}} \\ \hline
  {\footnotesize Computation Time }  & 3h25m & 5h34m & 12min & 13min & {\footnotesize { -}} & {\footnotesize { -}} & {\footnotesize { -}} & {\footnotesize { -}} \\ \hline \hline
\end{tabular}}\\
{\footnotesize Legend: s\textsc{gn}: $\varepsilon = 10^{-2}$, $\gamma = 0.1$, $\alpha = 0.47$, $B = 500$ iterations in total. Husbands: - same as wives. Coefficients for wives and husbands are estimated jointly. Full estimation results are in Table \ref{tab:est_hp_full}.}
\end{table}
Table \ref{tab:est_hp_short} replicates the first two columns of Tables 2 and 3 in \citet[pp1319-1321]{honore2018}. For brevity, only the first four coefficients are reported here. The estimates are similar for both specifications. However, the time required to compute the estimates is significantly reduced: the two specifications take 3.5 and 5.5 hours to estimate with their replication code. In comparison, Algorithm \ref{algo:sgn} finds estimates in less than 15 minutes using the same code. For comparison, for the $12$ coefficient model, the random-walk Metropolis-Hastings (MH) algorithm has yet to converge after $100000$ iterations (i.e. 10 hours, see Figure \ref{fig:HP_MCMC}, Appendix \ref{Apx:HP_add}) for the same distant starting value. MH converges more quickly from the same starting value as \citet{honore2018} (not reported). Table \ref{tab:est_hp_eps} in Appendix \ref{apx:extra_MC_Emp} provides additional results for $\varepsilon = 0.8 \cdot 10^{-2}, 2.11 \cdot 10^{-2}$. The latter is the MSE minimizing bandwidth. The range of bandwidths that provide accurate estimates is much narrower than in previous examples. Two features explain this issue: first, the optimal weighting matrix $W_n$ used for estimation has condition number $\kappa = 1.9 \cdot 10^{3}$, $4.4 \cdot 10^{3}$, for the smaller and larger specification, respectively. This is because several moments are strongly correlated. Second, the Jacobian estimated using Algorithm \ref{algo:sqn} at the solution has condition number $2 \cdot 10^{3}$, $5 \cdot 10^3$, indicating a relatively small $\underline{\sigma}$. Note that the dummies corresponding to $\ge$ 62 yrs-old and $\ge$ 65 yrs-old, and the corresponding moments, are somewhat redundant -- using a more parsimonious specification could help improve on both of these issues. 
\vspace{-0.0cm}

\section{Conclusion}
This paper shows, using an explicit algorithm, that faster GMM estimation is possible under standard econometric assumptions. For local optimization, the main idea is that smoothing only the Jacobian produces less smoothing bias -- allowing for larger bandwidths than existing methods which benefit optimization properties. For global optimization, introducing a global search step allows to adapt to the complexity of the estimation problem, without additional user intervention or input. 
\newpage
\baselineskip=14.0pt
\bibliographystyle{ecta}
\bibliography{refs}

\begin{appendices}
  \renewcommand\thetable{\thesection\arabic{table}}
  \renewcommand\thefigure{\thesection\arabic{figure}}
  \renewcommand{\theequation}{\thesection.\arabic{equation}}
  \renewcommand\thelemma{\thesection\arabic{lemma}}
  \renewcommand\thetheorem{\thesection\arabic{theorem}}
  \renewcommand\thedefinition{\thesection\arabic{definition}}
    \renewcommand\theassumption{\thesection\arabic{assumption}}
  \renewcommand\theproposition{\thesection\arabic{proposition}}
    \renewcommand\theremark{\thesection\arabic{remark}}
    \renewcommand\thecorollary{\thesection\arabic{corollary}}
\setcounter{equation}{0}
\setcounter{lemma}{0}
\clearpage \baselineskip=18.0pt
\appendix
\section{Preliminary Results for Section \ref{sec:main}} \label{apx:prelim}

\subsection{Preliminary Results for the Population Problem}
\begin{lemma} \label{lem:prelim_global}
  Suppose Assumptions \ref{ass:gn}-\ref{ass:sample_gn} hold. There exists $\overline{r}_g >0$ such that for any $\theta_b$ such that $\|\theta_b-\theta^\dagger\|_{G^\prime W G} \leq \overline{r}_g$, the following hold:
  \begin{align} 
    \|g(\theta_{b+1})\|_W &\leq (1-\overline{\gamma})\|g(\theta_{b})\|_W, \label{eq:contract}\\
    (1-\overline{\gamma}/2)\|\theta_b-\theta^\dagger\|_{G^\prime W G} &\leq \|g(\theta_b)\|_W \leq (1+\overline{\gamma}/2)\|\theta_b-\theta^\dagger\|_{G^\prime W G}, \label{eq:norm_equiv}
  \end{align}
  where $\theta_{b+1} = \theta_b - \gamma (G_b^\prime W G_b)^{-1}G_b^\prime W g(\theta_b)$. There also exists a $\underline{r}_g \in (0,\overline{r}_g]$ such that:
  \begin{align} \inf_{\|\theta-\theta^\dagger\|_{G^\prime W G} \geq \overline{r}_g} \|g(\theta)\|_W \geq (1+\frac{\overline{\gamma}}{2})(1-\overline{\gamma})\underline{r}_g. \label{eq:separ} \end{align}
\end{lemma}

\subsection{Preliminary Results for the Finite-Sample Problem}

\begin{lemma}[Deterministic bounds] \label{lem:dem_bounds} Suppose Assumption \ref{ass:gn} holds then: 
i. For any $(\theta_1,\theta_2)\in \Theta \time \Theta$ and $\varepsilon >0$: $\|G_{\varepsilon}(\theta_1)-G_{\varepsilon}(\theta_2)\| \leq L_G \|\theta_1-\theta_2\|,$
ii. For any $\theta \in \Theta$ and $\varepsilon >0$:
$\|G_{\varepsilon}(\theta)-G(\theta)\| \leq \varepsilon L_G M_{1,Z}$ where $M_{1,Z} = \int \|\phi^\prime(Z)\|dZ$.
\end{lemma}

To simplify notation, the following Lemma relies on the notation $\mathbb{P},\mathbb{E}$ instead of outer probabilities and expections $\mathbb{P}^*,\mathbb{E}^*$ \citep[][Ch1.2]{VanderVaart1996}. The results rely on tail bounds for empirical processes using backeting integrals \citep[][Ch2.14]{VanderVaart1996} under a $L_2$-smoothness condition.
\begin{lemma}[Stochastic bounds] \label{lem:stoch_bounds} Suppose Assumptions \ref{ass:gn}-\ref{ass:sample_gn} hold.\\  Let $C_\Theta = \int_0^1 \sqrt{1+ \log [N(x,\Theta,\|\cdot\|)]}dx$, there exists a universal constant $C>0$ such that:
\[ \mathbb{E}\left( \sup_{\|\theta_1-\theta_2\| \leq \delta } \sqrt{n} \big\|[\overline{g}_n(\theta_1)- g(\theta_1)]-[\overline{g}_n(\theta_2)- g(\theta_2)] \big\| \right) \leq C C_\Theta L_g \delta^\psi. \]
For any $c_n \geq 1$, the above inequality implies:
\begin{enumerate}  \setlength\itemsep{0em}
    \item[a.] Sample Moments:\\
    $\mathbb{P}\left( \sup_{\|\theta_1-\theta_2\| \leq \delta } \big\|[\overline{g}_n(\theta_1)- g(\theta_1)]-[\overline{g}_n(\theta_2)- g(\theta_2)] \big\|  \leq   c_n n^{-1/2} C_\Theta L_g \delta^\psi \right) \geq 1- C/c_n$,
    \item[b.] Smoothed moments, for any $\varepsilon >0$:\\
    $\mathbb{P}\left( \sup_{\|\theta_1-\theta_2\| \leq \delta } \big\|[\overline{g}_{n,\varepsilon}(\theta_1)- g_{\varepsilon}(\theta_1)]-[\overline{g}_{n,\varepsilon}(\theta_2)- g_{\varepsilon}(\theta_2)] \big\|  \leq   c_n n^{-1/2} C_\Theta L_g \delta^\psi \right) \geq 1- C/c_n$,
    \item[c.] Smoothed Jacobian, for any $\varepsilon >0$:\\
    $\mathbb{P}\left( \sup_{\|\theta_1-\theta_2\| \leq \delta } \big\|[\overline{G}_{n,\varepsilon}(\theta_1)- G_{\varepsilon}(\theta_1)]-[\overline{G}_{n,\varepsilon}(\theta_2)- G_{\varepsilon}(\theta_2)] \big\|  \leq   c_n \varepsilon^{-1}n^{-1/2} C_\Theta L_g M_{1,Z} \delta^\psi \right) \geq 1- C/c_n$, 
    where $M_{1,Z} = \int \|\phi^\prime(Z)\|dZ$.
\end{enumerate}
All three events a-c. hold jointly with the same probability bound $1-C/c_n$.
\end{lemma}

\begin{lemma}[Singular Values] \label{lem:singular} Suppose Assumptions \ref{ass:gn}-\ref{ass:sample_gn} hold.
  Let $\theta \in \Theta$ such that $\|\theta-\theta^\dagger\| \leq R_G$, where $R_G$ is defined in Lemma \ref{lem:cv_local}, then:\\
  $\sigma_{\min}[\overline{G}_{n,\varepsilon}(\theta)] \geq \underline{\sigma} - \left[\frac{c_n}{\varepsilon \sqrt{n}} C_\Theta L_g M_{1,Z} R_G^\psi +  \varepsilon L_G M_{1,Z}\right] := \underline{\sigma}_{n,\varepsilon}$,
  with probability $1-C/c_n$.
\end{lemma}

\begin{lemma} \label{lem:Taylor_Expand} Suppose Assumptions \ref{ass:gn}-\ref{ass:sample_gn} hold, then for any $\delta \geq 0$ and $\varepsilon>0$, $c_n \geq 1$. We have for $C$ and $C_\Theta$ as in Lemma \ref{lem:stoch_bounds}:
  \begin{align}
    \mathbb{P} &\Big(  \sup_{\|\theta_1-\theta_2\| \leq \delta} \|\overline{g}_n(\theta_1)-\overline{g}_n(\theta_2) - G_{n,\varepsilon}(\theta_1)(\theta_1-\theta_2)\|\notag \\ &\leq L_G \delta^2 + (1+M_{2,Z})L_g C_\Theta c_nn^{-1/2}[ \delta^\psi + \varepsilon^{\psi-1}\delta ] + L_G M_{1,Z} \varepsilon \delta \Big) \geq 1-C/c_n, \label{eq:Taylor_smoothed}
  \end{align}
  where $M_{2,Z} = \int_Z \|Z\|^2 \phi(Z)dZ$.
\end{lemma}

\begin{lemma} \label{lem:Reta} Suppose Assumptions \ref{ass:gn}-\ref{ass:sample_gn} hold. For $\eta \in (0,1)$ and $c_n \geq 1$, let:\\ 
  $R_{n}(\eta) := \frac{\eta \underline{\sigma}^2}{\kappa_W L_G} - C_a c_nn^{-1/2} - \frac{L_g}{L_G} C_\Theta (c_nn^{-1/2})^\psi$.
  Then uniformly in $\|\theta - \hat\theta_n\| \leq R_n(\eta)$, we have with probability $1-(1+C)/c_n$:
  \begin{align}
    (1-\eta)\|\theta-\hat\theta_n\|_{G^\prime W_n G} - \overline{\lambda}_W L_g C_\Theta (c_nn^{-1/2})^{1+\psi} &\leq
    \|\overline{g}_n(\theta)-\overline{g}_n(\hat\theta_n)\|_{W_n}\\ &\leq (1+\eta)\|\theta-\hat\theta_n\|_{G^\prime W_n G} + \overline{\lambda}_W L_g C_\Theta (c_nn^{-1/2})^{1+\psi}, \notag
  \end{align}
  where $G = G(\theta^\dagger)$.
\end{lemma}

\begin{lemma} \label{lem:separ_ns} Suppose Assumptions \ref{ass:gn}-\ref{ass:sample_gn} hold. For $\eta \in (0,1)$ and $R_n(\eta)$ defined in Lemma \ref{lem:Reta} let: $\overline{r}_n(\eta) = \frac{R_n(\eta)}{\sqrt{\kappa_G \kappa_W}} - C_ac_n n^{-1/2}$, and\\
$\underline{r}_{n,g}(\eta) = \frac{\delta(\overline{r}_{n,g}(\eta))}{\sqrt{\kappa_W}} - \overline{\lambda}_W^{1/2} c_nn^{-1/2} \left( L_g C_\Theta \text{diam}(\Theta)^\psi + \lambda_{\max}(\Sigma)^{1/2}p \right)$,
where $\delta(\cdot)$ is defined in Assumption \ref{ass:gn}v, $p = \text{dim}(\overline{g}_n(\theta))$, $\Sigma = \text{var}(\overline{g}_n(\theta^\dagger))$, $\kappa_G = \overline{\sigma}/\underline{\sigma}$ and $\overline{\sigma} = \sigma_{\max}[G(\theta^\dagger)]$. Then with probability $1-(1+C)/c_n$:
$\inf_{\|\theta-\hat\theta_n\|_{G^\prime W_n G} \geq \overline{r}_n(\eta)} \|\overline{g}_n(\theta)\|_{W_n} \geq \underline{r}_{n,g}(\eta).$
\end{lemma}

\begin{lemma} \label{lem:sample_obj_constract} Suppose Assumptions \ref{ass:gn}-\ref{ass:sample_gn} hold. Take $\eta\in(0,1)$ and let $x_b = \|\overline{g}_n(\theta_b)-\overline{g}_n(\hat\theta_n)\|_{W_n}$. For any $\varepsilon>0$, $c_n \geq 1$, we have:
\begin{align}
  &x_{b+1} \leq (1-\overline{\gamma})x_b + \frac{\underline{\sigma}_{n,\varepsilon}^{-3}}{(1-\eta)^2} \Delta_{2,n,\varepsilon}(x_b), \label{eq:contr_obj_ns}
\end{align}
uniformly in $\|\theta_b-\hat\theta_n\| \leq R_n(\eta)$, defined in Lemma \ref{lem:Reta}, with probability $1-(1+C)/c_n$, and:
$\Delta_{2,n,\varepsilon}(x_b)  = C_{4}  \left( \Gamma_{n,\varepsilon} + \varepsilon^{\psi-1}(c_nn^{-1/2})^{2} + [c_nn^{-1/2} + (c_nn^{-1/2})^2 \varepsilon^{-1}]x_b^\psi + (c_nn^{-1/2})x_b \right)$.
\end{lemma}

The following Lemma depends on $\varepsilon$ because it re-uses some of the preliminary and intermediate results for brievety. The assumptions on $\varepsilon$ could be relaxed at the cost of additional intermediate derivations.
\begin{lemma} \label{lem:norm_squared} 
  Suppose Assumptions \ref{ass:gn}-\ref{ass:sample_gn} hold, take $\tau$ and $\eta \in (0,1)$. Then for $c_nn^{-1/2}$ small enough, we have uniformly in $\|\theta-\hat\theta_n\| \leq R_n(\eta)$:
\begin{align*}
  (1-\tau)^2 \|\overline{g}_n(\theta)-\overline{g}_n(\hat\theta_n)\|^2_{W_n} - \Gamma^2_{2,n,\varepsilon}(\eta,\tau) &\leq \|\overline{g}_n(\theta)\|^2_{W_n} - \|\overline{g}_n(\hat\theta_n)\|^2_{W_n} \\ &\leq (1+\tau)^2 \|\overline{g}_n(\theta)-\overline{g}_n(\hat\theta_n)\|^2_{W_n} + \Gamma^2_{2,n,\varepsilon}(\eta,\tau),
\end{align*}
with probability $1-(1+C)/c_n$. The last term is given by:
\begin{align}
  \Gamma_{2,n,\varepsilon}(\eta,\tau) =  &\frac{ C_5 }{(1-\eta)^2[\overline{\gamma}-(\overline{\gamma}/2)^2]}\frac{1}{\min(\underline{\sigma}_{n,\varepsilon},\underline{\sigma}^2_{n,\varepsilon})}\Big( \Gamma_{n,\varepsilon} + (c_nn^{-1/2})^2 \varepsilon^{\psi-1} + (c_nn^{-1/2})^{1+\psi/2+\psi^2/2} \notag\\ &+ \varepsilon^{1/2}(c_nn^{-1/2})^{1+\psi/2} + (c_nn^{-1/2})^{(3+\psi)/2} \varepsilon^{(\psi-1)/2} \Big), \label{eq:Gamma2n} \end{align} 
and satisfies $n \Gamma_{2,n,\varepsilon}^2(\eta,\tau) = o(1)$ for any fixed $\eta,\tau \in (0,1)$ if $c_n=O(1)$, $\varepsilon = o(1)$, and $\sqrt{n}\varepsilon \to \infty$.
\end{lemma}

\section{Proofs for the Main Results} \label{apx:proofs_main}
The following derivations assume, without loss of generality, that $c_n n^{-1/2} \leq 1$ and $\varepsilon \leq 1$.

\paragraph{Proof of Theorem \ref{th:sgn_global00}.} Apply Theorem \ref{th:sgn_global} with $k = \sup_{n \geq 1} k_n$, finite and notice that $j/\log(n) \to \infty$ implies $j/\log(\tilde{\Gamma}_{n,\varepsilon}) \to \infty$, under the rate conditions for $\varepsilon$, so that (\ref{eq:asymequiv}) holds. By construction $\hat{\theta}_n$ is asymptotically Gaussian using a CLT for iid random variables with finite variance and Slutsky's Lemma, so that (\ref{eq:asymnorm}) follows directly from (\ref{eq:asymequiv}). \qed

\paragraph{Proof of Lemma \ref{lem:cv_local}.}
Take any $\theta_b$ such that $\|\theta_b - \theta^\dagger\| \leq R$ as defined in Lemma \ref{lem:cv_local}. Let $G_b = G(\theta_b)$. Since $g(\theta^\dagger)=0$, the next $\theta_{b+1}$ is such that:
\begin{align*}
    \theta_{b+1} - \theta^\dagger &= \theta_b - \theta^\dagger - \gamma \left( G_b^\prime W G_b \right)^{-1} G_b^\prime W [g(\theta_b) - g(\theta^\dagger)]\\
    &= (1-\gamma)(\theta_b - \theta^\dagger) - \gamma \left( G_b^\prime W G_b \right)^{-1} G_b^\prime W [G(\tilde\theta_b)-G_b](\theta_b-\theta^\dagger),
\end{align*}
for some intermediate value $\tilde\theta_b$. Now we have, by Lipschitz-continuity of $G$:
\begin{align*}
  \|\left( G_b^\prime W G_b \right)^{-1} G_b^\prime W [G(\tilde\theta_b)-G_b](\theta_b-\theta^\dagger)\| &\leq \frac{\sqrt{\overline{\lambda}_W}L_G}{\sqrt{\lambda_{min}[G_b^\prime W G_b]}}\|\theta_b-\theta^\dagger\|^2  \leq \frac{\sqrt{\kappa_W}L_G}{\underline{\sigma}}\|\theta_b-\theta^\dagger\|^2.
\end{align*}
Taking this back into the previous equality, we have:
\begin{align*}
  \|\theta_{b+1} - \theta^\dagger\| &\leq \left(1-\gamma + \gamma \frac{\sqrt{\kappa_W}L_G}{\underline{\sigma}}\|\theta_b-\theta^\dagger\| \right)\|\theta_b - \theta^\dagger\| \leq (1-\overline{\gamma})\|\theta_b - \theta^\dagger\|, \label{eq:contract_pop}
\end{align*}
for $\|\theta_b - \theta^\dagger\| \leq R$. Iterating over $b$, we find: $\|\theta_{b+1} - \theta^\dagger\| \leq (1-\overline{\gamma})^{b+1} \|\theta_0-\theta^\dagger\|$.
\qed

\paragraph{Proof of Lemma \ref{lem:cv_global}.} The proof relies on three properties derived in Lemma \ref{lem:prelim_global}. 
First, locally to $\theta^\dagger$, the iteration (\ref{eq:step1}) is a strict contraction for the objective function, i.e.:\footnote{The contraction is called strict because $0\leq (1-\overline{\gamma}) < 1$ implies a strict decrease in the loss function.}
\begin{align} 
  \|g(\theta_{b+1})\|_W \leq (1-\overline{\gamma})\|g(\theta_{b})\|_W, \tag{\ref{eq:contract}}
\end{align}
and $\|\theta_{b+1}-\theta^\dagger\| < \|\theta_b-\theta^\dagger\|$, cf. Lemma \ref{lem:cv_local}.

The second property involves the weighted norm:\footnote{In the weighted norm $G(\theta^\dagger)$ and $W$ do not vary with $\theta$.} $\|\theta-\theta^\dagger\|_{G^\prime W G}^2 := (\theta-\theta^\dagger)^\prime [G(\theta^\dagger)^\prime W G(\theta^\dagger)] (\theta-\theta^\dagger)$ which is closely related to $\|g(\theta)\|_{W}$ in step (\ref{eq:Glob_GN}). Under Assumption \ref{ass:gn}, there exists a $\overline{r}_g >0 $ such that, for any $\|\theta-\theta^\dagger\|_{G^\prime W G} \leq \overline{r}_g$, the following norm equivalence holds:
\begin{align} (1-\overline{\gamma}/2)\|\theta-\theta^\dagger\|_{G^\prime W G} \leq \|g(\theta)\|_W \leq (1+\overline{\gamma}/2)\|\theta-\theta^\dagger\|_{G^\prime W G}, \tag{\ref{eq:norm_equiv}} \end{align}
and (\ref{eq:contract}) also holds. The last property comes from the global identification condition. Under Assumption \ref{ass:gn}v, there exists another $\underline{r}_g \in (0,\overline{r}_g]$ such that:\footnote{Simply set $\underline{r}_g = [\inf_{\|\theta-\theta^\dagger\|_{G^\prime W G} \geq \overline{r}_g} \|g(\theta)\|_W]/[(1+\overline{\gamma}/2)(1-\overline{\gamma})] > 0$, by Assumption \ref{ass:gn}v.} 
\begin{align} \inf_{\|\theta-\theta^\dagger\|_{G^\prime W G} \geq \overline{r}_g} \|g(\theta)\|_W \geq (1+\overline{\gamma}/2)(1-\overline{\gamma})\underline{r}_g. \tag{\ref{eq:separ}} \end{align}

The three properties combined are enough to derive global convergence. Suppose at iteration $b$ that $\|\theta_b-\theta^\dagger\|_{G^\prime W G} \leq \overline{r}_g$, then $\|\theta_{b+1}-\theta^\dagger\|_{G^\prime W G}\leq  (1-\overline{\gamma}) \frac{1+\overline{\gamma}}{1-\overline{\gamma}} \|\theta_{b}-\theta^\dagger\|_{G^\prime W G} < \overline{r}_g$. 

If the global step (\ref{eq:Glob_GN}) finds $\|g(\theta^{b+1})\|_W < \|g(\theta_{b+1})\|_W$. Then $\|\theta^{b+1}-\theta^\dagger\|_{G^\prime W G} \leq \overline{r}_g$ from (\ref{eq:separ}). The local norm equivalence (\ref{eq:norm_equiv}) applies and:
\begin{align}
  \|\theta^{b+1}-\theta^\dagger\|_{G^\prime W G} \leq \frac{\|g(\theta^{b+1})\|_W}{1-\overline{\gamma}/2} < \frac{1+\overline{\gamma}/2}{1-\overline{\gamma}/2}(1-\overline{\gamma}) \underline{r}_g < \underline{r}_g,\label{eq:stable}
\end{align}
which is a strict contraction. This implies local stability: $\|\theta_{b+j}-\theta^\dagger\| \leq \underline{r}_g$ for all $j\geq 1$. Now, the contraction (\ref{eq:contract}) can be applied recursively with $j=1,2,\dots$ This is sufficient to prove global convergence under the weighted norm $\|\cdot\|_{G^\prime W G}$.

After $k$ iterations, with the choice of $k$ defined in the Lemma, we must have:
$\|\theta_k - \theta^\dagger\|_{G^\prime W G} \leq \underline{r}_g$, 
because there is at least one $\theta^\ell$ with $\ell \in \{0,\dots,k\}$ such that this inequality holds and the strict contraction property (\ref{eq:stable}) implies that after that $\ell$, the distance cannot be greater than $\underline{r}_g$. 
The worst-case is $\ell=k$. Then, apply (\ref{eq:stable}) from $b=k$ to $b=k+j$, to get that $\|\theta_{k+j}-\theta^\dagger\|_{G^\prime W G} \leq \underline{r}_g$ for any $j \geq 0$. Now iterate over (\ref{eq:contract}) to get $\|g(\theta_{b+j})\|_W \leq (1-\overline{\gamma})^j\|g(\theta_b)\|_W \leq (1-\overline{\gamma})^j(1+\overline{\gamma}/2)\underline{r}_g$, for any $j \geq 0$. Then, (\ref{eq:norm_equiv}) yields the result.
\qed
\paragraph{Proof of Lemma \ref{lem:smooth_foc}.} We have:
\begin{align}
    \|G_{n,\varepsilon}(\hat\theta_n)^\prime W_n \overline{g}_n(\hat\theta_n)\| 
    &\leq \|G(\theta^\dagger)^\prime W_n \overline{g}_n(\hat\theta_n) \| \label{eq:tobound1}
     \\ &+ \|[G_{n,\varepsilon}(\hat\theta_n)-G_\varepsilon(\hat\theta_n)]^\prime W_n \overline{g}_n(\hat\theta_n)\| 
     \label{eq:tobound2}
    \\ &+ \|[G_\varepsilon(\hat\theta_n)-G_\varepsilon(\theta^\dagger)]^\prime W_n \overline{g}_n(\hat\theta_n)\| 
    \label{eq:tobound3}
    \\ &+ \|[G_\varepsilon(\theta^\dagger)-G(\theta^\dagger)]^\prime W_n \overline{g}_n(\hat\theta_n)\|. \label{eq:tobound4}
\end{align}
In the following, each term on the right-hand side of the inequality will be bounded.

Let $p = \text{dim}(\overline{g}_n)$ and $\Sigma =\text{var}(g(\theta^\dagger;x_i))$, by construction of $\hat\theta_n$:
\[ \|\hat\theta_n-\theta^\dagger\| \leq \underline{\sigma}^{-1}\sqrt{\kappa_W}\|\overline{g}_n(\theta^\dagger)\| \leq \underline{\sigma}^{-1}\sqrt{\kappa_W \lambda_{\max}(\Sigma)p}c_n n^{-1/2} = C_a c_n n^{-1/2}, \]
with probability $1-1/c_n$. The last inequality follows from Markov's inequality: $\mathbb{P}(\|\Sigma^{-1/2}\overline{g}_n(\theta^\dagger)\| \geq \sqrt{p}c_nn^{-1/2}) \leq \mathbb{E}(\|\Sigma^{-1/2}\overline{g}_n(\theta^\dagger)\|^2)^{1/2}/[\sqrt{p}c_n] = 1/c_n$. Then, using the bounds in Lemma \ref{lem:stoch_bounds} and the Lipschitz continuity of $G$, we have:
\[ \|\overline{g}_n(\hat\theta_n) - \overline{g}_n(\theta^\dagger) - G(\theta^\dagger)(\hat\theta_n-\theta^\dagger)\| \leq L_GC_a^2 c_n^2 n^{-1} + C_\Theta L_gC_a^\psi(c_nn^{-1/2})^{1+\psi} \leq C_b (c_nn^{-1/2})^{1+\psi},\]
with probability $1-(1+C)/c_n$ since $\psi \in (0,1]$, with $C$ from Lemma \ref{lem:stoch_bounds}. Now note that: $G(\theta^\dagger)^\prime W_n[\overline{g}_n(\theta^\dagger) - G(\theta^\dagger)(\hat\theta_n-\theta^\dagger)]=0$, 
by construction of $\hat\theta_n$ and using projection arguments. This implies that:
\begin{align} \|G(\theta^\dagger)^\prime W_n \overline{g}_n(\hat\theta_n)\| \leq \overline{\sigma}\overline{\lambda}_WC_b (c_nn^{-1/2})^{1+\psi} = C_{(\ref{eq:tobound1})} (c_nn^{-1/2})^{1+\psi}, \tag{\ref{eq:tobound1}}
\end{align}
with probability $1-(1+C)/c_n$ for some constant $C_{(\ref{eq:tobound1})}$, where $\overline{\sigma}=\sigma_{\max}[G(\theta^\dagger)]$. Now, in order to bound (\ref{eq:tobound2}), use:
\[ \|[G_{n,\varepsilon}(\hat\theta_n)-G_\varepsilon(\hat\theta_n)]^\prime W_n \overline{g}_n(\hat\theta_n)\| \leq \overline{\lambda}_W \sup_{\theta \in \Theta}\|G_{n,\varepsilon}(\theta)-G_\varepsilon(\theta)\| \times \|\overline{g}_n(\hat\theta_n)\|.\]
Using the previous bounds, we have: 
\begin{align*} 
    \|\overline{g}_n(\hat\theta_n)\| &\leq  \|\overline{g}_n(\hat\theta_n) - \overline{g}_n(\theta^\dagger) - G(\theta^\dagger)(\hat\theta_n-\theta^\dagger) \| + \|\overline{g}_n(\theta^\dagger) + G(\theta^\dagger)(\hat\theta_n-\theta^\dagger)\|\leq C_c c_nn^{-1/2},
\end{align*}
with probability $1-(1+C)/c_n$ for some $C_c$ which depends on $C_a$ and $C_b$.
We can also bound the difference between the sample and population smoothed Jacobian:
\begin{align*}
    &\sup_{\theta \in \Theta}\|G_{n,\varepsilon}(\theta)-G_\varepsilon(\theta)\|\\ &\leq \frac{1}{\varepsilon} \sup_{\theta \in \Theta}\| \overline{g}_n(\theta)-g(\theta) \| \int_Z \|\phi^\prime(Z)\|dZ\\
    &\leq \frac{1}{\varepsilon} \left(\sup_{\theta \in \Theta}\| [\overline{g}_n(\theta)-g(\theta)]-[\overline{g}_n(\theta^\dagger)-g(\theta^\dagger)] \| + \|\overline{g}_n(\theta^\dagger)-g(\theta^\dagger)\| \right) \int_Z \|\phi^\prime(Z)\|dZ\\
    &\leq \frac{1}{\varepsilon}\left( C_\Theta L_g \text{diam}(\Theta)c_n n^{-1/2} +  \sqrt{\lambda_{\max}(\Sigma)p}c_nn^{-1/2} \right)M_{1,Z},
\end{align*}
with probability $1-C/c_n$. Putting these bounds together, we get:
\begin{align} \|[G_{n,\varepsilon}(\hat\theta_n)-G_\varepsilon(\hat\theta_n)]^\prime W_n \overline{g}_n(\hat\theta_n)\| \leq \frac{C_{(\ref{eq:tobound2})}}{\varepsilon} (c_n n^{-1/2})^{2+\psi}, \tag{\ref{eq:tobound2}}
\end{align}
with probability $1-(1+C)/c_n$. 
For (\ref{eq:tobound3}), the bound for $\overline{g}_n(\hat\theta_n)$ can be used in tandem with the Lipschitz-continuity of the smoothed Jacobian and the bound on $\|\hat\theta_n-\theta^\dagger\|$, this yields:
\begin{align*}
    \|[G_\varepsilon(\hat\theta_n)-G_\varepsilon(\theta^\dagger)]^\prime W_n \overline{g}_n(\hat\theta_n)\| &\leq \overline{\lambda}_W L_G C_a C_c (c_n n^{-1/2})^{2+\psi} = C_{(\ref{eq:tobound3})}  (c_n n^{-1/2})^{2+\psi}, \tag{\ref{eq:tobound3}}
\end{align*}
with probability $1-(1+C)/c_n$. For the last (\ref{eq:tobound4}), we have
\begin{align*}
    \|[G_\varepsilon(\theta^\dagger)-G(\theta^\dagger)]^\prime W_n \overline{g}_n(\hat\theta_n)\| &\leq \overline{\lambda}_W C_c L_G M_{1,Z} \varepsilon c_n n^{-1/2} = C_{(\ref{eq:tobound4})} \varepsilon c_nn^{-1/2}, \tag{\ref{eq:tobound4}}
\end{align*}
with probability $1-(1+C)/c_n$. Combining all the bounds into the smoothed first-order condition, we have:
\begin{align}
    \|G_{n,\varepsilon}(\hat\theta_n)^\prime W_n \overline{g}_n(\hat\theta_n)\| \leq C_{(\ref{eq:smooth_foc})} (c_n n^{-1/2})^{1+\psi} \left( 1 + \frac{c_n n^{-1/2}}{\varepsilon} + \frac{\varepsilon}{(c_n n^{-1/2})^{\psi}} \right), \label{eq:smooth_foc}
\end{align}
with probability $1-(1+C)/c_n$ using $C_{(\ref{eq:smooth_foc})} = \max(C_{(\ref{eq:tobound1})}+C_{(\ref{eq:tobound3})},C_{(\ref{eq:tobound2})},C_{(\ref{eq:tobound4})})$ and assuming (without loss of generality) that $(c_n n^{-1/2})^{2+\psi} \leq (c_n n^{-1/2})^{1+\psi}$, i.e. $c_nn^{-1/2}\leq 1$. \qed

Proposition \ref{prop:local_conv0} is implied by the following Proposition which makes more explicit some parts of the result.
\begin{proposition} \label{prop:local_conv} Suppose Assumptions \ref{ass:gn}-\ref{ass:sample_gn} hold. Take $\gamma \in (0,1)$ and $\overline{\gamma} \in (0,\gamma)$, $c_n \geq 1$ as in Lemma \ref{lem:smooth_foc}, and $R_G$ as in Lemma \ref{lem:cv_local}. Uniformly in $\theta_{b} \in \Theta$ such that $\|\theta_b - \theta^\dagger\| \leq R_{n,G}$:
  \begin{align*}
\|\theta_{b+1}-\hat\theta_n\| \leq &\left(1-\gamma + \gamma \underline{\sigma}_{n,\varepsilon}^{-1} \sqrt{\kappa_W} [L_G\|\theta_b - \hat\theta_n\| + M_{1,Z}\varepsilon] \right)\|\theta_b - \hat\theta_n\| \\ &+ \gamma \Delta_{n,\varepsilon}(\|\theta_b-\hat\theta_n\|)  
\end{align*}
with probability $1-(1+C)/c_n$, where $R_{n,G}:=R_G - C_a c_n n^{-1/2}$, $\underline{\sigma}_{n,\varepsilon} = \underline{\sigma} - C_\sigma[ \frac{c_n n^{-1/2}}{\varepsilon} + \varepsilon]$, $M_{1,Z} = \int \|\phi^\prime (Z)\|dZ$. $C_a = \underline{\sigma}^{-1}\sqrt{\kappa_W \lambda_{\max}(\Sigma)p}$ and $C_\sigma \geq 0$ is given in Lemma \ref{lem:singular}. The remainder term $\Delta_{n,\varepsilon}$ is
\begin{align*}
  \Delta_{n,\varepsilon}(\|\theta_b-\hat\theta_n\|) \leq \frac{C_2}{\underline{\sigma}_{n,\varepsilon}^2} \left( \Gamma_{n,\varepsilon} +  \frac{(c_n n^{-1/2})^2}{\varepsilon}\|\theta_b-\hat\theta_n\|^\psi + \frac{c_n n^{-1/2}}{\varepsilon}\|\theta_b-\hat\theta_n\| \right).
\end{align*} 
Furthermore, if $\theta_b$ is such that:
\begin{align*}
  \|\theta_b-\hat\theta_n\| \leq \left( \frac{\gamma- \overline{\gamma}}{\gamma} - M_{1,Z}\varepsilon \right) \frac{\underline{\sigma}_{n,\varepsilon}}{\sqrt{\kappa_W}L_G} := R_{n,\varepsilon}, \label{eq:radius}
\end{align*}
then, with the same probability $1-(1+C)/c_n$, uniformly in $\|\theta_b - \hat\theta_n\|\leq \min(R_{n,G},R_{n,\varepsilon})$:
\begin{align}
  \|\theta_{b+1}-\hat\theta_n\| \leq (1-\overline{\gamma})\|\theta_b - \hat\theta_n\| + \gamma \Delta_{n,\varepsilon}(\|\theta_b-\hat\theta_n\|). \tag{\ref{eq:contract_ns}}
\end{align}
\end{proposition}

\paragraph{Proof of Proposition \ref{prop:local_conv}}
Let $G_b = G_{n,\varepsilon}(\theta_b)$ with $\|\theta_b - \hat\theta_n\| \leq R_G - C_a c_n n^{-1/2}$. Using the constants from Lemma \ref{lem:smooth_foc}, we have with a probability of at least $1-1/c_n$ that $\|\hat\theta_n-\theta^\dagger\|\leq R_G$. By definition of $\theta_{b+1}$, we have:
\begin{align*} 
  \theta_{b+1} &= \theta_b - \gamma (G_b^\prime W_n G_b)^{-1}G_b^\prime W_n \overline{g}_n(\theta_b),
\end{align*}
which we can re-write as:
\begin{align} 
  \hspace{-0.35cm}\theta_{b+1}-\hat\theta_n -(1-\gamma)(\theta_b-\hat\theta_n) &= - \gamma (G_b^\prime W_n G_b)^{-1}G_b^\prime W_n \left( \overline{g}_n(\theta_b) - \overline{g}_n(\hat\theta_n) - G_b (\theta_b-\hat\theta_n)\right) \label{eq:toboundp1}\\
  &+ \gamma (G_b^\prime W_n G_b)^{-1}G_b^\prime W_n \overline{g}_n(\hat\theta_n). \label{eq:toboundp2}
\end{align}
The proof boils down to finding bounds for (\ref{eq:toboundp1}) and (\ref{eq:toboundp2}). First, using the singular value bound $\underline{\sigma}_{n,\varepsilon}$ from Lemma \ref{lem:singular}, we have:
\begin{align}
  \|(\ref{eq:toboundp2})\| &\leq \gamma \underline{\sigma}_{n,\varepsilon}^{-2}\ \underline{\lambda}_W^{-1}\|G_{n,\varepsilon}(\hat\theta_n)^\prime W_n \overline{g}_n(\hat\theta_n)\| \label{eq:toboundp2_1}\\
  &+\gamma \underline{\sigma}_{n,\varepsilon}^{-2}\ \kappa_W\|G_{n,\varepsilon}(\hat\theta_n)-G_b\| \times \| \overline{g}_n(\hat\theta_n)\| \label{eq:toboundp2_2}
\end{align}
Lemma \ref{lem:smooth_foc} gives the bound $\|(\ref{eq:toboundp2_1})\| \leq \gamma \underline{\sigma}_{n,\varepsilon}^{-2}\ \underline{\lambda}_W^{-1} \Gamma_{n,\varepsilon}$, with probability $1-(1+C)/c_n$, for the $C$ in Lemma \ref{lem:stoch_bounds}. A bound $\| \overline{g}_n(\hat\theta_n)\| \leq C_c c_nn^{-1/2}$ with probability $1-(1+C)/c_n$, is derived in the proof of Lemma \ref{lem:smooth_foc}. It remains to bound $\|G_{n,\varepsilon}(\hat\theta_n)-G_{n,\varepsilon}(\theta_b)\|$. Using Lemma \ref{lem:stoch_bounds}:
\begin{align*}
    \|(\ref{eq:toboundp2_2})\| &\leq \frac{\gamma \kappa_W}{\underline{\sigma}_{n,\varepsilon}^{2}}C_c c_nn^{-1/2} \left( \|G_\varepsilon(\theta_b)-G_\varepsilon(\hat\theta_n)\| + \|[G_\varepsilon(\theta_b)-G_\varepsilon(\hat\theta_n)]-[G_{n,\varepsilon}(\theta_b)-G_{n,\varepsilon}(\hat\theta_n)]\| \right)\\
    &\leq \frac{\gamma \kappa_W}{\underline{\sigma}_{n,\varepsilon}^{2}}C_c c_nn^{-1/2} \left( L_G \|\theta_b-\hat\theta_n\| + C_\Theta L_g M_{1,Z} \frac{c_n n^{-1/2}}{\varepsilon}\|\theta_b-\hat\theta_n\|^\psi \right),
\end{align*}
with probability $1-C/c_n$. Putting the bounds together, we have:
\[ \|(\ref{eq:toboundp2})\| \leq \frac{\gamma C_{(\ref{eq:toboundp2})}}{\underline{\sigma}_{n,\varepsilon}^2} \left(\Gamma_{n,\varepsilon} + c_nn^{-1/2}\|\theta_b-\hat\theta_n\| + \frac{(c_nn^{-1/2})^2}{\varepsilon}\|\theta_b-\hat\theta_n\|^\psi \right), \]
with probability $1-(1+C)/c_n$.\footnote{The probability does not change when combining the bounds together because they all are derived from the same two events, one is the empirical process bound used for the results in Lemma \ref{lem:stoch_bounds}, and the other is a bound based on Markov's inequality used in Lemma \ref{lem:smooth_foc}.}
For the next part, consider $\|(\ref{eq:toboundp1})\| \leq \gamma \underline{\sigma}_{n,\varepsilon}^{-1}\sqrt{\kappa_W} \|\overline{g}_n(\theta_b)-\overline{g}_n(\hat\theta_n) - G_b (\theta_b-\hat\theta_n)\|$ so we can focus on bounding the difference found in the norm: 
\begin{align}
  \|\overline{g}_n(\theta_b)-\overline{g}_n(\hat\theta_n) - G_b (\theta_b-\hat\theta_n)\| &\leq \|\overline{g}_n(\theta_b)-\overline{g}_n(\hat\theta_n) - g(\theta_b)-g(\hat\theta_n)\| \label{eq:toboundp1_1}\\
  &+ \|g(\theta_b)-g(\hat\theta_n) - G(\theta_b)(\theta_b-\hat\theta_n)\| \label{eq:toboundp1_2}\\
  &+ \| G(\theta_b) - G_\varepsilon(\theta_b)\| \times \|(\theta_b-\hat\theta_n)\| \label{eq:toboundp1_3}\\
  &+ \| G_{n,\varepsilon}(\theta_b) - G_\varepsilon(\theta_b)\| \times \|(\theta_b-\hat\theta_n)\|. \label{eq:toboundp1_4}
\end{align}
Lemma \ref{lem:stoch_bounds} gives $\|(\ref{eq:toboundp1_1})\| \leq C_\Theta L_g c_n n^{-1/2}\|\theta_b - \hat\theta_n\|^\psi$, with probability $1-C/c_n$. Lipschitz continuity of $G$ implies $\|(\ref{eq:toboundp1_2})\| \leq L_G \|\theta_b - \hat\theta_n\|^2$, which plays the same role for convergence as in Lemma \ref{lem:cv_local}. Lemma \ref{lem:dem_bounds} gives $\|(\ref{eq:toboundp1_3})\| \leq M_{1,Z}L_G \varepsilon \|\theta_b-\hat\theta_n\|$. For the last term:
\begin{align*}
  \| G_{n,\varepsilon}(\theta_b) - G_\varepsilon(\theta_b)\| &\leq \| [G_{n,\varepsilon}(\theta_b) - G_\varepsilon(\theta_b)]-[ G_{n,\varepsilon}(\theta^\dagger) - G_\varepsilon(\theta^\dagger)]\|\\
  &+ \| G_{n,\varepsilon}(\theta^\dagger) - G_\varepsilon(\theta^\dagger) - \frac{1}{\varepsilon}[ \overline{g}_n(\theta^\dagger) - g(\theta^\dagger) ] \|
  + \frac{\|\overline{g}_n(\theta^\dagger)\|}{\varepsilon}\\
  &\leq C_\Theta L_g M_{1,Z}R_G^\psi\frac{c_n n^{-1/2}}{\varepsilon} + C_\Theta L_g \varepsilon^{\psi-1} c_n n^{-1/2} + \sqrt{p\lambda_{\max}(\Sigma)}c_nn^{-1/2},
\end{align*}
with probability $1-(1+C)/c_n$. For $\varepsilon \leq 1$ this implies $\|(\ref{eq:toboundp1_4})\| \leq C_{(\ref{eq:toboundp1_4})} \frac{c_nn^{-1/2}}{\varepsilon} \|\theta_b-\hat\theta_n\|$, with the same probability.
Putting everything together, we have:
\begin{align}
\|\theta_{b+1}-\hat\theta_n -(1-\gamma)(\theta_b-\hat\theta_n)\| \leq \gamma\frac{\sqrt{\kappa_W}L_G}{\underline{\sigma}_{n,\varepsilon}} \left( \|\theta_b-\hat\theta_n\|+M_{1,Z}\varepsilon \right) \|\theta_b-\hat\theta_n\| + \gamma \Delta_{n,\varepsilon}, \label{eq:contract_ineq}
\end{align}
with probability $1-(1+C)/c_n$ where, after some simplifications:
\[ \Delta_{n,\varepsilon} \leq \frac{C_2}{\underline{\sigma}_{n,\varepsilon}^2} \left( \Gamma_{n,\varepsilon} +  \frac{(c_nn^{-1/2})^2}{\varepsilon}\|\theta_b-\hat\theta_n\|^\psi + \frac{c_nn^{-1/2}}{\varepsilon}  \|\theta_b-\hat\theta_n\| \right). \]
\qed

Theorem \ref{th:sgn_local0} in the main text, is implied by the following Theorem, which makes more explicit parts of the result.

\begin{theorem} \label{th:sgn_local} Assume, without loss of generality, that $R_{n,\varepsilon} \leq R_{n,G}$. Take $\gamma \in (0,1)$, $\overline{\gamma} \in (0,\gamma)$, and $\tau \in (0,1)$. Suppose Assumptions \ref{ass:gn}-\ref{ass:sample_gn} hold, and assume that $\varepsilon$, $c_nn^{-1/2}$ are small enough that the following two inequalities hold: 
  \begin{align}
     \Delta_{n,\varepsilon}(R_{n,\varepsilon}) &\leq  \frac{\overline{\gamma}}{\gamma}R_{n,\varepsilon}, \label{eq:contract_cond}\\
    \frac{(c_n n^{-1/2})^2}{\varepsilon}+\frac{c_n n^{-1/2}}{\varepsilon} &< \frac{\overline{\gamma}\underline{\sigma}^2\tau}{\gamma C_2}. \label{eq:n_cn_eps1}
  \end{align}
  Then, with probability $1-(1+C)/c_n$, uniformly in $\|\theta_0-\hat\theta_n\| \leq R_{n,\varepsilon}$, for all $b \geq 0$:
  \begin{align} \|\theta_{b}-\hat\theta_n\| \leq &\Big(1-\overline{\gamma} + \tau\overline{\gamma} \Big)^b \|\theta_0-\hat\theta_n\| \notag\\ &+  \frac{\gamma}{\overline{\gamma}(1-\tau)}\frac{C_2}{\underline{\sigma}_{n,\varepsilon}^2} \left(\Gamma_{n,\varepsilon}  + C_{n,\varepsilon} \left[ \frac{(c_nn^{-1/2})^2}{\varepsilon} \right]^{\frac{1}{1-\psi}} \right), \tag{\ref{eq:th1_main}} 
  \end{align}
  setting $[ \frac{(c_nn^{-1/2})^2}{\varepsilon}]^{\frac{1}{1-\psi}}=0$ and $C_{n,\varepsilon}=0$ if $\psi=1$, while $C_{n,\varepsilon} = \left( \frac{\overline{\gamma}\underline{\sigma}^2\tau}{\gamma C_2} - \frac{c_nn^{-1/2}}{\varepsilon} \right)^\frac{\psi}{\psi-1}$ if $\psi<1$.
  Suppose that $\varepsilon = o(1)$ and $\sqrt{n}\varepsilon \to +\infty$, then for $c_n = O(1)$: 
  \begin{align*} 
    b \geq \frac{\log(\Gamma_{n,\varepsilon})-\log(\|\theta_0-\hat\theta_n\|)}{\log(1-\overline{\gamma} + \tau \overline{\gamma})} \Rightarrow \sqrt{n}\|\theta_{b}-\hat\theta_n\| = o_p(1). 
  \end{align*}
  \end{theorem}

\paragraph{Proof of Theorem \ref{th:sgn_local}} If (\ref{eq:contract_cond}) holds, then Proposition \ref{prop:local_conv} implies that for $\|\theta_0-\hat\theta_n\| \leq R_{n,\varepsilon}$, with probability $1-(1+C)/c_n$, we have by recursion for all $b\geq 0$:
\[ \|\theta_{b+1}-\hat\theta_n\| \leq (1-\overline{\gamma}) \|\theta_b-\hat\theta_n\| + \gamma\Delta_{n,\varepsilon}(R_{n,\varepsilon}) \leq R_{n,\varepsilon}, \]
since $\Delta_{n,\varepsilon}$ is increasing so that we have stability of the contraction property.

Let $a_n = (c_n n^{-1/2})^2/\varepsilon$ and $b_n = (c_nn^{-1/2})/\varepsilon$. Let $x_b = \|\theta_b-\hat\theta_n\|$, then equation (\ref{eq:contract_ns}) can be re-written as:
\[ x_{b+1} \leq (1-\overline{\gamma})x_b + \gamma \frac{C_2}{\underline{\sigma}_{n,\varepsilon}^2} \left( \Gamma_{n,\varepsilon} + a_n x_b^\psi + b_n x_b \right). \]
For $\psi=1$, condition (\ref{eq:n_cn_eps1}) implies that $a_n+b_n \leq \tau \overline{\gamma}$ and:
\[ x_{b+1} \leq (1-\overline{\gamma} + \tau \overline{\gamma})x_b + \gamma \frac{C_2}{\underline{\sigma}_{n,\varepsilon}^2}  \Gamma_{n,\varepsilon}, \]
from which (\ref{eq:th1_main}) immediately follows. For $\psi<1$, using $a_n x_b^\psi = (a_n x_b^{\psi-1})x_b$, either:
\begin{itemize}
  \item $x_b$ is such that $x_b \geq a_n^{\frac{1}{1-\psi}} C_{n,\varepsilon}^{1/\psi}$. Then we have $1-\overline{\gamma} + \gamma \frac{C_2}{\underline{\sigma}_{n,\varepsilon}^2}(b_n + a_n x_b^{\psi-1}) \leq 1-\overline{\gamma} + \tau \overline{\gamma}$ which implies:
  $x_{b+1} \leq (1-\overline{\gamma} + \tau \overline{\gamma})x_b + \gamma \frac{C_2}{\underline{\sigma}_{n,\varepsilon}^2}  \Gamma_{n,\varepsilon},$
  \item or $x_b$ is such that  $x_b \leq a_n^{\frac{1}{1-\psi}} C_{n,\varepsilon}^{1/\psi}$. Then we have $1-\overline{\gamma} + \gamma \frac{C_2}{\underline{\sigma}_{n,\varepsilon}^2}b_n \leq 1-\overline{\gamma} + \tau \overline{\gamma}$, which implies:
  $x_{b+1} \leq (1-\overline{\gamma} + \tau \overline{\gamma})x_b + \gamma \frac{C_2}{\underline{\sigma}_{n,\varepsilon}^2} \left( \Gamma_{n,\varepsilon} + C_{n,\varepsilon} a_n^{\frac{1}{1-\psi}} \right).$
\end{itemize}
Majoring these two inequalities yields (\ref{eq:th1_main}) for $\psi<1$. 
Take $b  \geq \frac{\log(\Gamma_{n,\varepsilon})-\log(\|\theta_0-\hat\theta_n\|)}{\log(1-\overline{\gamma} + \tau \overline{\gamma})}$, then 
\begin{align*}
  &(1-\overline{\gamma})^b \|\theta_0-\hat\theta_n\| \leq \Gamma_{n,\varepsilon}
  \Rightarrow \|\theta_b-\hat\theta_n\| \leq (1+\gamma \frac{C_2}{\underline{\sigma}_{n,\varepsilon}^2}) \Gamma_{n,\varepsilon} + \gamma \frac{C_2}{\underline{\sigma}_{n,\varepsilon}^2} C_{n,\varepsilon} a_n^{\frac{1}{1-\psi}},
\end{align*}
with probability $1-(1+C)/c_n$. If $\varepsilon = o(1)$ and $\sqrt{n} \varepsilon \to \infty$, then $\sqrt{n}\Gamma_{n,\varepsilon} = o(1)$ and $\sqrt{n}a_n^{\frac{1}{1-\psi}}=o(1)$ for $c_n = O(1)$ which implies the desired result. \qed\\

The following considers global convergence of the combined local and global search steps. The term $\tilde\Gamma_{n,\varepsilon}$ which will determine the estimation accuracy is more complicated than $\Gamma_{n,\varepsilon}$ found in Theorem \ref{th:sgn_local0}, equation (\ref{eq:th1_main}):
\begin{align*}
  \tilde\Gamma_{n,\varepsilon} &= \Gamma_{n,\varepsilon} + (c_nn^{-1/2})^2\varepsilon^{\psi-1} + (c_nn^{-1/2})^{1+\psi/2+\psi^2/2}  + \sqrt{\varepsilon}(c_nn^{-1/2})^{1+\psi/2}\\ & + (c_nn^{-1/2})^{(3+\psi)/2}\varepsilon^{(\psi-1)/2}+ [a_n^{1+\psi/(1-\psi)} +  c_nn^{-1/2} a_n^{1/(1-\psi)}]\max[C_n(\tau,\tilde\gamma),C_n(\tau,\tilde\gamma)^{\psi}], 
\end{align*}
where $a_n = c_nn^{-1/2} + (c_nn^{-1/2})^2 \varepsilon^{-1}$ for $\psi<1$ and $a_n=0$ otherwise, and for $\tau,\tilde\gamma \in (0,1)$: $C_n(\tau,\tilde\gamma) = (\tau\tilde\gamma - c_nn^{-1/2})^{\frac{1}{\psi-1}}$ if $\psi <1$, and $C_n(\tau,\tilde\gamma)= 0$ otherwise. As in Theorem \ref{th:sgn_local0}, it still holds that: $\sqrt{n}\tilde\Gamma_{n,\varepsilon} = o(1)$ when $\varepsilon = o(1)$ and $\sqrt{n}\varepsilon \to \infty$.
\begin{theorem} \label{th:sgn_global} Suppose Assumptions \ref{ass:gn}-\ref{ass:sample_gn} hold. Take $\gamma \in (0,1)$, $\overline{\gamma}\in(0,\gamma)$, and pick $\tau \in (0,1)$ small enough to satisfy: \begin{align} \frac{1+\tau}{1-\tau}\frac{1+\overline{\gamma}/2}{1-\overline{\gamma}/2}(1-\overline{\gamma}) < 1.  \label{eq:small_tau}\end{align} 
  Let $\tilde \gamma \in (0,1)$ be such that: $1-\tilde{\gamma} := \frac{1+\tau}{1-\tau}(1-\overline{\gamma})$.  There exists $\underline{r}_{n,g}(\overline{\gamma}/2)$ with $\liminf_{n \to \infty} \underline{r}_{n,g}(\overline{\gamma}/2) >0$ and $k_n \geq 0$ satisfying: $\sup_{\theta \in \Theta} \left( \inf_{ 0 \leq \ell \leq k_n} \|\theta^\ell - \theta\|_{G^\prime W_n G} \right) \leq \underline{r}_{n,g}$,  such that for $c_nn^{-1/2},\varepsilon$ small enough, and $b = k_n + j$, with $j \geq 0$:
\begin{align}
  \|\theta_b- \hat\theta_n\|_{G^\prime W_n G} &\leq (1-\tilde \gamma + \tau \tilde \gamma)^{j} \frac{1+\overline{\gamma}/2}{1-\overline{\gamma}/2} \underline{r}_{n,g} \notag\\&+ \frac{\tilde{C}_2}{(1-\overline{\gamma}/2)^3(1-\tau)\tilde \gamma [\overline{\gamma}-(\overline{\gamma}/2)^2]}\frac{1}{\min(1,\underline{\sigma}_{n,\varepsilon},\underline{\sigma}_{n,\varepsilon}^2)}\tilde{\Gamma}_{n,\varepsilon}, \label{eq:global_local_cv_rate}
\end{align}
with probability $1-(1+C)/c_n$. Suppose that $\varepsilon = o(1)$, and $\sqrt{n}\varepsilon \to \infty$, then for $c_n = O(1)$:
\begin{align}
    j \geq \frac{ \log(\Gamma_{n,\varepsilon}) - \log\left( \frac{1+\overline{\gamma}/2}{1-\overline{\gamma}/2}\underline{r}_{n,g}(\overline{\gamma}/2) \right) }{\log(1-\tilde \gamma + \tau \tilde \gamma)} \Rightarrow \sqrt{n}\|\theta_b- \hat\theta_n\|_{G^\prime W_n G} = o_p(1). \label{eq:cv_global_ns}
\end{align}
\end{theorem}

\paragraph{Proof of Theorem \ref{th:sgn_global}:}
The main steps broadly follow those in Lemma \ref{lem:cv_global} with some additional terms that account for non-smoothness and sampling uncertainty. In the following Lemmas \ref{lem:Reta}, \ref{lem:separ_ns}, \ref{lem:sample_obj_constract}, and
\ref{lem:norm_squared} will be applied with $\eta = \overline{\gamma}/2 \in (0,1)$.

Take $k_n \geq 1$ such that: $\inf_{0 \leq j \leq k_n} \sup_{\theta - \theta^j} \|\theta-\theta^j\|_{G^\prime W_n G} \leq \underline{r}_{n,g}(\overline{\gamma}/2)$, defined in Lemma \ref{lem:separ_ns}. As in the proof of Lemma \ref{lem:cv_global} assume, without loss of generality, that for $b = k_n$ we have $\|\hat\theta_n - \theta_b\| \leq \underline{r}_{n,g}(\overline{\gamma}/2)$. Using $\underline{r}_{n,g}(\overline{\gamma}/2) \leq \underline{\sigma}\underline{\lambda}_W^{1/2} R_n(\overline{\gamma}/2)$ we also have $\|\theta_b-\hat\theta_n\| \leq R_n(\overline{\gamma}/2)$ so that Lemmas \ref{lem:Reta} and \ref{lem:sample_obj_constract} can be applied. Let $x_b = \|\overline{g}_n(\theta_b) - \overline{g}_n(\hat\theta_n)\|_{W_n}$, we have:
\begin{align} x_{b+1} \leq (1-\overline{\gamma})x_b + \Delta_{2,n,\varepsilon}(x_b), \label{eq:contract_global_b}\end{align}
with probability $1-(1+C)/c_n$. Using the reverse triangular inequality, this implies that:
\begin{align*}
  \|\overline{g}_n(\theta_{b+1})\|_{W_n} \leq (1-\overline{\gamma})x_b + \Delta_{2,n,\varepsilon}(x_b) + \|\overline{g}_n(\hat\theta_n)\|_{W_n}, 
\end{align*}
with probability $1-(1+C)/c_n.$ And now Lemma \ref{lem:Reta} applied to $x_b$ yields:
\begin{align}
  \|\overline{g}_n(\theta_{b+1})\|_{W_n} &\leq (1+\overline{\gamma}/2)(1-\overline{\gamma})\underline{r}_{n,g}(\overline{\gamma}/2) \notag \\ &+ \Delta_{2,n,\varepsilon}(x_b) + \|\overline{g}_n(\hat\theta_n)\|_{W_n} + \overline{\lambda}_W L_g C_\Theta (c_nn^{-1/2})^{1+\psi}, \label{eq:main_global_ns}
\end{align}
with probability $1-(1+C)/c_n.$ Suppose $\theta^{b+1}$ is such that:
$\|\overline{g}_n(\theta^{b+1})\|_{W_n} \leq \|\overline{g}_n(\theta_{b+1})\|_{W_n}$, 
then $\theta^{b+1}$ satisfies the same inequality (\ref{eq:main_global_ns}). Notice that $(1+\overline{\gamma}/2)(1-\overline{\gamma}) <1$, $\liminf_{n \to \infty} \underline{r}_{n,g}(\overline{\gamma}/2) >0$, since $\overline{\gamma}$ is fixed but $( \Delta_{2,n,\varepsilon}(x_b) + \|\overline{g}_n(\hat\theta_n)\|_{W_n} + \overline{\lambda}_W L_g C_\Theta (c_nn^{-1/2})^{1+\psi} ) \leq ( \Delta_{2,n,\varepsilon}(x_b) + C_c c_nn^{} + \overline{\lambda}_W L_g C_\Theta (c_nn^{-1/2})^{1+\psi} )$ for which the $\limsup_{n\to\infty}$ is zero when $c_nn^{-1/2} \to 0$, $\varepsilon \to 0$, and $\sqrt{n} \varepsilon/c_n \to \infty$. This implies that for $c_nn^{-1/2}$, $\varepsilon$ small enough and $\sqrt{n} \varepsilon/c_n$ large enough, we have:
$\|\overline{g}_n(\theta^{b+1})\|_{W_n} \leq \underline{r}_{n,g}(\overline{\gamma}/2)$,
with probability $1-(1+C)/c_n$. Now apply Lemma \ref{lem:separ_ns} to find that it implies $\|\theta^{b+1} - \hat\theta_n\|_{G^\prime W_n G} \leq \overline{r}_{n,g}(\overline{\gamma}/2)$ with the same probability, which in turn implies $\|\theta^{b+1} - \hat\theta_n\| \leq R_n(\overline{\gamma}/2)$. Lemma \ref{lem:Reta} now applies to $\theta^{b+1}$:
\begin{align}
  (1-\overline{\gamma}/2)\|\theta^{b+1}-\hat\theta_n\|_{G^\prime W_n G} \leq \|\overline{g}_n(\theta^{b+1})-\overline{g}_n(\hat\theta_n)\|_{W_n} + \overline{\lambda}_W L_g C_\Theta (c_nn^{-1/2})^{1+\psi}, \label{eq:b_plus_1s}
\end{align}
again with probability $1-(1+C)/c_n$. The main issue here is that $\|\overline{g}_n(\theta^{b+1})\|_{W_n} \leq \|\overline{g}_n(\theta_{b+1})\|_{W_n}$ does not directly imply an ordering between $\|\overline{g}_n(\theta^{b+1})-\overline{g}_n(\hat\theta_n)\|_{W_n}$ and $x_{b+1}$. Pick $\tau \in (0,1)$; for $c_nn^{-1/2}$ small enough, Lemma \ref{lem:norm_squared} applies and we have:
\[ (1-\tau)\|\overline{g}_n(\theta^{b+1}) - \overline{g}_n(\hat\theta_n)\|_{W_n} \leq (1+\tau)\|\overline{g}_n(\theta_{b+1}) - \overline{g}_n(\hat\theta_n)\|_{W_n} + \sqrt{2}\Gamma_{2,n,\varepsilon}(\overline{\gamma}/2,\tau), \]
with probability $1-(1+C)/c_n$, $\Gamma_{2,n,\varepsilon}$ is made explicit in (\ref{eq:Gamma2n}). Plug this back into (\ref{eq:contract_global_b}):
\begin{align*}
  (1-\tau)(1-\overline{\gamma}/2)\|\theta^{b+1}-\hat\theta_n\|_{G^\prime W_n G} &\leq (1+\tau)(1+\overline{\gamma}/2)(1-\overline{\gamma})\underline{r}_{n,g}(\overline{\gamma}/2)\\ 
  &+ \sqrt{2}\Gamma_{2,n,\varepsilon}(\overline{\gamma}/2,\tau) + \overline{\lambda}_W L_g C_\Theta (c_nn^{-1/2})^{1+\psi} + \Delta_{2,n,\varepsilon}(x_b),
\end{align*}
with probability $1-(1+C)/c_n.$ When $\tau \in (0,1)$ is small enough that:
$\frac{1+\tau}{1-\tau}\frac{1+\overline{\gamma}/2}{1-\overline{\gamma}/2}(1-\overline{\gamma}) <1$,
using the same arguments as above, we have for $c_nc^{-1/2}$, $\varepsilon$ small enough and $\sqrt{n}\varepsilon/c_n$ large enough that $\|\theta^{b+1}-\hat\theta_n\|_{G^\prime W_n G} \leq \underline{r}_{n,g}(\overline{\gamma}/2)$ with probability $1-(1+C)/c_n$. This implies that with the same probability $\|\theta_{b+j}-\hat\theta_n\| \leq \underline{r}_{n,g}(\overline{\gamma}/2)$, for all $j \geq 0$. Combine (\ref{eq:contract_global_b}) and (\ref{eq:b_plus_1s}) to find that even after setting $\theta_{b+1}=\theta^{b+1}$ we have:
\begin{align}
  x_{b+1} \leq \frac{1+\tau}{1-\tau}(1-\overline{\gamma})  x_b + \Delta_{2,n,\varepsilon}(x_b) + \Gamma_{2,n,\varepsilon}(\overline{\gamma}/2,\tau) + \overline{\lambda}_W L_g C_\Theta (c_nn^{-1/2})^{1+\psi},
\end{align}
with probability $1-(1+C)/c_n$, where:
$1-\tilde \gamma := \frac{1+\tau}{1-\tau}(1-\overline{\gamma}) < 1$, 
by assumption about $\tau$. Now we will use similar steps as in the proof of Theorem \ref{th:sgn_local} to bound $\Delta_{2,n,\varepsilon}(x_b)$ and get a $(1-\tilde{\gamma} + \tau \tilde{\gamma})$ convergence rate. Set $a_n = [c_nn^{-1/2} + (c_nn^{-1/2})^2 \varepsilon^{-1}]$ and $b_n = c_nn^{-1/2}$. We have:
\[ \Delta_{2,n,\varepsilon}(x_b) = C_4 \left( \Gamma_{n,\varepsilon} + \varepsilon^{\psi-1}(c_nn^{-1/2})^2 + a_n x_b^\psi + b_n x_b \right). \]
If $\psi = 1$ and $a_n + b_n \leq \tau \tilde{\gamma}$, then we get:
\begin{align*}
  x_{b+1} \leq (1-\tilde \gamma + \tau \tilde \gamma)  x_b + C_4 \left( \Gamma_{n,\varepsilon} + \varepsilon^{\psi-1}(c_nn^{-1/2})^2 \right) +  \Gamma_{2,n,\varepsilon}(\overline{\gamma}/2,\tau) + \overline{\lambda}_W L_g C_\Theta (c_nn^{-1/2})^{1+\psi},
\end{align*}
with probability $1-(1+C)/c_n$, as desired. If $\psi < 1$, as in the proof of Theorem \ref{th:sgn_local} there are two cases:
\begin{itemize}
  \item $x_b \geq a_n^{\frac{1}{1-\psi}} \left( \frac{\tau \tilde \gamma - b_n}{C_4}\right)^{\frac{1}{\psi-1}}$, then $a_n x_b^{\psi} + b_n x_b \leq \tau \tilde{\gamma} x_b$,
  \item $x_b \leq a_n^{\frac{1}{1-\psi}} \left( \frac{\tau \tilde \gamma - b_n}{C_4}\right)^{\frac{1}{\psi-1}}$, then $[a_n^{1 + \frac{\psi}{1-\psi}} + b_n a_n^{\frac{1}{1-\psi}}] \max( C_n(\tau),C_n(\tau)^\psi)$.
\end{itemize}
Where $C_n(\tau) = \left( \frac{\tau \tilde \gamma - b_n}{C_4}\right)^{\frac{1}{\psi-1}}$ when $\psi <1$, $C_n(\tau)=0$ when $\psi =1$. A common upper-bound:
\begin{align*}
  x_{b+1} \leq (1-\tilde \gamma + \tau \tilde \gamma)  x_b &+ C_4 \left( \Gamma_{n,\varepsilon} + \varepsilon^{\psi-1}(c_nn^{-1/2})^2 \right) +  \Gamma_{2,n,\varepsilon}(\overline{\gamma}/2,\tau) + \overline{\lambda}_W L_g C_\Theta (c_nn^{-1/2})^{1+\psi}\\
  &+ [a_n^{1 + \frac{\psi}{1-\psi}} + b_n a_n^{\frac{1}{1-\psi}}] \max( C_n(\tau),C_n(\tau)^\psi),
\end{align*}
with probability $1-(1+C)/c_n$. Iterate the above inequality over $j=0,\dots$ to find for $b = k_n + j$:
\begin{align*}
  x_{b+j} \leq (1-\tilde \gamma + \tau \tilde \gamma)^j x_{k_n}  &+ \frac{1}{(1-\tau)\tilde\gamma}\Big[ C_4 \left( \Gamma_{n,\varepsilon} + \varepsilon^{\psi-1}(c_nn^{-1/2})^2 \right) +  \Gamma_{2,n,\varepsilon}(\overline{\gamma}/2,\tau) \\
  &+ \overline{\lambda}_W L_g C_\Theta (c_nn^{-1/2})^{1+\psi} + [a_n^{1 + \frac{\psi}{1-\psi}} + b_n a_n^{\frac{1}{1-\psi}}] \max( C_n(\tau),C_n(\tau)^\psi) \Big],
\end{align*}
with probability $1-(1+C)/c_n$. Apply Lemma \ref{lem:Reta} to both sides of the inequality:
\begin{align*}
  \|\theta_{b}-\hat\theta_n\|_{G^\prime W_n G} &\leq (1-\tilde \gamma + \tau \tilde \gamma)^j \frac{1+\overline{\gamma}/2}{1-\overline{\gamma}/2} \underline{r}_{n,g}(\overline{\gamma}/2)\\  &+ \frac{(1-\overline{\gamma}/2)^{-1}}{(1-\tau)\tilde\gamma}\Big[ C_4 \left( \Gamma_{n,\varepsilon} + \varepsilon^{\psi-1}(c_nn^{-1/2})^2 \right) +  \Gamma_{2,n,\varepsilon}(\overline{\gamma}/2,\tau) \\
  &+ \overline{\lambda}_W L_g C_\Theta (c_nn^{-1/2})^{1+\psi} + [a_n^{1 + \frac{\psi}{1-\psi}} + b_n a_n^{\frac{1}{1-\psi}}] \max( C_n(\tau),C_n(\tau)^\psi) \Big]\\
  &+ 2 (1-\overline{\gamma}/2)^{-1}\overline{\lambda}_W L_g C_\Theta (c_nn^{-1/2})^{1+\psi},
\end{align*}
with probability $1-(1+C)/c_n$. This inequality simplifies to (\ref{eq:global_local_cv_rate}) using $\tilde{\Gamma}_{n,\varepsilon}$ which is given in the main text. The constant $\tilde{C}_2$ combines those in the above inequality and $C_5$ in (\ref{eq:Gamma2n}). Just as in the proof of Theorem \ref{th:sgn_local}, when
$j \geq \frac{ \log(\Gamma_{n,\varepsilon}) - \log\left( \frac{1+\overline{\gamma}/2}{1-\overline{\gamma}/2}\underline{r}_{n,g}(\overline{\gamma}/2) \right) }{\log(1-\tilde\gamma + \tau \tilde \gamma)}$,
the leading term is less than $\Gamma_{n,\varepsilon} = o(n^{-1/2})$. Also, under the stated assumptions: $\Gamma_{2,n,\varepsilon}(\overline{\gamma}/2,\tau) = o(n^{-1/2})$, $a_n^{1 + \frac{\psi}{1-\psi}} = o(n^{-1/2})$, $b_n = o(n^{-1/2})$, and $b_n a_n^{\frac{1}{1-\psi}} = o(n^{-1/2})$ which implies the desired result: $\sqrt{n}\|\theta_{b}-\hat\theta_n\|_{G^\prime W_n G} = o_p(1)$ and concludes the proof.
\qed

\section{Benchmark: the Population Problem} \label{sec:popu}
The following derives results similar to Section \ref{sec:main} for the population objective $\|g(\theta)\|_W^2$ without smoothing. The results highlight some of the tradeoffs that are made when setting tuning parameters without the complications of sampling uncertainty.
\paragraph{Local Convergence.} Take a starting value $\theta_0 \in \Theta$ and iterate:
\begin{align}
  \theta_{b+1} = \theta_b - \gamma (G_b^\prime W G_b)^{-1} G_b^\prime W g(\theta_b),\quad b=0,1,\dots \tag{\ref{eq:step1}}
\end{align}
where $G_b = G(\theta_b)$. Because $G(\theta^\dagger)$ has full rank and $G$ is continuous, the smallest singular value $\sigma_{\min}[G(\theta)]$ is strictly positive in a neighborhood of $\theta=\theta^\dagger$. The following Lemma proves the local convergence of $\theta_b$ using this local identification condition. 
\begin{lemma} \label{lem:cv_local} Suppose Assumption \ref{ass:gn} i-iv, vi hold. Let $0<\underline{\sigma} < \sigma_{\min}[G(\theta^\dagger)]$, and $R_G>0$ be such that $\|\theta-\theta^\dagger\|\leq R_G \Rightarrow \underline{\sigma} \leq \sigma_{\min}[G(\theta)]$. Pick any $\gamma \in (0,1)$ and $\overline{\gamma} \in (0,\gamma)$, then for any $\theta_0 \in \Theta$ such that: $ \|\theta_0 - \theta^\dagger\| \leq R := \min\left( R_G, [\gamma-\overline{\gamma}]\frac{\underline{\sigma}}{\gamma L_G \sqrt{\kappa_W}} \right),$ we have:
\[ \|\theta_b - \theta^\dagger\| \leq (1-\overline{\gamma})^b \|\theta_0-\theta^\dagger\|. \]
\end{lemma}
Lemma \ref{lem:cv_local} relates to results for Gauss-Newton applied to non-linear systems of equations. The main differences here compared to e.g. \citet[Th11.2]{nocedal-wright:06} is that $g$ can be overdetermined, which gives prominence to the weighting matrix, and the choice of learning rate is linked to the area $R$ and the feasible rate $\overline{\gamma} \in (0,\gamma)$ of convergence. 

\paragraph{Global Convergence.} For an arbitrary starting value $\theta_0$, Lemma \ref{lem:cv_local} does not guarantee convergence to $\theta^\dagger$. Using a covering sequence $(\theta^b)_{b \geq 0}$, the following augments the local search with a global grid search step:
\begin{align}
  &\theta_{b+1} = \theta_b - \gamma (G_b^\prime W G_b)^{-1}G_b^\prime W g(\theta_b) \tag{\ref{eq:step1}}\\
  &\text{if } \|g(\theta^{b+1})\|_W < \|g(\theta_{b+1})\|_W, \text{ set } \theta_{b+1} = \theta^{b+1}.  \label{eq:Glob_GN}
\end{align}
Step (\ref{eq:Glob_GN}) resets the local search (\ref{eq:step1}) when $\theta^{b+1}$ strictly reduces the objective function. The following Lemma shows that this is sufficient to achieve global convergence.

\begin{lemma} \label{lem:cv_global} Suppose Assumption \ref{ass:gn} holds. Take $\gamma \in (0,1)$, $\overline{\gamma} \in (0,\gamma)$. There is a $\underline{r}_g > 0$ such that for $k \geq 0$ satisfying: $D(\theta^1,\dots,\theta^k) = \sup_{\theta \in \Theta} \left( \inf_{0\leq \ell \leq k} \| \theta^\ell - \theta \|_{G^\prime W G} \right) \leq \underline{r}_g$,
  and any $b = k + j$ with $j \geq 0$, we have:
  \[ \|\theta_b -\theta^\dagger\|_{G^\prime W G} \leq \left(1 - \overline{\gamma}\right)^j\frac{1+\overline{\gamma}/2}{1-\overline{\gamma}/2}\underline{r}_g. \]
\end{lemma}
The constant $\underline{r}_g$, constructed in the proof, depends on $R$ from Lemma \ref{lem:cv_local} as well as $W$, the local behaviour of $G$ around $\theta^\dagger$, and the global identification condition through $\delta(\cdot)$. It is small enough that local convergence holds, as in Lemma \ref{lem:cv_local}, and there is no value away from $\theta^\dagger$ that improves the fit of the model. After those $k$ iterations, the global step can only bind with a $\theta^{b+1}$ that is locally convergent. The algorithm transitions from a slow global to a fast local convergence to the solution.

\newpage

\begin{titlingpage} 
  \emptythanks
  \title{ {Supplement to\\ \lQ {\bf Noisy, Non-Smooth, Non-Convex Estimation of Moments Conditions Models}''}}
  \author{Jean-Jacques Forneron\thanks{Department of Economics, Boston University, 270 Bay State Rd, MA 02215 Email: jjmf@bu.edu}}
  \setcounter{footnote}{0}
  \setcounter{page}{0}

  \clearpage 
  \maketitle 
  \thispagestyle{empty} 
  \begin{center}
  This Supplemental Material consists of Appendices \ref{apx:prelim_proofs}, \ref{apx:prelim_extensions}, \ref{apx:proofs_prelim_extensions}, \ref{apx:proofs_extensions}, \ref{apx:extra_MC_Emp}, \ref{apx:comp}, \ref{apx:cover}, and \ref{sec:Code} to the main text.
  \end{center}
\end{titlingpage}

\setcounter{page}{1}

\section{Proofs for the Preliminary Results} \label{apx:prelim_proofs}
\subsection{Preliminary Results for the Population Problem}
\paragraph{Proof of Lemma \ref{lem:prelim_global}.}
First, we will derive (\ref{eq:contract}). Take $\theta_b$ such that $\|\theta_b-\theta^\dagger\|\leq R$, as defined in Lemma \ref{lem:cv_local}, and let $G_b = G(\theta_b)$, we have for some intermediate value $\tilde \theta_b$ between $\theta_b$ and $\theta_{b+1}$:\footnote{First notice that: $\theta_{b+1}-\theta_b = -\gamma (B_b^\prime W G_b)^{-1}G_b^\prime W g(\theta_b)$. Then, the mean-value theorem implies: $g(\theta_{b+1})-g(\theta_b) = G(\tilde\theta_b)(\theta_{b+1}-\theta_b)$.}
\begin{align}
  g(\theta_{b+1}) - (1-\gamma)g(\theta_b) &=  \gamma \left( I_d - G(\tilde \theta_b)(G_b^\prime W G_b)^{-1} G_b^\prime W  \right) g(\theta_b) \notag\\
  &= \gamma \left( I_d - G_b(G_b^\prime W G_b)^{-1} G_b^\prime W  \right) g(\theta_b) \label{eq:term1}\\
  &+ \gamma \left( G_b - G(\tilde\theta_b) \right) (G_b^\prime W G_b)^{-1} G_b^\prime W   g(\theta_b) \label{eq:term2}.
\end{align}
To bound (\ref{eq:term1}), notice that $[I_d - G_b(G_b^\prime W G_b)^{-1} G_b^\prime W]G_b=0$ so that:
\begin{align*}
  (\ref{eq:term1}) &= \gamma \left( I_d - G_b(G_b^\prime W G_b)^{-1} G_b^\prime W  \right) [g(\theta_b)-G_b(\theta_b-\theta^\dagger)]\\
  &= \gamma \left( I_d - G_b(G_b^\prime W G_b)^{-1} G_b^\prime W  \right) [G(\bar \theta_b)-G_b](\theta_b-\theta^\dagger),
\end{align*}
for another intermediate value $\bar \theta_b$ between $\theta_b$ and $\theta^\dagger$. By Lipschitz continuity:
\begin{align*}
  \|(\ref{eq:term1})\| &\leq \gamma L_G\|\theta_b-\theta^\dagger\|^2 \leq \left(\frac{\gamma L_G}{\underline{\sigma}}\|\theta_b-\theta^\dagger\|\right)\|g(\theta_b)\|,
\end{align*}
using $\|g(\theta_b)\| = \|G(\bar{\theta}_b)(\theta_b-\theta^\dagger)\| \geq \underline{\sigma}\|\theta_b-\theta^\dagger\|$.
To bound (\ref{eq:term2}), Lemma \ref{lem:cv_local} implies that $\|\tilde \theta_b-\theta^\dagger\| \leq \|\theta_b-\theta^\dagger\|$, then using the Lipschitz continuity of $G$:
\begin{align*}
  \|(\ref{eq:term2})\| &\leq \left( \frac{\gamma L_G\kappa_W}{\underline{\sigma}}\|\theta_b-\theta^\dagger\| \right)\|g(\theta_b)\|.
\end{align*}
Putting together the bounds, we have:
\begin{align*}
  \|g(\theta_{b+1})\|_W &\leq \left(1-\gamma + \frac{\gamma \overline{\lambda}_W L_G}{\underline{\sigma}} [ \kappa_W + 1 ] \|\theta_b-\theta^\dagger\| \right)\|g(\theta_b)\|_W\\
  &\leq (1-\overline{\gamma})\|g(\theta_b)\|_W,\quad \text{ if } \|\theta_b-\theta^\dagger\| \leq \underline{\sigma}\frac{\gamma - \overline{\gamma}}{\gamma \overline{\lambda}_W L_G[\kappa_W + 1]}.
\end{align*}
This is the desired result (\ref{eq:contract}). 
For (\ref{eq:norm_equiv}), note that a mean-value expansion $\|g(\theta)\|_W = \|G(\tilde\theta)(\theta-\theta^\dagger)\|_W$ and the reverse triangular inequality imply:
\begin{align*}
  \|g(\theta)\|_W &\geq \|G(\theta^\dagger)(\theta-\theta^\dagger)\|_W -  \|[G(\theta^\dagger)-G(\tilde \theta)](\theta-\theta^\dagger)\|_W\\
  &\geq \|\theta-\theta^\dagger\|_{G^\prime W G} - \overline{\lambda}_W L_G \|\theta-\theta^\dagger\|^2 \geq \left( 1-\frac{\overline{\lambda}_W^2L_G}{\underline{\sigma}}\|\theta-\theta^\dagger\| \right)\|\theta-\theta^\dagger\|_{G^\prime W G},
\end{align*}
pick $\|\theta-\theta^\dagger\| \leq \frac{\overline{\gamma}}{2}\frac{\underline{\sigma}}{\overline{\lambda}_W^2 L_G}$ to get the lower bound in (\ref{eq:norm_equiv}). Repeat the same steps with the triangular inequality to get the upper bound. 
The $\overline{r}_g$ can then be explicitly derived from the inequalities above using the norm equivalence between $\|\cdot\|$ and $\|\cdot\|_{G^\prime W G}$. 
Finally, for (\ref{eq:separ}) the global identification condition, Assumption \ref{ass:gn}v, implies
\[ \inf_{\|\theta-\theta^\dagger\|_{G^\prime W G} \geq \overline{r}_g} \|g(\theta)\|_W \geq \inf_{\|\theta-\theta^\dagger\| \geq \underline{\sigma}\sqrt{\underline{\lambda}_W}\overline{r}_g } \|g(\theta)\|_W = \delta(\underline{\sigma}\sqrt{\underline{\lambda}_W}\overline{r}_g),\]
and set $\underline{\delta} = \delta(\underline{\sigma}\sqrt{\underline{\lambda}_W}\overline{r}_g)>0$. The local norm equivalence (\ref{eq:norm_equiv}) implies for $\|\theta - \theta^\dagger\|_{G^\prime W G} \leq \underline{r}_g \leq \overline{r}_g$ that we have $\|g(\theta)\|_W \leq (1+\overline{\gamma}/2)(1-\overline{\gamma})\underline{r}_g \leq \underline{\delta}$ if $\underline{r}_g \leq \min(\frac{\underline{\delta}}{(1+\overline{\gamma}/2)(1-\overline{\gamma})},\overline{r}_g)$ which yields the desired result and concludes the proof.
\qed
\subsection{Preliminary Results for the Finite-Sample Problem}
\paragraph{Proof of Lemma \ref{lem:identities}}
\begin{itemize} \setlength\itemsep{0em}
  \item[i.] By construction $\overline{g}_{n,\varepsilon}(\theta) = \int_Z \overline{g}_n(\theta + \varepsilon Z)\phi (Z)dZ$, applying the change of variable $u = \theta + \varepsilon Z$, we get the identity: $\int_Z \overline{g}_n(\theta + \varepsilon Z)\phi (Z)dZ = \varepsilon^{-1} \int_u \overline{g}_n(u)\phi (\frac{u-\theta}{\varepsilon})du$,
  \item[ii.] Applying Leibniz's rule to the first identity, we get: $\partial_\theta \overline{g}_{n,\varepsilon}(\theta) = - \varepsilon^{-2}\int_u \overline{g}_n(u)\phi^\prime(\frac{u-\theta}{\varepsilon})du$. Re-apply the change of variable $u = \theta + \varepsilon Z$ to get the second identity: $\partial_\theta \overline{g}_{n,\varepsilon}(\theta) =-\frac{1}{\varepsilon}\int_Z \overline{g}_n(\theta + \varepsilon Z)\phi^\prime (Z)dZ$,
  \item[iii.] The first part follows from the same derivations as above, the second follows from Liebniz's rule applied directly to $G_\varepsilon(\theta)$.
\end{itemize}\qed

\paragraph{Proof of Lemma \ref{lem:dem_bounds}} 
\begin{itemize}  \setlength\itemsep{0em}
    \item[i.] We can write: $G_\varepsilon(\theta) = \int_Z G(\theta+\varepsilon Z) \phi(Z)dZ.$     Now pick any two $(\theta_1,\theta_2)\in \Theta \time \Theta$, we have:
    \[ \|G_{\varepsilon}(\theta_1)-G_{\varepsilon}(\theta_2)\| \leq \int_Z \|G(\theta_1+\varepsilon Z)-G(\theta_2+\varepsilon Z)\| \phi(Z)dZ \leq L_G \|\theta_1-\theta_2\|. \]
    \item[ii.] Using the definition of $G_\varepsilon$ and $G$, we have:
    \begin{align*}
        \|G_\varepsilon(\theta)-G(\theta)\| &= \|\int_Z [G(\theta+\varepsilon Z)-G(\theta) ]\phi(Z)dZ \| \leq \varepsilon L_G \int_Z \|Z\|\phi(Z)dZ = \varepsilon L_G M_{1,Z}.
    \end{align*}
\end{itemize} \qed

\paragraph{Proof of Lemma \ref{lem:stoch_bounds}.} The first inequality is a consequence of Theorem 2.14.2 in \citet{VanderVaart1996}.
\begin{enumerate} \setlength\itemsep{0em}
    \item[a.] Inequality a. follows from Markov's inequality and the first inequality.
    \item[b.] For inequality b., note that for any $\|\theta_1-\theta_2\| \leq \delta$ we have:
    \begin{align*}
        &\|[\overline{g}_{n,\varepsilon}(\theta_1)- g_{\varepsilon}(\theta_1)]-[\overline{g}_{n,\varepsilon}(\theta_2)- g_{\varepsilon}(\theta_2)]\|\\ &\quad= \|\int_Z \left( [\overline{g}_{n}(\theta_1+\varepsilon Z)- g(\theta_1+\varepsilon Z)]-[\overline{g}_{n}(\theta_2 + \varepsilon Z)- g(\theta_2 + \varepsilon Z)] \right) \phi(Z)dZ\|\\
        &\quad\leq \int_Z \sup_{\|\theta_1-\theta_2\|\leq \delta}\|[\overline{g}_{n}(\theta_1)- g(\theta_1)]-[\overline{g}_{n}(\theta_2)- g(\theta_2)] \| \phi(Z)dZ\\
        &\quad= \sup_{\|\theta_1-\theta_2\|\leq \delta}\|[\overline{g}_{n}(\theta_1)- g(\theta_1)]-[\overline{g}_{n}(\theta_2)- g(\theta_2)] \|,
    \end{align*}
    since $\int \phi(Z)dZ = 1.$ Then inequality a. yields the desired result.
    \item[c.] For inequality c., note that for any $\|\theta_1-\theta_2\| \leq \delta$ we have:
    \begin{align*}
        &\|[\overline{G}_{n,\varepsilon}(\theta_1)- G_{\varepsilon}(\theta_1)]-[\overline{G}_{n,\varepsilon}(\theta_2)- G_{\varepsilon}(\theta_2)]\|\\ &\quad= \| -\frac{1}{\varepsilon}\int_Z \left( [\overline{g}_{n}(\theta_1+\varepsilon Z)- g(\theta_1+\varepsilon Z)]-[\overline{g}_{n}(\theta_2 + \varepsilon Z)- g(\theta_2 + \varepsilon Z)] \right) \phi^\prime(Z)dZ\|\\
        &\quad\leq \frac{1}{\varepsilon}\int_Z \sup_{\|\theta_1-\theta_2\|\leq \delta}\|[\overline{g}_{n}(\theta_1)- g(\theta_1)]-[\overline{g}_{n}(\theta_2)- g(\theta_2)] \| \times \|\phi^\prime(Z)\|dZ\\
        &\quad= \sup_{\|\theta_1-\theta_2\|\leq \delta}\|[\overline{g}_{n}(\theta_1)- g(\theta_1)]-[\overline{g}_{n}(\theta_2)- g(\theta_2)] \| \frac{M_{1,Z}}{\varepsilon},
    \end{align*} 
    where $M_{1,Z} = \int\|\phi^\prime (Z)\| dZ$, the result then follows from inequality a.
\end{enumerate} 
For the final statement, simply note that all three derivations require bounding the same supremum, i.e. the same event.\qed

\paragraph{Proof of Lemma \ref{lem:singular}.} Note that $\overline{G}_{n,\varepsilon}(\theta) = \overline{G}_{n,\varepsilon}(\theta) - G(\theta) + G(\theta)$, using Weyl's inequality for singular values \citep[see Problem III.6.5 in][]{bhatia2013}, we have:
\[ \sigma_{\min}[G(\theta)] \leq \sigma_{\min}[\overline{G}_{n,\varepsilon}(\theta)] + \sigma_{\max}[\overline{G}_{n,\varepsilon}(\theta)-G(\theta)], \]
which implies:
\[ \sigma_{\min}[\overline{G}_{n,\varepsilon}(\theta)] \geq \underline{\sigma} - \sigma_{\max}[\overline{G}_{n,\varepsilon}(\theta)-G(\theta)], \]
since $\sigma_{\min}[G(\theta)] \geq \underline{\sigma}$ for $\|\theta - \theta^\dagger\| \leq R_G$. Note that $\sigma_{\max}[\overline{G}_{n,\varepsilon}(\theta)-G(\theta)] = \|\overline{G}_{n,\varepsilon}(\theta)-G(\theta)\|$ (spectral norm). By the triangular inequality and Lemmas \ref{lem:dem_bounds}, \ref{lem:stoch_bounds}:
\begin{align*}
    \|\overline{G}_{n,\varepsilon}(\theta)-G(\theta)\| &\leq \|\overline{G}_{n,\varepsilon}(\theta)-G_\varepsilon(\theta)\| + \|G_{\varepsilon}(\theta)-G(\theta)\|\\
    &\leq  \frac{c_n}{\varepsilon \sqrt{n}} C_\Theta L_g M_{1,Z} R_G^\psi +  \varepsilon L_G M_{1,Z},
\end{align*}
with probability $1-C/c_n$. Putting everything together yields the desired result.\qed

\paragraph{Proof of Lemma \ref{lem:Taylor_Expand}:}
The result holds for $\delta=0$, so we can work with $\delta >0$. Take any two $\theta_1,\theta_2$; we have: 
\begin{align}
  \|[\overline{g}_n(\theta_1) - \overline{g}_n(\theta_2)]-G_{n,\varepsilon}(\theta_1)(\theta_1-\theta_2)\|
  &\leq \|[\overline{g}_n(\theta_1) - \overline{g}_n(\theta_2)]-G(\theta_1)(\theta_1-\theta_2)\| \label{eq:e1}\\
  &+ \| G_\varepsilon(\theta_1) - G(\theta_1)\| \times \|\theta_1-\theta_2\| \label{eq:e2}\\
  &+ \| G_{n,\varepsilon}(\theta_1) - G_\varepsilon(\theta_1) \| \times \|\theta_1-\theta_2\|. \label{eq:e3}
\end{align}
Using Lemma \ref{lem:stoch_bounds} and the Lipschitz continuity of $G$, we have:
\[ \|(\ref{eq:e1})\| \leq L_gC_\Theta c_nn^{-1/2}\|\theta_1-\theta_2\|^\psi + L_G\|\theta_1-\theta_2\|^2,  \]
with probability $1-C/c_n$. Also, we have $\|(\ref{eq:e2})\| \leq \varepsilon L_G M_{1,Z}\|\theta_1-\theta_2\|$. Using the integral representation, we have - after a mean-zero adjustement:
\[ G_{n,\varepsilon}(\theta_1) - G_\varepsilon(\theta_1) = \frac{1}{\varepsilon} \int ([\overline{g}_{n}(\theta_1+\varepsilon Z)-\overline{g}_{n}(\theta_1)]-[g(\theta_1+\varepsilon Z)-g(\theta_1)])Z\phi(Z)dZ, \]
from which we deduce, using Lemma \ref{lem:stoch_bounds} again, that:
\[ \|(\ref{eq:e3})\| \leq L_g C_\Theta \left[ \int \|Z\|^{1+\psi}  \phi(Z)dZ \right] c_n n^{-1/2}\varepsilon^{\psi-1}\|\theta_1-\theta_2\|,  \]
with probability $1-C/c_n$. Combine the three bounds together to get the desired result. Note that $ \int \|Z\|^{1+\psi}  \phi(Z)dZ \leq  \int (1+\|Z\|^{2})  \phi(Z)dZ = 1+M_{2,Z}$
\qed

\paragraph{Proof of Lemma \ref{lem:Reta}:} The proof will only focus on the lower bound, since the derivations are similar for the upper bound. Using the reverse triangular inequality:
\begin{align*}
  \|\overline{g}_n(\theta) - \overline{g}_n(\hat\theta_n)\|_{W_n} \geq \|\theta-\hat\theta_n\|_{G^\prime W_n G} - \|\overline{g}_n(\theta) - \overline{g}_n(\hat\theta_n) - G(\theta^\dagger) (\theta-\hat\theta_n)\|_{W_n}.
\end{align*}
Using Lemma \ref{lem:stoch_bounds} and the Lipschitz continuity of $G$, we have:
\begin{align*} &\|\overline{g}_n(\theta) - \overline{g}_n(\hat\theta_n) - G(\theta^\dagger) (\theta-\hat\theta_n)\|_{W_n}\\ &\leq \overline{\lambda}_W \left( L_g C_\Theta c_n n^{-1/2}\|\theta - \hat\theta_n\|^\psi + L_G \|\theta-\hat\theta_n\|\Big[\|\theta-\hat\theta_n\| + C_a c_n n^{-1/2}\Big] \right), \end{align*}
with probability $1-(1+C)/c_n$. There are two cases. If $\|\theta-\hat\theta_n\| \leq c_nn^{-1/2}$, then:
\begin{align*}
  &\|\overline{g}_n(\theta) - \overline{g}_n(\hat\theta_n) - G(\theta^\dagger) (\theta-\hat\theta_n)\|_{W_n}\\ &\leq \underline{\sigma}^{-2} \kappa_W L_G [ \|\theta-\hat\theta_n\| + C_a c_nn^{-1/2}  ]\|\theta-\hat\theta_n\|_{G^\prime W_n G} + \overline{\lambda}_W L_g C_\Theta (c_nn^{-1/2})^{1+\psi}\\
  &\leq \eta \|\theta-\hat\theta_n\|_{G^\prime W_n G} + \overline{\lambda}_W L_g C_\Theta (c_nn^{-1/2})^{1+\psi},
\end{align*}
if $\|\theta-\hat\theta_n\| \leq R_n(\eta)$. Otherwise, when $\|\theta-\hat\theta_n\| \geq c_nn^{-1/2}$ then
\begin{align*}
  &\|\overline{g}_n(\theta) - \overline{g}_n(\hat\theta_n) - G(\theta^\dagger) (\theta-\hat\theta_n)\|_{W_n}\\ &\leq \underline{\sigma}^{-2} \kappa_W L_G [ \|\theta-\hat\theta_n\| + C_a c_nn^{-1/2} + L_g C_\Theta (c_nn^{-1/2})^\psi  ]\|\theta-\hat\theta_n\|_{G^\prime W_n G}\\
  &\leq \eta \|\theta-\hat\theta_n\|_{G^\prime W_n G},
\end{align*}
if $\|\theta-\hat\theta_n\| \leq R_n(\eta)$. Combining the bounds for a given value of $\eta$ yields the desired results.
\qed

\paragraph{Proof of Lemma \ref{lem:separ_ns}}
For any $r>0$, 
\[\|\theta-\hat\theta_n\|_{G^\prime W_n G} \geq r \Rightarrow \|\theta-\theta^\dagger\| \geq \frac{r}{\overline{\sigma}\overline{\lambda}_W^{1/2}} - \|\theta^\dagger - \hat\theta_n\|. \]
Pick $r = \underline{\sigma} \underline{\lambda}_W^{1/2} R_n(\eta)$, this implies:
\[  \|\theta-\theta^\dagger\| \geq \frac{R_n(\eta)}{\sqrt{\kappa_G \kappa_W}} - \|\theta^\dagger - \hat\theta_n\| \geq \frac{R_n(\eta)}{\sqrt{\kappa_G \kappa_W}} - C_ac_n n^{-1/2} := \overline{r}_{n,g}(\eta), \]
with probability $1-1/c_n$, where $C_a$ is defined in the proof of Lemma \ref{lem:smooth_foc}. Notice that $\overline{r}_{n,g}(\eta) \to \overline{r}_{g}(\eta) >0$ as $n \to \infty$.
Making use of (reverse) triangular inequalities, we have with probability $1-(1+C)/c_n$:
\begin{align*}
  \inf_{\|\theta-\hat\theta_n\|_{G^\prime W_n G} \geq \underline{\sigma}\underline{\lambda}_W^{1/2}R_n(\eta)}\|\overline{g}_n(\theta)\|_{W_n} &\geq  \inf_{\|\theta-\theta^\dagger\| \geq \overline{r}_{n,g}(\eta) }\|\overline{g}_n(\theta)\|_{W_n}\\
  &\geq \frac{\delta(\overline{r}_{n,g}(\eta))}{\sqrt{\kappa_W}} - \overline{\lambda}_W^{1/2} c_nn^{-1/2} \left( L_g C_\Theta \text{diam}(\Theta)^\psi + \lambda_{\max}(\Sigma)^{1/2}p \right)\\
  &:= \underline{r}_{n,g}(\eta),
\end{align*}
where the last terms are derived using Lemma \ref{lem:stoch_bounds} and the bound for $\|\overline{g}_n(\theta^\dagger)\|$ in the proof of Lemma \ref{lem:smooth_foc}. By continuity of $\delta$, $\underline{r}_{n,g}(\eta) \to \underline{r}_{g}(\eta)>0$ as $n \to \infty$ and $c_nn^{-1/2} \to 0$.
\qed

\paragraph{Proof of Lemma \ref{lem:sample_obj_constract}:} Take $\theta_b$ as described in the Lemma, recall that we have $\theta_{b+1}-\theta_b = -\gamma (G_b^\prime W_n G_b)^{-1} G_b^\prime W_n \overline{g}_n(\theta_b)$, where $G_b = G_{n,\varepsilon}(\theta_b)$. The main idea will be to use the identity:
\begin{align} \|\overline{g}_n(\theta)-\overline{g}_n(\hat\theta_n)\|_{W_n}^2 = \|\overline{g}_n(\theta)\|_{W_n}^2 - \|\overline{g}_n(\hat\theta_n)\|_{W_n}^2 - 2\left( \overline{g}_n(\theta) - \overline{g}_n(\hat\theta_n) \right)^\prime W_n \overline{g}_n(\hat\theta_n), \label{eq:identity} \end{align} 
in tandem with the following inequality:
\begin{align}
  \|\overline{g}_n(\theta_{b+1})-\overline{g}_n(\hat\theta_n)\|_{W_n} &\leq (1-\gamma)\|\overline{g}_n(\theta_{b})-\overline{g}_n(\hat\theta_n)\|_{W_n}\label{eq:contr_a1} \\ &+ \|\overline{g}_n(\theta_{b+1})-\overline{g}_n(\hat\theta_n) - (1-\gamma)[\overline{g}_n(\theta_{b})-\overline{g}_n(\hat\theta_n)] \|_{W_n}. \label{eq:diff_b1}
\end{align}
From Lemma \ref{lem:Taylor_Expand}, we have for some $C_{(\ref{eq:Taylor_smoothed})} >0$:
\begin{align*}
  \|\overline{g}_n(\theta_{b+1}) - \overline{g}_n(\theta_{b}) - G_{n,\varepsilon}(\theta_b)(\theta_{b+1}-\theta_b)\|_{W_n} &\leq \overline{\lambda}_W L_G [\|\theta_{b+1}-\theta_b\|^2 + M_{2,Z}\varepsilon \|\theta_{b+1}-\theta_b\|] \\ &+ \overline{\lambda}_W C_{(\ref{eq:Taylor_smoothed})} c_nn^{-1/2}[ \|\theta_{b+1}-\theta_b\|^\psi + \varepsilon^{\psi-1}\|\theta_{b+1}-\theta_b\| ],
\end{align*} 
with probability $1-C/c_n$. Note that for $\|\theta_b-\hat\theta_n\| \leq R_{n,\varepsilon}$, we have:
\[ \|\theta_{b+1}-\theta_b\| \leq \underline{\sigma}_{n,\varepsilon}^{-1}\sqrt{\kappa_W} [\|\overline{g}_n(\theta_b)-\overline{g}_n(\hat\theta_n)\|_{W_n} + \|\overline{g}_n(\hat\theta_n)\|_{W_n}],\]
with probability $1-C/c_n$. We also have $\|\overline{g}_n(\hat\theta_n)\|_{W_n} \leq C_a c_nn^{-1/2}$, with probability $1-1/c_n$.  Now notice that:
\begin{align*}
  &\|(\ref{eq:diff_b1})\|_{W_n} = \| \overline{g}_n(\theta_{b+1}) - \overline{g}_n(\theta_{b}) + \gamma[ \overline{g}_n(\theta_{b}) - \overline{g}_n(\hat\theta_{n})]  \|_{W_n} \\ &\leq \| \overline{g}_n(\theta_{b+1}) - \overline{g}_n(\theta_{b}) -  G_{n,\varepsilon}(\theta_{b+1}-\theta_b)  \|_{W_n}  + \gamma \|(I-P_b)[\overline{g}_n(\theta_b)-\overline{g}_n(\hat\theta_n)]\|_{W_n} + \gamma \|P_b \overline{g}_n(\hat\theta_n)\|_{W_n},
\end{align*}
where $P_b = G_b(G_b^\prime W_n G_b )^{-1}G_b^\prime W_n $ is an orthogonal projection matrix. By orthogonality:
\begin{align*}
  \|(I-P_b)[\overline{g}_n(\theta_b)-\overline{g}_n(\hat\theta_n)]\|_{W_n} &= \|(I-P_b)[\overline{g}_n(\theta_b)-\overline{g}_n(\hat\theta_n) - G_b(\theta_b-\hat\theta_n)]\|_{W_n}\\
  &\leq \|\overline{g}_n(\theta_b)-\overline{g}_n(\hat\theta_n) - G_b(\theta_b-\hat\theta_n)\|_{W_n}\\
  &\leq \overline{\lambda}_W L_G (\|\theta_{b}-\hat\theta_n\|^2 + \varepsilon M_{2,Z}\|\theta_{b}-\hat\theta_n\|)\\
  &+ \overline{\lambda}_W C_{(\ref{eq:Taylor_smoothed})} c_nn^{-1/2}[ \|\theta_{b}-\hat\theta_n\|^\psi + \varepsilon^{\psi-1}\|\theta_{b}-\hat\theta_n\| ],
\end{align*} 
with probability $1-(1+C)/c_n$. Then, using Lemma \ref{lem:stoch_bounds} we have:
\[  \| G_{n,\varepsilon}(\theta_b)- G_{n,\varepsilon}(\hat\theta_n)\|  \leq  L_g C_\Theta M_{1,Z} c_n n^{-1/2}\varepsilon^{-1}\|\theta_b-\hat\theta_n\|^\psi + L_G\|\theta_b-\hat\theta_n\|,  \]
with probability $1-(1+C)/c_n$. Together with Lemma \ref{lem:smooth_foc}, this implies that:
\begin{align*}
  \|P_g \overline{g}_n(\hat\theta_n)\|_{W_n} \leq \underline{\sigma}_{n,\varepsilon}^{-2}\sqrt{\kappa_W}[\Gamma_{n,\varepsilon} + C_a c_nn^{-1/2}( L_g C_\Theta M_{1,Z} c_n n^{-1/2}\varepsilon^{-1}\|\theta_b-\hat\theta_n\|^\psi + L_G\|\theta_b-\hat\theta_n\| ) ],
\end{align*}
with probability $1-(1+C)/c_n$. Lemma \ref{lem:Reta} implies that for $\eta \in (0,1)$ and $\|\theta_b-\hat\theta_n\| \leq R_n(\eta)$, we have with probability $1-(1+C)/c_n$:
\[ \|\theta_b-\hat\theta_n\|_{G^\prime W_n G} \leq \frac{\underline{\sigma}_{n,\varepsilon}^{-1} \underline{\lambda}_W^{-1/2}}{1-\eta}\left( \|\overline{g}_n(\theta_b) - \overline{g}_n(\hat\theta_n)\|_{W_n} + \overline{\lambda}_W L_g C_\Theta (c_nn^{-1/2})^{1+\psi} \right). \]
Let $x_b = \|\overline{g}_n(\theta_b)-\overline{g}_n(\hat\theta_n)\|_{G^\prime W_n G}$. After grouping terms appropriately, the above inequalities imply that (\ref{eq:contr_a1})-(\ref{eq:diff_b1}) can be re-written, for $\|\theta_b-\hat\theta_n\|\leq R_n(\eta)$, as:
\begin{align*}
  x_{b+1} &\leq (1-\gamma)x_b + 3 L_G \underline{\sigma}_{n,\varepsilon}^{-1}\sqrt{\kappa_W} \left( \overline{\lambda}_W + \frac{\gamma \underline{\sigma}_{n,\varepsilon}^{-1}\sqrt{\kappa_W}}{(1-\eta)^2}  \right)x_b^2 + \overline{\lambda}_W L_G \underline{\sigma}_{n,\varepsilon}^{-1}\sqrt{\kappa_W} \left( 1 + \frac{\gamma}{1-\eta} \right) \varepsilon x_b\\
  &+ C_{(\ref{eq:diff_b1})} \frac{\underline{\sigma}_{n,\varepsilon}^{-3}}{(1-\eta)^2} \left( \Gamma_{n,\varepsilon} + \varepsilon^{\psi-1}(c_nn^{-1/2})^{2} + [c_nn^{-1/2} + (c_nn^{-1/2})^2 \varepsilon^{-1}]x_b^\psi + (c_nn^{-1/2})x_b \right),
\end{align*}
with probability $1-(1+C)/c_n$ when assuming (without loss of generality) that $\overline{\lambda}_W \geq 1$ and $\underline{\sigma}_{n,\varepsilon} \leq 1$. Lemma \ref{lem:Reta} also implies that:
\[ x_b^2 \leq [(1+\eta)\|\theta_b-\hat\theta_n\|_{G^\prime W_n G}] x_b + \overline{\lambda}_W L_g C_\Theta (c_nn^{-1/2})^{1+\psi} x_b, \]
with probability $1-(1+C)/c_n$. For $\varepsilon >0$ small enough, let $R_{2,n,\varepsilon}(\eta)>0$ be such that:
\[ 3(1+\eta) L_G \underline{\sigma}_{n,\varepsilon}^{-1}\sqrt{\kappa_W} \left( \overline{\lambda}_W + \frac{\gamma \underline{\sigma}_{n,\varepsilon}^{-1}\sqrt{\kappa_W}}{(1-\eta)^2}  \right)R_{2,n,\varepsilon}(\eta) + \overline{\lambda}_W L_G \underline{\sigma}_{n,\varepsilon}^{-1}\sqrt{\kappa_W} \left( 1 + \frac{\gamma}{1-\eta} \right) \varepsilon \leq \overline{\gamma}-\gamma, \]
then for $\|\theta_b-\hat\theta_n\|_{G^\prime W_n G} \leq R_{2,n,\varepsilon}$ (which is bounded below when $\varepsilon \searrow 0$), we have:
\[ x_{b+1} \leq (1-\overline{\gamma})x_b + C_{(\ref{eq:contr_obj_ns})} \frac{\underline{\sigma}_{n,\varepsilon}^{-3}}{(1-\eta)^2} \left( \Gamma_{n,\varepsilon} + \varepsilon^{\psi-1}(c_nn^{-1/2})^{2} + [c_nn^{-1/2} + (c_nn^{-1/2})^2 \varepsilon^{-1}]x_b^\psi + (c_nn^{-1/2})x_b \right), \]
with probability $1-(1+C)/c_n$ which implies the desired result.
\qed

\paragraph{Proof of Lemma \ref{lem:norm_squared}:}
Start with the identity:
\begin{align}
  \|\overline{g}_n(\theta)-\overline{g}_n(\hat\theta_n)\|^2_{W_n} = \|\overline{g}_n(\theta)\|^2_{W_n} - \|\overline{g}_n(\hat\theta_n)\|^2_{W_n} - 2 \left( \overline{g}_n(\theta)-\overline{g}_n(\hat\theta_n) \right)^\prime W_n \overline{g}_n(\hat\theta_n). \label{eq:idt}
\end{align}
which holds for any $\theta \in \Theta$. Now use Lemmas \ref{lem:smooth_foc} and \ref{lem:Taylor_Expand} to bound the last term:
\begin{align*}
  &\|\left( \overline{g}_n(\theta)-\overline{g}_n(\hat\theta_n) \right)^\prime W_n \overline{g}_n(\hat\theta_n)\|\\ &\leq \|\theta-\hat\theta_n\| \times \|G_{n,\varepsilon}(\hat\theta_n)^\prime W_n \overline{g}_n(\hat\theta_n)\|
  + \overline{\lambda}_W C_a c_nn^{-1/2}\| \overline{g}_n(\theta)-\overline{g}_n(\hat\theta_n) - G_{n,\varepsilon}(\hat\theta_n)(\theta-\hat\theta_n) \|\\
  &\leq \Gamma_{n,\varepsilon}\|\theta-\hat\theta_n\| + \overline{\lambda}_W C_a c_nn^{-1/2}\| \overline{g}_n(\theta)-\overline{g}_n(\hat\theta_n) - G_{n,\varepsilon}(\hat\theta_n)(\theta-\hat\theta_n) \|,
\end{align*}
with probability $1-(1+C)/c_n$. Lemma \ref{lem:Taylor_Expand} implies that with probability $1-C/c_n$:
\begin{align*}
  \| \overline{g}_n(\theta)-\overline{g}_n(\hat\theta_n) - G_{n,\varepsilon}(\hat\theta_n)(\theta-\hat\theta_n) \| &\leq L_g C_\Theta \|\theta-\hat\theta_n\|^2 + L_G M_{1,Z} \varepsilon \|\theta-\hat\theta_n\|\\ &+ L_g C_\Theta c_nn^{-1/2}[ \|\theta-\hat\theta_n\|^\psi + \varepsilon^{\psi-1}\|\theta-\hat\theta_n\| ].
\end{align*}
Since $\|\theta-\hat\theta_n\| \leq R_n(\eta)$ by assumption, Lemma \ref{lem:Reta} implies that:
\[ \| \theta - \hat\theta_n \| \leq \frac{\underline{\sigma}_{n,\varepsilon}^{-1}\underline{\lambda}_W^{-1/2}}{1-\eta} \left( \|\overline{g}_n(\theta)-\overline{g}_n(\hat\theta_n)\|_{W_n} + \overline{\lambda}_W L_g C_\Theta (c_nn^{-1/2})^{1+\psi} \right), \]
with probability $1-(1+C)/c_n$. Let $x = \|\overline{g}_n(\theta)-\overline{g}_n(\hat\theta_n)\|_{W_n}$, grouping the bounds together:
\begin{align*}
  &\|\left( \overline{g}_n(\theta)-\overline{g}_n(\hat\theta_n) \right)^\prime W_n \overline{g}_n(\hat\theta_n)\|\\ &\leq C_{(\ref{eq:idt})}\frac{\underline{\sigma}_{n,\varepsilon}^{-2}}{(1-\eta)^2} \Big( c_nn^{-1/2} x^2 + [\Gamma_{n,\varepsilon} + (c_nn^{-1/2})^2 \varepsilon^{\psi-1}]x + (c_nn^{-1/2})^2 x^\psi\\ &+ (c_nn^{-1/2})^{1+\psi}\Gamma_{n,\varepsilon} + (c_nn^{-1/2})^{2+\psi+\psi^2} + \varepsilon (c_nn^{-1/2})^{2+\psi} + (c_nn^{-1/2})^{3+\psi} \varepsilon^{\psi-1} \Big),
\end{align*}
with probability $1-(1+C)/c_n$. Using the same $x$ as above, we have:\footnote{Here we will use $\min[1-(1-\tau)^2,(1+\tau)^2-1] \geq 2\tau -\tau^2$ to derive the $(1-\tau)^2$ lower and a $(1+\tau)^2$ upper bounds we want in the Lemma.}
\begin{align}
&\Big| \|\overline{g}_n(\theta)\|^2_{W_n} - \|\overline{g}_n(\hat\theta_n)\|^2_{W_n} - x^2 \Big| \leq  ( 2\tau -\tau^2 ) x^2 \notag\\ 
&+ C_{(\ref{eq:R1})}\Big[\frac{\underline{\sigma}_{n,\varepsilon}^{-4}}{(1-\eta)^4} \frac{1}{\overline{\gamma}-[\overline{\gamma}/2]^2} \left( \Gamma_{n,\varepsilon}^2 + (c_nn^{-1/2})^4 \varepsilon^{2\psi-2} + (c_nn^{-1/2})^{2 + 2 \psi} \right) \label{eq:R1}\\
& + \frac{\underline{\sigma}_{n,\varepsilon}^{-2}}{(1-\eta)^2} \left( (c_nn^{-1/2})^{1+\psi}\Gamma_{n,\varepsilon} + (c_nn^{-1/2})^{2+\psi+\psi^2} + \varepsilon (c_nn^{-1/2})^{2+\psi} + (c_nn^{-1/2})^{3+\psi} \varepsilon^{\psi-1} \right)\Big] \notag
\end{align}
with probability $1-(1+C)/c_n$ if $2C_{(\ref{eq:idt})} \underline{\sigma}_{n,\varepsilon}^{-2}(1-\eta)^{-2} c_nn^{-1/2} \leq (2\tau - \tau^2)/3$, which holds for $c_n n^{-1/2}$ sufficiently small. Equation (\ref{eq:Gamma2n}) bounds the last two lines  in (\ref{eq:R1}).
Putting everything together, we get:
\begin{align*}
  (1-\tau)^2 \|\overline{g}_n(\theta)-\overline{g}_n(\hat\theta_n)\|^2_{W_n} - \Gamma^2_{2,n,\varepsilon}(\eta,\tau) &\leq \|\overline{g}_n(\theta)\|^2_{W_n} - \|\overline{g}_n(\hat\theta_n)\|^2_{W_n} \\ &\leq (1+\tau)^2 \|\overline{g}_n(\theta)-\overline{g}_n(\hat\theta_n)\|^2_{W_n} + \Gamma^2_{2,n,\varepsilon}(\eta,\tau) ,
\end{align*}
with probability $1-(1+C)/c_n$, for $\|\theta-\hat\theta_n\| \leq R_n(\eta)$, or  $\|\theta-\hat\theta_n\|_{G^\prime W_n G} \leq  \underline{\sigma} \underline{\lambda}_W^{1/2} R_n(\eta)$. The result that $n \Gamma_{2,n,\varepsilon}^2(\eta,\tau) = o(1)$ follows from $\sqrt{n}\Gamma_{n,\varepsilon}=o(1)$ under the restrictions on $\varepsilon$.
\qed

\section{Preliminary Results for Section \ref{sec:extend}} \label{apx:prelim_extensions}

\begin{lemma} \label{lem:bound_est} Suppose Assumptions \ref{algo:sgn}-\ref{ass:sample_gn} hold. For $L \geq 1$, define:
  \begin{align*}
    \hat{G}_L(\theta_b) = \frac{1}{L} \sum_{\ell=0}^{L-1} \frac{1}{\varepsilon} [ \overline{g}_n(\theta_{b-\ell} + \varepsilon Z_{b-\ell}) - \overline{g}_n(\theta_{b-\ell}) ]Z_{b-\ell}^\prime,  \quad 
    \tilde{G}_L(\theta_b) = \frac{1}{L} \sum_{\ell=0}^{L-1} G(\theta_{b-\ell})Z_{b-\ell} Z_{b-\ell}^\prime
  \end{align*}
  Take $c_n$ and $b_{\max} \geq 1$. Let $t_n = \log(c_n) + \log(b_{\max}+L+1)$, we have:
  \begin{align}
    \sup_{0 \leq b \leq b_{\max}} \|\hat{G}_L(\theta_b) - \tilde{G}_L(\theta_b)\| \leq \left(L_g C_\Theta c_n n^{-1/2} \varepsilon^{\psi-1}+ \varepsilon L_G\right)\left[ d_\theta + 2 d_\theta t_n +\sqrt{2 d_\theta t_n} \right]^{3/2}, \label{eq:bound_est1}
  \end{align}  
with probability $1-(2+C)/c_n$.
\end{lemma}

\begin{lemma} \label{lem:tail_singular} Suppose Assumptions \ref{algo:sgn}-\ref{ass:sample_gn} hold. For $L \geq 1$, define:
  \begin{align*}
    \tilde{G}_L(\theta_b) = \frac{1}{L} \sum_{\ell=0}^{L-1} G(\theta_{b-\ell}) Z_{b-\ell}Z_{b-\ell}^\prime, \quad G_L(\theta_b) = \frac{1}{L} \sum_{\ell=0}^{L-1} G(\theta_{b-\ell})
  \end{align*}
  Take $c_n$ and $b_{\max} \geq 1$, let $d = d_\theta + d_g$. Let $t_n = [\log(c_n) + d \log(9) + \log(b_{\max}+1)]/L$ $\tilde{t}_n = C_Z[\log(c_n) + \log(b_{\max}+L+1) + \log(2)]L^{-1/2}$, for a constant $C_Z$ which only depends on $d_\theta$, $\mu_{d_\theta} = \mathbb{E}(\|ZZ^\prime -I_{d_\theta}\|)$ where $Z \sim \mathcal{N}(0,I_{d_\theta})$ and $\overline{\sigma}_n = \sigma_{\max}[G(\theta^\dagger)] + L_G C_a c_n n^{-1/2}$, we have:
  \begin{align}
    \sigma_{\max}[\overline{G}_L(\theta_b) - G_L(\theta_b)] \leq 4 \overline{\sigma}_n \left( t_n + \sqrt{t_n} \right) + L_G\left(\max_{-L+1 \leq \ell \leq 0} \|\theta_{b-\ell}-\hat\theta_n\|\right)[\mu_{d_\theta} + \tilde{t}_n], \label{eq:bound_est2}
  \end{align}
  with probability $1-3/c_n$.
  \end{lemma}

\section{Proofs for the Preliminary Results for Section \ref{sec:extend}} \label{apx:proofs_prelim_extensions}

\paragraph{Proof of Lemma \ref{lem:bound_est}:} Take $b \geq 0$ and $\ell \geq 0$. For any $Z_{b-\ell} \in \mathbb{R}^{d_\theta}$ we have:
\begin{align*}
  \| \frac{1}{\varepsilon}[\overline{g}_n(\theta_{b-\ell} + \varepsilon Z_{b-\ell}) - \overline{g}_n(\theta_{b-\ell}) - \varepsilon G(\theta_{b-\ell})Z_{b-\ell}]Z_{b-\ell}^\prime \| \leq L_g C_\Theta c_n n^{-1/2} \varepsilon^{\psi-1} \|Z_{b-\ell}\|^{1+\psi} + \varepsilon L_G \| Z_{b-\ell} \|^3,
\end{align*}  
uniformly in $\theta_{b-\ell} \in \Theta$, with probability $1-(1+C)/c_n$. The inequality relies on Lemma \ref{lem:stoch_bounds} and the Liptschitz-continuity assumption. Note that $\|Z_{b-\ell}\|^2 \sim \chi^2_{d_\theta}$ so that Lemma 1 in \citet{laurent2000} implies that, for any $t >0$:
\[ \mathbb{P}(\|Z_{b-\ell}\|^2 \geq d_\theta + 2d_\theta t + \sqrt{2d_\theta t}) \leq \exp(-t). \]
Pick $t_n = \log(b_{\max} + L + 1) + \log(c_n)$, we have:
\[ \sup_{-L + 1 \leq \ell \leq b_{\max} } \|Z_{\ell}\|^2 \leq d_\theta + 2d_\theta t_n + \sqrt{2d_\theta t_n}, \]
with probability $1-1/c_n$. Since $d_\theta + 2d_\theta t_n + \sqrt{2d_\theta t_n} \geq 1$, this yields the bound:
\begin{align*}
  &\sup_{ 0\leq b \leq b_{\max}, -L+1 \leq \ell \leq 0 }\| \frac{1}{\varepsilon}[\overline{g}_n(\theta_{b-\ell} + \varepsilon Z_{b-\ell}) - \overline{g}_n(\theta_{b-\ell}) - \varepsilon G(\theta_{b-\ell})Z_{b-\ell}]Z_{b-\ell}^\prime \|\\ &\quad \quad \leq \left( L_gC_\Theta c_n n^{-1/2} \varepsilon^{\psi-1}+ \varepsilon L_G\right)\left[ d_\theta + 2 d_\theta t_n +\sqrt{2 d_\theta t_n} \right]^{3/2},
\end{align*}  
with probability $1-(2+C)/c_n$. Bound the average with the $\sup$ and we get the desired result (\ref{eq:bound_est1}).
\qed

\paragraph{Proof of Lemma \ref{lem:tail_singular}:} The result will be derived using an $\varepsilon$-net argument \citep[see][Section 4.2]{vershynin2018} combined with an exponential tail inequality. Notice that:
\[ \overline{G}_L(\theta_b) - G_L(\theta_b) = \frac{1}{L} \sum_{\ell=0}^{L-1} G(\theta_{b-\ell})[Z_{b-\ell}Z_{b-\ell}^\prime - I_{d_\theta}], \]
where by Gaussianity of $Z_{b-\ell}$, the $Z_{b-\ell}Z_{b-\ell}^\prime$ are iid Wishart distributed. Using the triangular inequality on the spectral norm and the Lipschitz-continuity of the Jacobian, we have:
\begin{align*}
  \sigma_{\max}[\overline{G}_L(\theta_b) - G_L(\theta_b)] &\leq \sigma_{\max}\left[\frac{1}{L} \sum_{\ell=0}^{L-1} G(\hat\theta_n)[Z_{b-\ell}Z_{b-\ell}^\prime - I_{d_\theta}]\right] \\ &+ \frac{1}{L} \sum_{\ell=0}^{L-1}L_G \|\theta_{b-\ell}-\hat\theta_n\| \times \|Z_{b-\ell}Z_{b-\ell}^\prime - I_{d_\theta}\|.
\end{align*}
The following derives the bounds for these two terms. Let $S^{d_\theta-1}$ and $S^{d_g-1}$ be respectively the unit sphere in $\mathbb{R}^{d_\theta}$ and $\mathbb{R}^{d_g}$, i.e. $x \in S^{d_\theta-1}$ implies $x \in \mathbb{R}^{d_\theta}$ and $\|x\|=1$. For any given $b \in \{0,\dots,b_{\max}\}$ and matrix $A$ with dimensions $d_g \times d_\theta$, the largest singular value satisfies:
\[ \sigma_{\max}[ A ] = \sup_{(x,y)\in S^{d_\theta-1} \times S^{d_g-1}} y^\prime A x. \]
Pick any two $(x,y)\in S^{d_\theta-1} \times S^{d_g-1}$, $b \geq 0, \ell \geq 0$ and a $t \in \mathbb{R}$ such that $t y^\prime x < 1/2$:
\begin{align*}
  \mathbb{E} \left( \exp[ t y^\prime G(\hat\theta_n) [Z_{b-\ell}Z_{b-\ell}^\prime - I_d] x ]\right) &= \mathbb{E} \left( \exp[ \text{trace}(t x y^\prime G(\hat\theta_n) Z_{b-\ell}Z_{b-\ell}^\prime )\right) \exp(-ty^\prime G(\hat\theta_n) x)\\
  &= [ \text{det}( I - 2 t x y^\prime G(\hat\theta_n) ) ]^{-1/2} \exp(-ty^\prime G(\hat\theta_n) G(\hat\theta_n) x)\\
  &= [  1 - 2 t y^\prime G(\hat\theta_n) x  ]^{-1/2} \exp(-ty^\prime G(\hat\theta_n) x),
\end{align*}
where the second equality follows from the formula for the moment generating function of a Wishart distribution \citep[][Section 8]{muirhead1982}, the third equality follows from Sylvester's determinant identity. Note that $y^\prime G(\hat\theta_n) x \leq \sigma_{\max}[G(\hat\theta_n)]$ for any unitary $x,y$. Let $\overline{\sigma}_n = \sigma_{\max}[G(\hat\theta_n)]$. Using an inequality from the proof of Lemma 1 in \citet{laurent2000}, we have for $u = t y^\prime G(\hat\theta_n)x \leq \overline{\sigma}_n t < 1/2$, $t \geq 0$:
\[ \log\left[ \mathbb{E} \left( \exp[ t y^\prime G(\hat\theta_n) [Z_{b-\ell}Z_{b-\ell}^\prime - I_d] x ]\right) \right] \leq \frac{u^2}{1-2u} \leq \frac{t^2 \overline{\sigma}_n^2 }{1-2t \overline{\sigma}_n},\]
summing over $\ell=0$ to $L-1$, yields:
\[ \sum_{\ell=0}^{L-1} \log\left[ \mathbb{E} \left( \exp[ t y^\prime G(\hat\theta_n) [Z_{b-\ell}Z_{b-\ell}^\prime - I_d] x ]\right) \right] \leq \frac{t^2 L\overline{\sigma}_n^2 }{1-2t \overline{\sigma}_n} = \frac{t^2 2L\overline{\sigma}_n^2 }{2(1-2t \overline{\sigma}_n)}.\]
From \citet{birge1998}, this implies the following inequality for any $t >0$:
\[ \mathbb{P}\left( \sum_{\ell =0}^{L-1} y^\prime G(\hat\theta_n) [Z_{b-\ell}Z_{b-\ell}^\prime - I_d] x \geq 2\overline{\sigma}_n [t + \sqrt{L t}] \right) \leq \exp(-t). \]

Now this will be combined with an $\varepsilon$-net argument. Using \citet{vershynin2018}, Problem 4.4.3, for any $\varepsilon$-nets $N_1,N_2$ of the spheres $S^{d_\theta-1}$ and $S^{d_g-1}$ we have when $\varepsilon < 1/2$:
\[ \sigma_{\max}[A] \leq \sup_{(x,y)\in N_1 \times N_2} \frac{1}{1-2\varepsilon} y^\prime A x. \]
The cardinality of $N_1$ and $N_2$ is at most $(2/\varepsilon +1)^{d_\theta}$ and $(2/\varepsilon +1)^{d_g}$, respectively \citep[Corollary 4.2.13]{vershynin2018}. Pick $\varepsilon = 1/4$, the cardinality of $N_1 \times N_2$ is at most $9^d$ with $d = d_\theta + d_g$. Then, for $0 \leq b \leq b_{\max}$, we have:
\begin{align*} 
  &\mathbb{P}\left( \sup_{0 \leq b \leq b_{\max}} \left[\sup_{(x,y)\in S^{d_\theta-1} \times S^{d_g -1}}\sum_{\ell =0}^{L-1} y^\prime G(\hat\theta_n) [Z_{b-\ell}Z_{b-\ell}^\prime - I_d] x\right] \geq 4\overline{\sigma}_n [t_n + \sqrt{t_n}] \right) \\ &\leq (b_{\max}+1)9^d\exp(-L t_n) = \frac{1}{c_n},
\end{align*}
for $t_n  = [\log(c_n) + d \log(9) + \log(b_{\max}+1)]/L$. This yields the first bound:
\[ \sup_{0 \leq b \leq b_{\max}} \sigma_{\max}\left[\frac{1}{L} \sum_{\ell=0}^{L-1} G(\hat\theta_n)[Z_{b-\ell}Z_{b-\ell}^\prime - I_{d_\theta}]\right] \leq 4 \overline{\sigma}_n \left( t_n + \sqrt{t_n} \right), \]
with probability $1-1/c_n$. Note that the probability is conditional on the sample data. We also have $\overline{\sigma}_n \leq \sigma_{\max}(G(\theta^\dagger)) + L_G\|\hat\theta_n - \theta^\dagger\| \leq \overline{\sigma} + L_G C_a c_n n^{-1/2}$ with probability $1-1/c_n$. We can set $\overline{\sigma}_n = \overline{\sigma} + L_G C_a c_n n^{-1/2}$ in the bound and we get the same result with unconditional probability $1-2/c_n$.

To bound $\sigma_{\max}(Z_{b-\ell}Z_{b-\ell}^\prime - I_{d_\theta}) = \|Z_{b-\ell}Z_{b-\ell}^\prime - I_{d_\theta}\|$, follow the same steps as above replacing $G(\hat\theta_n)$ with $I_{d_\theta}$ and taking $y \in S^{d_\theta-1}$, and we get for $t>0$:
\[ \mathbb{P} \left( \|Z_{b-\ell}Z_{b-\ell}^\prime - I_{d_\theta}\|  \geq 4 (t + \sqrt{t})  \right) \leq \exp(-t), \]
from which we deduce that the distribution of $\|Z_{b-\ell}Z_{b-\ell}^\prime - I_{d_\theta}\|$ is sub-exponential. Let $\mu_{d_\theta} = \mathbb{E}(\|Z_{b-\ell}Z_{b-\ell}^\prime -I_d\|)$, apply Bernstein's inequality to find:\footnote{See e.g. \citet{vershynin2018}, Theorem 2.8.1, centering with $\mu_{d_\theta}$ ensures the r.v. has mean zero.}
\[ \mathbb{P}\left( \left| \sum_{\ell=0}^{L-1} [\|Z_{b-\ell}Z_{b-\ell}^\prime -I_d\| - \mu_{d_\theta}] \right|  > t \right) \leq 2\exp\left( - c \min\left( \frac{t^2}{L C_Z^2 }, \frac{t}{C_Z} \right) \right),\]
where $c$ is an absolute constant and $C_Z$ is the Orlicz norm of $\|Z_{b-\ell}Z_{b-\ell}^\prime -I_d\| - \mu_{d_\theta}$. Take:
\[ \tilde{t}_n = C_Z \frac{\log(c_n) + \log(2) + \log(b_{\max}+L+1)}{\sqrt{L}}, \]
then with probability $1-1/c_n$, we have:
\[ \sup_{-L+1 \leq \ell \leq b_{\max}} \frac{1}{L} \sum_{\ell=0}^{L-1} \|Z_{b-\ell}Z_{b-\ell}^\prime -I_d\| \leq \mu_{d_\theta} + \tilde{t}_n, \]
uniformly in $b \in \{0,\dots,b_{\max}\}$. Now apply this to get:
\[ \frac{1}{L} \sum_{\ell=0}^{L-1}[\|\theta_{b-\ell} - \hat\theta_n\| \times \|Z_{b-\ell}Z_{b-\ell}^\prime -I_d\|] \leq L_G \left(\max_{0\leq \ell \leq L-1}  \|\theta_{b-\ell} - \hat\theta_n\|\right) \left( \mu_{d_\theta} + \tilde{t}_n \right),   \]
uniformly in $b \in \{0,\dots,b_{\max}\}$, with probability $1-1/c_n$.
Pick the same $t_n = \log(c_n) + \log(b_{\max}+L+1) + d\log(9)$ for both bounds and we get:
\begin{align*}
  \sigma_{\max}[\overline{G}_L(\theta_b) - G_L(\theta_b)] \leq 4 \overline{\sigma}_n \left( t_n + \sqrt{ t_n } \right) + L_G \left(\max_{0\leq \ell \leq L-1}  \|\theta_{b-\ell} - \hat\theta_n\|\right) \left( \mu_{d_\theta} + \tilde{t}_n \right), 
\end{align*}
with probability $1-3/c_n$. This is the desired result (\ref{eq:bound_est2}).
\qed

\section{Proofs for Section \ref{sec:extend}} \label{apx:proofs_extensions}

\paragraph{Proof of Proposition \ref{prop:local_conv2}:}
Take $\theta_{b+1}$ as described in (\ref{eq:step1}'). As in the proof of Proposition \ref{prop:local_conv}, we have when $\|\theta_b-\hat\theta_n\| \leq R_{n,G}$:
\begin{align*} 
  &\|\theta_{b+1} - \hat\theta_n - (1-\gamma)(\theta_b-\hat\theta_n) - \alpha(\theta_b-\theta_{b-1})\|\\ &\leq (\ref{eq:toboundp1}) + (\ref{eq:toboundp2}) \leq (\ref{eq:contract_ineq}) = \gamma \frac{\sqrt{\kappa_W}L_G}{\underline{\sigma}_{n,\varepsilon}}\left( \|\theta_b-\hat\theta_n\| + M_{1,Z}\varepsilon \right)\|\theta_b-\hat\theta_n\| + \gamma \Delta_{n,\varepsilon},
\end{align*}
with probability $1-(1+C)/c_n$, with the same $\Delta_{n,\varepsilon}$ used in Proposition \ref{prop:local_conv}. The left-hand-side can be re-written in companion form as:
\begin{align*} 
  &\|\theta_{b+1} - \hat\theta_n - (1-\gamma)(\theta_b-\hat\theta_n) - \alpha(\theta_b-\theta_{b-1})\| = \| (\boldsymbol{\theta_{b+1}} - \boldsymbol{\hat\theta_n}) - A(\gamma,\alpha)(\boldsymbol{\theta_b} - \boldsymbol{\hat\theta_n}) \|.
\end{align*}
Now, apply the reverse triangular inequality and plug-in the above inequality to get:
\begin{align*} 
  \| \boldsymbol{\theta_{b+1}} - \boldsymbol{\hat\theta_n} \| &\leq  \lambda_{\max}[A(\gamma,\alpha)]\| \boldsymbol{\theta_{b}} - \boldsymbol{\hat\theta_n} \| \\
  &+ \gamma \frac{\sqrt{\kappa_W}L_G}{\underline{\sigma}_{n,\varepsilon}}\left( \|\theta_b-\hat\theta_n\| + M_{1,Z}\varepsilon \right)\|\theta_b-\hat\theta_n\| + \gamma \Delta_{n,\varepsilon},
\end{align*}
with probability $1-(1+C)/c_n$, with $\lambda_{\max}[A(\gamma,\alpha)] = 1-\gamma(\alpha)$ by definition. Now pick $\|\theta_b - \hat\theta_n\| \leq R_{n,\varepsilon}(\alpha)$, we have:
\begin{align*} 
  \| \boldsymbol{\theta_{b+1}} - \boldsymbol{\hat\theta_n}\| &\leq  (1-\overline{\gamma})\| \boldsymbol{\theta_{b}} - \boldsymbol{\hat\theta_n} \|  + \gamma \Delta_{n,\varepsilon},
\end{align*}
with probability $1-(1+C)/c_n$ which concludes the proof.
\qed

\paragraph{Proof of Proposition \ref{prop:local_conv3}:}
Similar to Propositions \ref{prop:local_conv} and \ref{prop:local_conv2} we can write with $\hat{G}_b = \hat{G}_L(\theta_b)$:
\begin{align}
  \|\theta_{b+1}-\hat\theta_n - &(1-\gamma)(\theta_b-\hat\theta_n)\| \notag\\ \leq &\gamma\sqrt{\kappa_W} \sigma_{\min}[\hat{G}_b]^{-1}\| \overline{g}_n(\theta_b)-\overline{g}_n(\hat\theta_n) - G(\hat\theta_n)(\theta_b-\hat\theta_n) \| \label{eq:tbb0}\\
  &+ \gamma \sigma_{\min}[\hat{G}_b]^{-2} \underline{\lambda}_W^{-1}\|\hat{G}_b^\prime W_n \overline{g}_n(\hat\theta_n)\| \label{eq:tbb1}\\
  &+  \gamma \sqrt{\kappa_W} \sigma_{\min}[\hat{G}_b]^{-1}\| \hat{G}_b-G(\hat\theta_n)\| \times \|\theta_b - \hat\theta_n\|. \label{eq:tbb12}
\end{align}
First, a lower bound for the least singular value of $\hat{G}_b$ is required. 
Weyl's inequality implies $\sigma_{\min}(\hat{G}_b) \geq \sigma_{\min}[G(\theta^\dagger)] - \|G(\theta^\dagger) - \hat{G}_b\|$. Using the triangular inequality, Lipschitz-continuity of $G$, and Lemmas \ref{lem:smooth_foc}, \ref{lem:tail_singular} and \ref{lem:bound_est}:
\begin{align*}
  \|G(\theta^\dagger) - \hat{G}_b\| &\leq \|G_L(\theta_b) - G(\hat\theta_n)\| + \|G(\hat\theta_n)-G(\theta^\dagger)\| + \|\hat{G}_b - \tilde G_b\| + \|G_L(\theta_b) - \tilde G_b\| \\ &\leq L_G (\max_{-L+1 \leq \ell \leq 0}\|\theta_{b-\ell}-\hat\theta_n\|) + L_G C_a c_n n^{-1/2} + \|(\ref{eq:bound_est1})\| + \|(\ref{eq:bound_est2})\|,
\end{align*}
where (\ref{eq:bound_est1}), (\ref{eq:bound_est2}) are given in Lemmas \ref{lem:tail_singular} and \ref{lem:bound_est}. They imply that 
\begin{align*} \|(\ref{eq:bound_est1})\| + \|(\ref{eq:bound_est2})\| 
  &\leq C_{(\ref{eq:bound_est1})}\left( c_nn^{-1/2}\varepsilon^{\psi-1} + \varepsilon \right)\delta_n^{3/2} + C_{(\ref{eq:bound_est2})}(1+c_nn^{-1/2})\delta_n L^{-1/2} \\ &+ L_G \mu_{d_\theta} (\max_{-L+1 \leq \ell \leq 0} \|\theta_{b-\ell}-\hat\theta_n\|),
\end{align*}
with probability $1-(5 + C)/c_n$ when each $\|\theta_{b-\ell}-\hat\theta_n\| \leq R_G$, and $\delta_n = \log(c_n) + \log(b_{\max}+L+1) \geq 1$. Pick $0<R_{G,2}\leq R_G$ such that:
\[ \sigma_{\min}[G(\theta^\dagger)] - L_G(1+\mu_{d_\theta})R_{G,2} = \underline{\sigma}/2. \]
Note that $R_{G,2}$ only depends on $G$ and $d_\theta$. Putting everything together, for  $\max_{-L+1 \leq \ell \leq 0}\|\theta_{b-\ell}-\hat\theta_n\| \leq R_{G,2}$, there is a constant $C_{\sigma,2}>0$ such that:
\begin{align}
\sigma_{\min}(\hat G_b) \geq \underline{\sigma}/2 - C_{\sigma,2}\left( c_n n^{-1/2}\varepsilon^{\psi-1} + \varepsilon + L^{-1/2} \right) \delta_n^{3/2} := \hat{\underline{\sigma}}_{n,\varepsilon}, \label{eq:sing_bnd_hat}
\end{align}
with probability $1-(5+C)/c_n$. 
This allows to handle the singular value appearing several times in (\ref{eq:tbb1}). Next, the bound for (\ref{eq:tbb0}) can be derived using the same steps used in the proof of Proposition \ref{prop:local_conv}:
\[ \| \overline{g}_n(\theta_b)-\overline{g}_n(\hat\theta_n) - G(\hat\theta_n)(\theta_b-\hat\theta_n) \| \leq  L_G \|\theta_b- \hat\theta_n\|^2 + L_g C_{\Theta}c_nn^{-1/2}\|\theta_b- \hat\theta_n\|^{\psi}, \]
with probability $1-(1+C)/c_n$. Next, to bound (\ref{eq:tbb1}) consider:
\begin{align}
  \|\hat{G}_b^\prime W_n\overline{g}_n(\hat\theta_n)\| &\leq \overline{\lambda}_W\|\hat{G}_b - G(\theta^\dagger)\| \times \|\overline{g}_n(\hat\theta_n)\| \label{eq:tbb2}\\
  &+ \|G(\theta^\dagger)^\prime W_n \overline{g}_n(\hat\theta_n) \| \label{eq:tbb3}
\end{align}
From the proof of Lemma \ref{lem:smooth_foc}, we have:
\begin{align*}
  \|(\ref{eq:tbb2})\| \leq \overline{\lambda}_W\|\hat{G}_b - G(\theta^\dagger)\| C_c c_n n^{-1/2}, \quad
  \|(\ref{eq:tbb3})\| \leq C_{(\ref{eq:tobound1})}(c_nn^{-1/2})^{1+\psi}, \notag
\end{align*}
with probability $1-(1+C)/c_n$. Using the derivations for the singular value above, we further get:
\[ \|(\ref{eq:tbb2})\| \leq C_{(\ref{eq:tbb2})} c_nn^{-1/2} \left[  \left(c_nn^{-1/2}\varepsilon^{\psi-1} + \varepsilon + L^{-1/2} \right)\delta_n^{3/2} + \max_{-L+1 \leq \ell \leq 0} \|\theta_{b-\ell}-\hat\theta_n\| \right], \]
with probability $1-(5+C)/c_n$. Finally, for (\ref{eq:tbb12}), we can use:
\[ \|\hat{G}_b - G(\hat\theta_n)\| \leq L_G \mu_{d_\theta}(\max_{-L+1 \leq \ell \leq 0}\|\theta_{b-\ell}-\hat\theta_n\|) +  C_{(\ref{eq:tbb12})} \left( c_n n^{-1/2}\varepsilon^{\psi-1} + \varepsilon + L^{-1/2} \right)\delta_n^{3/2} , \]
with probability $1-(5+C)/c_n$. 

Now take $\mathcal{E}_b = (\max_{-L+1 \leq \ell \leq 0} \|\theta_{b-\ell}-\hat\theta_n\|) \leq R_{G,2} - C_a c_nn^{-1/2}$, so that $\|\theta_{b-\ell}-\theta^\dagger\|\leq R_G$ for all $\ell \in \{ -L+1,\dots,0 \}$. Combine the bounds to find:
\begin{align*}
  \|\theta_{b+1}-\hat\theta_n\| \leq \left(1-\gamma + \gamma\hat{\underline{\sigma}}_{n,\varepsilon}^{-1}\sqrt{\kappa_W}L_G\left[  \|\theta_b - \hat\theta_n\| + \mu_{d_\theta} \mathcal{E}_b   \right] \right)\|\theta_b- \hat\theta_n\| + \frac{\gamma}{\hat{\underline{\sigma}}_{n,\varepsilon}^2} \hat{\Delta}_{n,\varepsilon}(\|\theta_b-\hat\theta_n\|,\mathcal{E}_b),
\end{align*}
with probability $1-(5+C)/c_n$. The last term is given by:
\[ \hat{\Delta}_{n,\varepsilon}(\|\theta_b-\hat\theta_n\|,\mathcal{E}_b) = C_3 \left( \hat{\Gamma}_{n,\varepsilon} + \delta_n^{3/2}[c_nn^{-1/2}\varepsilon^{\psi-1} + \varepsilon + L^{-1/2}]\mathcal{E}_b + c_nn^{-1/2}\|\theta_b-\hat\theta_n\|^\psi   \right), \]
using: $\hat{\Gamma}_{n,\varepsilon} = c_nn^{-1/2} \left[ (c_nn^{-1/2})^\psi + \delta_n^{3/2}(c_nn^{-1/2}\varepsilon^{\psi-1} + \varepsilon + L^{-1/2}) \right].$
Noting that $\|\theta_b- \hat\theta_n\| \leq \mathcal{E}_b$, this implies that \[ \mathcal{E}_b \leq \frac{\overline{\gamma}-\gamma}{\gamma} \frac{\hat{\underline{\sigma}}_{n,\varepsilon}}{L_G \sqrt{\kappa_W}(1+\mu_{d_\theta})} \Rightarrow  \|\theta_{b+1}-\hat\theta_n\| \leq (1-\overline{\gamma})\|\theta_b-\hat\theta_n\| + \frac{\gamma}{\hat{\underline{\sigma}}_{n,\varepsilon}^2} \hat{\Delta}_{n,\varepsilon}(\|\theta_b-\hat\theta_n\|,\mathcal{E}_b), \]
with probability $1-(5+C)/c_n$, which is the desired contraction result.
\qed

\section{Additional Results for Section \ref{sec:MC_Emp}} \label{apx:extra_MC_Emp}
\subsection{Aiyagari Model} \label{apx:Aiyagari}
\paragraph{Additional implementation details}
Consumers face an exogenous income flow and choose consumption $c_t$ to maximize the expected intertemporal utility under a borrowing constraint:
$\max \mathbb{E}_t \left( \sum_{j=0}^\infty \beta^j u(c_{t+j}) \right) \text{ s.t. } a_{t + j +1} + c_{t + j} \leq y_{t + j} + (1+r)a_{t + j}, a_{t + j} \geq \underline{a}$, 
where $a_{t+j}\geq 0$ are future saving at period $t+j$ for $j=0,1,\dots$, $r \geq 0$ is the interest rate, and $\underline{a} \leq 0$ is the borrowing constraint. The utility function is $u(c) = (c^\gamma -1)/(1-\gamma)$, where $\gamma > 0$ measures risk-aversion. Finally, income $y_{t+j}$ follows an AR(1) process in logarithms: $\log(y_{t+1}) = \mu + \rho [ \log(y_{t}) - \mu ] + \sigma e_t, \quad e_t \overset{iid}{\sim} \mathcal{N}(0,1)$.

To solve the model, the income process is discretized using the quadrature method of \citet{tauchen1991} on a 15 point grid. Then the policy function is computed using value function iterations over a discretized state-space of 300 points.\footnote{There are many other solution methods for this type of model. Value function iterations can be applied to wide range of models but also make estimation very difficult. As such, it makes for a good benchmark to evaluate Algorithm \ref{algo:sgn}.} A panel of consumers is generated $(a_{it},y_{it})$ with $i = 1,\dots n$, $n = 10000$, $t=1,2$.\footnote{A burn-in of $248$ time periods is discarded for each individual to approximate the ergodic distribution.} The estimation matches the $\tau=0.2,0.3,\dots,0.7,0.8$ level quantiles of the asset distribution $(y_{i2})$ between sample and simulated data, as well as the OLS AR(1) estimates from regressing $\log(y_{i2})$ on $\log(y_{i1})$. Interest rates are fixed to a $5\%$ annual rate. The true value is $\theta^\dagger = (\beta^\dagger,\gamma^\dagger,\mu^\dagger,\rho^\dagger,\sigma^\dagger) = (0.97,3,\log(6.5),0.7,0.2)$.

\begin{figure}[ht] \caption{Aiyagari Model: Iteration of Algorithm \ref{algo:sgn}} \label{fig:Aiyagari1} \centering
  \includegraphics[scale=0.52]{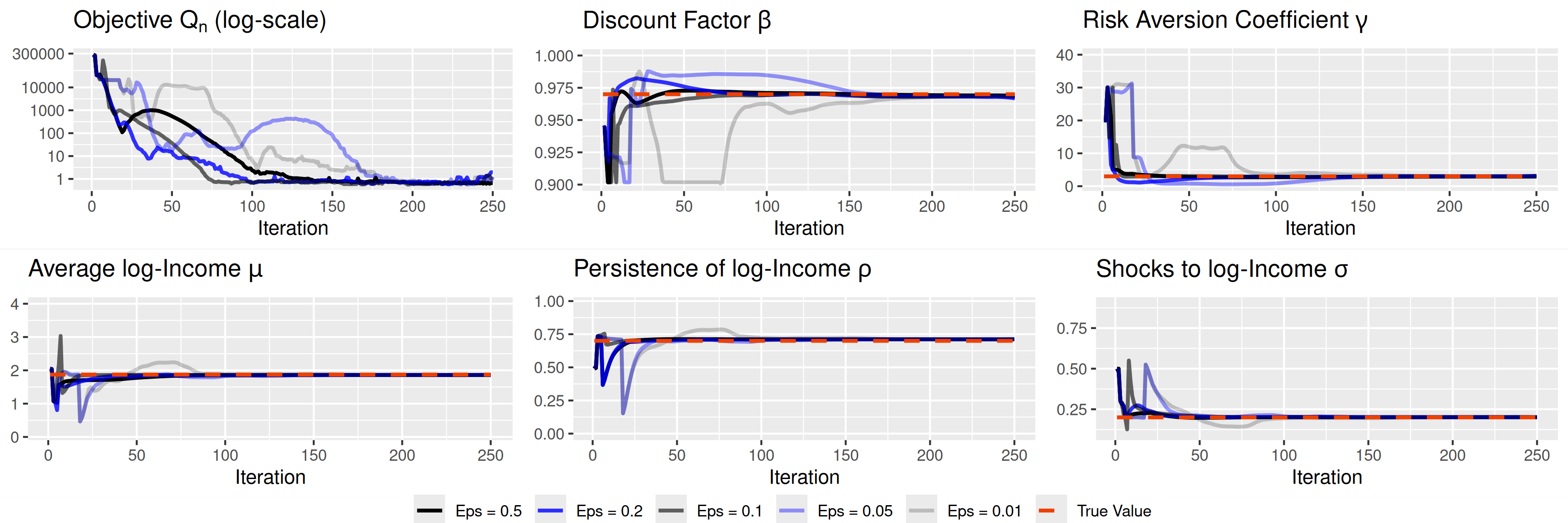}\\
  {\footnotesize \textbf{Legend:} $n=10000$, $T=2$. Algorithm \ref{algo:sgn} uses $\gamma = 0.1$, $\alpha=0.47$.}
\end{figure}

\begin{table}[ht] \caption{Aiyagari Model: Estimates with Different $\gamma,\alpha$}
  \label{tab:Aiyagari_tab2}
  \centering \setlength\tabcolsep{2.5pt}
  \renewcommand{\arraystretch}{0.9} 
  {\small 
  \begin{tabular}{cc|c|ccccc|c}
    \hline \hline
    & & $Q_n$ & $\beta$ & $\gamma$ & $\mu$ & $\rho$ & $\sigma$ & Time  \\ \hline
     &  & - & 0.970 & 3.000 & 1.872 & 0.700 & 0.200 & (hr:mn)\\ \hline
    $\gamma$ & $\alpha$ & \multicolumn{6}{c}{Algorithm \ref{algo:sgn}}\\
    \hline
    $0.1$ & $0.47$ & 0.551 & 0.968 & 3.031 & 1.860 & 0.712 & 0.201 & 00:34 \\ 
    $0.1$ & $0.0$ & 0.561 & 0.968 & 3.025 & 1.860 & 0.712 & 0.201 & 00:34 \\ 
    $0.2$ & $0.0$ & 0.557 & 0.968 & 3.030 & 1.860 & 0.712 & 0.201 & 00:34 \\ 
    $0.4$ & $0.0$ & 0.560 & 0.968 & 3.029 & 1.860 & 0.712 & 0.201 & 00:34\\ 
    $0.01$ & $0.0$ & 594.199 & 0.981 & 1.663 & 1.627 & 0.644 & 0.263 & 00:34 \\ 
    $0.01$ & $0.81$ & 0.681 & 0.970 & 2.894 & 1.858 & 0.711 & 0.201 & 00:34 \\ 
     \hline \hline
  \end{tabular}
  }\\
{\footnotesize \textbf{Legend:} Algorithm \ref{algo:sgn} $\varepsilon = 0.1$, $B = 250$ iterations in total. }
\end{table}

\subsection{Interdependent Durations} \label{Apx:HP_add}

\begin{figure}[ht] \caption{Interdependent Duration Estimates: MCMC and s\textsc{gn}} \label{fig:HP_MCMC}
  \centering
  \includegraphics[scale=0.45]{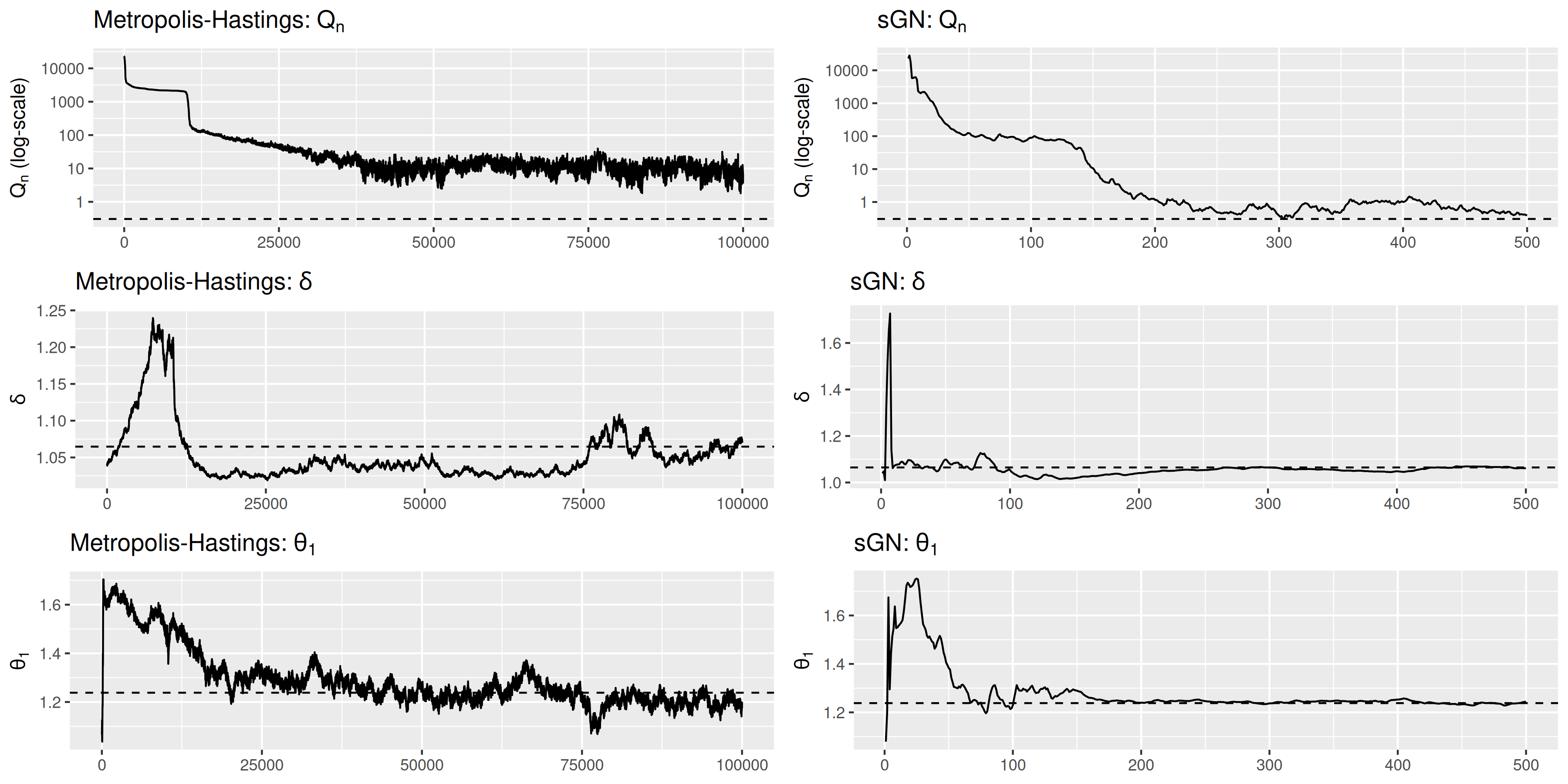}\\
  {\footnotesize Legend: s\textsc{gn}: $\varepsilon = 10^{-2}$, $\gamma = 0.1$, $\alpha = 0.47$, $B = 500$ iterations in total. MCMC: $100000$ iterations, same starting value, random-walk tuned to target $\approx 38\%$ acceptance rate around the solution $\hat\theta_n$.}
\end{figure}


\begin{table} \caption{Interdependent Duration Estimates: \citet{honore2018} and s\textsc{gn}}
  \label{tab:est_hp_full}
  \centering \setlength\tabcolsep{2.5pt}
  \renewcommand{\arraystretch}{0.9} 
  {\small \begin{tabular}{l|cc|aa||cc|aa} \hline \hline
  & \multicolumn{4}{c||}{Coefficients for Wives} & \multicolumn{4}{c}{Coefficients for Husbands}\\ \hline
  & \multicolumn{2}{c|}{\footnotesize Honor\'e \& de Paula} & \multicolumn{2}{c||}{s\textsc{gn}} & \multicolumn{2}{c|}{\footnotesize Honor\'e \& de Paula} & \multicolumn{2}{c}{s\textsc{gn}}\\ \hline
  \multirow{2}{*}{\footnotesize $\delta$} & 
  $1.052$ & $1.064$ & $1.065$ & $1.066 $ & 1.052 & 1.064 & $1.065$ & $1.066$ \\ 
  & {\footnotesize { $(0.039) $}} & {\footnotesize { $(0.042) $}} & 
  {\footnotesize { $(0.039) $}} & {\footnotesize { $ (  0.037) $}}
  & {\footnotesize {$ (  0.039) $}} & {\footnotesize {$ (  0.042) $}} & {\footnotesize {$ (  0.039) $}} & {\footnotesize {$ (  0.037) $}} \\ \hline
  \multirow{2}{*}{\footnotesize $\theta_1$} & 
  $1.244$ & $1.244$ & $1.238$ & $1.224$ & $1.169$ & $1.218$ & $1.179 $ & $1.188$ \\ 
  & {\footnotesize { $(0.054) $}} & {\footnotesize { $(0.054) $}} & 
  {\footnotesize { $(0.055) $}} & {\footnotesize { $ (  0.050) $}}& 
  {\footnotesize { $(0.043) $}} & {\footnotesize { $ (  0.058) $}}& {\footnotesize {$ (  0.043) $}} & {\footnotesize {$ (  0.040) $}}\\ \hline
  \multirow{2}{*}{\footnotesize $\ge$ 62 yrs-old} & 
  $10.640$ & $13.446$ & $11.194$ & $11.602 $ & $31.532$ & $39.824$ & $32.817$ & $34.268$\\ 
  & {\footnotesize { $(5.916) $}} & {\footnotesize { $(5.694) $}} & 
  {\footnotesize { $(7.818) $}} & {\footnotesize { $ (  5.692) $}}& {\footnotesize { $(11.356) $}} & {\footnotesize { $ (11.372) $}}& 
  {\footnotesize {$ (  8.131) $}} & {\footnotesize {$ (  7.672) $}}\\ \hline
  \multirow{2}{*}{\footnotesize $\ge$ 65 yrs-old} & 
  $10.036$ & $12.326$ & $10.613$ & $11.721 $ & $25.696$ & $29.254$ & $26.352$ & $26.000$\\ 
 & {\footnotesize { $ ( 11.555) $}} & {\footnotesize { $ ( 7.495) $}} & 
 {\footnotesize { $ ( 10.067) $}} & {\footnotesize { $ ( 10.897) $}}& {\footnotesize { $ ( 9.497) $}} & {\footnotesize { $ ( 11.229) $}}& {\footnotesize {$ ( 13.215) $}} & {\footnotesize {$ ( 14.289) $}}\\ \hline
 \multirow{2}{*}{\footnotesize Constant}& 
 $-5.786$ & $-5.790$ & $-5.781$ & $-5.664$ & $-5.587 $ & $-5.449$ & $-5.642 $ & $-5.320 $ \\
   & {\footnotesize { $ (  0.225) $}} & {\footnotesize { $ (  0.276) $}} & 
   {\footnotesize { $ (  0.226) $}} & {\footnotesize { $ (  0.250) $}}& {\footnotesize { $ (  0.231) $}} & {\footnotesize { $ (  0.266) $}}& {\footnotesize {$ (  0.189) $}} & {\footnotesize {$ (  0.207) $}}\\ \hline 
   \multirow{2}{*}{\footnotesize Age Diff.}& 
   $-0.074$ & $-0.075$ & $-0.074$ & $-0.073$ & $0.021$ & $0.025$ & $0.021$ & $0.025 $\\ 
  & {\footnotesize { $ (  0.016) $}} & {\footnotesize { $ (  0.016) $}} & {\footnotesize { $ (  0.016) $}} & {\footnotesize { $ (  0.016) $}}& {\footnotesize { $ (  0.008) $}} & {\footnotesize { $ (  0.007) $}}& {\footnotesize {$ (  0.008) $}} & {\footnotesize {$ (  0.008) $}}\\ \hline 
  \multirow{2}{*}{\footnotesize Non-Hisp. Black}&  
  & $-0.149$ & $ $ & $   -0.159 $ & & $-0.203$ & & $-0.185$ \\ 
  & {\footnotesize { $  $}} & {\footnotesize { $ (0.153) $}} & {\footnotesize { $  $}} & {\footnotesize { $ (  0.156) $}}& {\footnotesize { $  $}} & {\footnotesize { $ (  0.155) $}}& {\footnotesize {$ $}} & {\footnotesize {$ (  0.162) $}}\\  \hline 
  \multirow{2}{*}{\footnotesize Other race} & $          $ & $   -0.649 $ & $          $ & $   -0.634 $ & & $-0.151$ & & $-0.197$ \\ 
 & {\footnotesize { $ $}} & {\footnotesize { $ (  0.337) $}} & {\footnotesize { $ $}} & {\footnotesize { $ (  0.308) $}}& {\footnotesize { $ $}} & {\footnotesize { $ (  0.287) $}}& {\footnotesize {$ $}} & {\footnotesize {$ (  0.289) $}}\\  \hline 
 \multirow{2}{*}{\footnotesize Hispanic}& $          $ & $   -0.490 $ & $          $ & $   -0.494 $ & & $-0.626$ & & $-0.633$\\ 
   & {\footnotesize { $ $}} & {\footnotesize { $ (  0.192) $}} & {\footnotesize { $ $}} & {\footnotesize { $ (  0.189) $}}& {\footnotesize {$ $}} & {\footnotesize {$ (  0.180) $}}& {\footnotesize {$ $}} & {\footnotesize {$ (  0.178) $}}\\  \hline 
   \multirow{2}{*}{\footnotesize High school or GED}& & $0.052$ & $          $ & $    0.029 $ & & $-0.109$ & & $-0.106$ \\ 
   & {\footnotesize { $ $}} & {\footnotesize { $ (  0.158) $}} & {\footnotesize { $ $}} & {\footnotesize { $ (  0.158) $}}& {\footnotesize { $ $}} & {\footnotesize { $ (  0.118) $}}& {\footnotesize { $ $}} & {\footnotesize { $ (  0.128) $}}\\  \hline 
  \multirow{2}{*}{\footnotesize Some college}& $          $ & $   -0.131 $& $          $ & $   -0.141 $ & & $-0.357$ & & $-0.343$ \\ 
  & {\footnotesize { $           $}} & {\footnotesize { $ (  0.169) $}} & {\footnotesize { $           $}} & {\footnotesize { $ (  0.170) $}}& {\footnotesize {$           $}} & {\footnotesize {$ (  0.133) $}}& {\footnotesize {$           $}} & {\footnotesize {$ (  0.144) $}}\\  \hline 
  \multirow{2}{*}{\footnotesize College or above}& $          $ & $   -0.052 $ & $          $ & $   -0.072$ & & $-0.522$ & & $-0.502$ \\ 
  & {\footnotesize { $           $}} & {\footnotesize { $ (  0.189) $}} & {\footnotesize { $           $}} & {\footnotesize { $ (  0.192) $}}& {\footnotesize {$           $}} & {\footnotesize {$ (  0.128) $}}& {\footnotesize {$           $}} & {\footnotesize {$ (  0.135) $}}\\  \hline 
  \multirow{2}{*}{\footnotesize NE}& $          $ & $0.002$ & $          $ & $   -0.041 $ & & $0.060$ & & $0.043$ \\ 
  & {\footnotesize { $           $}} & {\footnotesize { $ (  0.146) $}} & {\footnotesize { $           $}} & {\footnotesize { $ (  0.149) $}}& {\footnotesize {$           $}} & {\footnotesize {$ (  0.122) $}}& {\footnotesize {$           $}} & {\footnotesize {$ (  0.133) $}}\\  \hline 
  \multirow{2}{*}{\footnotesize SO}& $          $ & $    0.065 $ & $          $ & $    0.027 $ & & $-0.219$ & & $-0.239$\\ 
  & {\footnotesize { $           $}} & {\footnotesize { $ (  0.115) $}} & {\footnotesize { $           $}} & {\footnotesize { $ (  0.115) $}}& {\footnotesize {$           $}} & {\footnotesize {$ (  0.106) $}}& {\footnotesize {$           $}} & {\footnotesize {$ (  0.111) $}}\\  \hline 
  \multirow{2}{*}{\footnotesize WE}& $          $ & $    0.217 $ & $          $ & $    0.188 $ &  & $0.066$ &  & $0.067$\\ 
  & {\footnotesize { $           $}} & {\footnotesize { $ (  0.145) $}} & {\footnotesize { $           $}} & {\footnotesize { $ (  0.150) $}}& {\footnotesize {$           $}} & {\footnotesize {$ (  0.121) $}}& {\footnotesize {$           $}} & {\footnotesize {$ (  0.131) $}}\\  \hline 
  \multirow{2}{*}{\footnotesize $\tau$} & $    0.526 $ & $    0.429 $ & $    0.416 $ & $    0.427 $ & {\footnotesize { -}} & {\footnotesize { -}} & {\footnotesize { -}} & {\footnotesize { -}} \\ 
  & {\footnotesize { $ (  0.399) $}} & {\footnotesize { $ (  0.371) $}} & {\footnotesize { $ (  0.296) $}} & {\footnotesize { $ (  0.300) $}} & {\footnotesize { -}} & {\footnotesize { -}} & {\footnotesize {-}} & {\footnotesize {-}}\\   \hline
  {\footnotesize Starting Obj. Value} & $93.70$ & $89.77$ & $2.10^4$ & $5.10^4$ & {\footnotesize { -}} & {\footnotesize { -}} & {\footnotesize { -}} & {\footnotesize { -}} \\ 
  {\footnotesize Final Obj. Value } & $0.470$ & $0.758$ & $0.306$ & $0.364$ & {\footnotesize { -}} & {\footnotesize { -}} & {\footnotesize { -}} & {\footnotesize { -}} \\    \hline
  {\footnotesize Number of Coef. } & $12$ & $30$ & $12$ & $30$ &  {\footnotesize { -}} & {\footnotesize { -}} & {\footnotesize { -}} & {\footnotesize { -}} \\ 
  {\footnotesize Number of Obs.} & $     1227 $ & $     1227 $ & $     1227 $ & $     1227 $ &  {\footnotesize { -}} & {\footnotesize { -}} & {\footnotesize { -}} & {\footnotesize { -}} \\ \hline
  {\footnotesize Computation Time }  & 3h25m & 5h34m & 12min & 13min & {\footnotesize { -}} & {\footnotesize { -}} & {\footnotesize { -}} & {\footnotesize { -}} \\ \hline \hline
\end{tabular}}\\
{\footnotesize Legend: s\textsc{gn}: $\varepsilon = 10^{-2}$, $\gamma = 0.1$, $\alpha = 0.47$, $B = 500$ iterations in total. Husbands: - same as wives. Coefficients for wives and husbands are estimated jointly.}
\end{table}

\begin{table} \caption{Interdependent Duration Estimates: Different Bandwidths $\varepsilon$}
  \label{tab:est_hp_eps}
  \centering \setlength\tabcolsep{2.5pt}
  \renewcommand{\arraystretch}{0.9} 
  {\small \begin{tabular}{l|cc|aa||cc|aa} \hline \hline
  & \multicolumn{4}{c||}{Coefficients for Wives} & \multicolumn{4}{c}{Coefficients for Husbands}\\ \hline
  & \multicolumn{2}{c|}{\footnotesize {$\varepsilon = 0.8 \cdot 10^{-2}$}} & \multicolumn{2}{c||}{ \footnotesize {$\varepsilon = 2.11 \cdot 10^{-2}$} } & \multicolumn{2}{c|}{\footnotesize {$\varepsilon = 0.8 \cdot 10^{-2}$}} & \multicolumn{2}{c}{ \footnotesize {$\varepsilon = 2.11 \cdot 10^{-2}$} }\\ \hline
  \multirow{2}{*}{\footnotesize $\delta$} & 
  $1.061$& $1.067$ & $1.064$ & $1.066$ & $1.061$ & $1.067$ & $1.064$ & $1.066$ \\ 
  & {\footnotesize { $ (  0.039) $}} & {\footnotesize { $ (  0.037) $}} 
  & {\footnotesize { $(0.039)$}} & {\footnotesize { $(0.037)$}}
  & {\footnotesize {$(0.039)$}} & {\footnotesize {$(0.037)$}} 
  & {\footnotesize {$(0.039)$}} & {\footnotesize {$(0.037)$}} \\ \hline
  \multirow{2}{*}{\footnotesize $\theta_1$} & 
  $1.239$& $1.234$ & $1.241$ & $1.233$ & $1.180$ & $1.190$ & $1.176$ & $1.189$ \\ 
  & {\footnotesize { $ (  0.055) $}} & {\footnotesize { $ (  0.050) $}} 
  & {\footnotesize { $(0.055)$}} & {\footnotesize { $(0.050)$}}
  & {\footnotesize {$(0.043)$}} & {\footnotesize {$(0.040)$}}
  & {\footnotesize {$(0.043)$}} & {\footnotesize {$(0.040)$}}\\ \hline
  \multirow{2}{*}{\footnotesize $\ge$ 62 yrs-old} & 
  $10.417$& $12.347$ & $11.469$ & $12.272$ & $33.007$ & $34.674$ & $32.881$ & $34.406$\\ 
  & {\footnotesize { $ (  7.818) $}} & {\footnotesize { $ (  5.692) $}} 
  & {\footnotesize { $(7.818)$}} & {\footnotesize { $(5.692)$}}
  & {\footnotesize {$(8.131)$}} & {\footnotesize {$(7.672)$}}
  & {\footnotesize {$(8.131)$}} & {\footnotesize {$(7.672)$}}\\ \hline
  \multirow{2}{*}{\footnotesize $\ge$ 65 yrs-old} & 
  $11.029$& $12.260$ & $11.413$ & $11.441$ & $25.179$ & $25.930$ & $24.768$ & $25.950$\\ 
 & {\footnotesize { $ ( 10.067) $}} & {\footnotesize { $ ( 10.897) $}} 
 & {\footnotesize { $ ( 10.067)$}} & {\footnotesize { $ ( 10.897)$}}
 & {\footnotesize {$( 13.215)$}} & {\footnotesize {$( 14.289)$}}
 & {\footnotesize {$ ( 13.215)$}} & {\footnotesize {$ ( 14.289)$}}\\ \hline
 \multirow{2}{*}{\footnotesize Constant}& 
 $-5.774$& $-5.691$ & $-5.791$ & $-5.771$ & $-5.639$ & $-5.333$ & $-5.623$ & $-5.328$ \\
 & {\footnotesize { $ (  0.226) $}} & {\footnotesize { $ (  0.250) $}} 
 & {\footnotesize { $(0.226)$}} & {\footnotesize { $(0.250)$}}
 & {\footnotesize {$(0.189)$}} & {\footnotesize {$(0.207)$}}
 & {\footnotesize {$(0.189)$}} & {\footnotesize {$(0.207)$}}\\ \hline 
   \multirow{2}{*}{\footnotesize Age Diff.}& 
   $-0.074$& $-0.073$ & $-0.075$ & $-0.074$ & $0.021$ & $0.025$ & $0.021$ & $0.025$\\ 
  & {\footnotesize { $ (  0.016) $}} & {\footnotesize { $ (  0.016) $}} 
  & {\footnotesize { $(0.016)$}} & {\footnotesize { $(0.016)$}}
  & {\footnotesize {$(0.008)$}} & {\footnotesize {$(0.008)$}}
  & {\footnotesize {$(0.008)$}} & {\footnotesize {$(0.008)$}}\\ \hline 
  \multirow{2}{*}{\footnotesize Non-Hisp. Black}&  $ $& $-0.143$ & $  $ & $-0.160$ & $  $ & $-0.173$ & $  $ & $-0.183$ \\ 
  & {\footnotesize { } } & {\footnotesize { $ (  0.156) $}} 
  & {\footnotesize { $    $}} & {\footnotesize { $(0.156)$}}
  & {\footnotesize {$    $}} & {\footnotesize {$(0.162)$}}
  & {\footnotesize {$    $}} & {\footnotesize {$(0.162)$}}\\  \hline 
  \multirow{2}{*}{\footnotesize Other race} & $ $ & $-0.640$ & $  $ & $-0.614$ & $  $ & $-0.165$ & $  $ & $-0.172$ \\ 
 & {\footnotesize { $ $}} & {\footnotesize { $ (  0.308) $}} 
 & {\footnotesize { $    $}} & {\footnotesize { $(0.308)$}}
 & {\footnotesize {$    $}} & {\footnotesize {$(0.289)$}}
 & {\footnotesize {$    $}} & {\footnotesize {$(0.289)$}}\\  \hline 
 \multirow{2}{*}{\footnotesize Hispanic}& $ $& $-0.493$ & $  $ & $-0.482$ & $  $ & $-0.629$ & $  $ & $-0.616$\\ 
   & {\footnotesize { $ $}} & {\footnotesize { $ (  0.189) $}} 
   & {\footnotesize { $    $}} & {\footnotesize { $(0.189)$}}
   & {\footnotesize {$    $}} & {\footnotesize {$(0.178)$}}
   & {\footnotesize {$    $}} & {\footnotesize {$(0.178)$}}\\  \hline 
   \multirow{2}{*}{\footnotesize High school or GED}& $ $ & $0.009$ & $  $ & $0.033$ & $  $ & $-0.102$ & $  $ & $-0.103$ \\ 
   & {\footnotesize { $ $}} & {\footnotesize { $ (  0.158) $}} 
   & {\footnotesize { $    $}} & {\footnotesize { $(0.158)$}}
   & {\footnotesize {$    $}} & {\footnotesize {$(0.128)$}}
   & {\footnotesize {$    $}} & {\footnotesize {$(0.128)$}}\\  \hline 
  \multirow{2}{*}{\footnotesize Some college}& $ $& $-0.159$ & $  $ & $-0.141$ & $  $ & $-0.327$ & $  $ & $-0.334$ \\ 
  & {\footnotesize { $ $}} & {\footnotesize { $ (  0.170) $}} 
  & {\footnotesize { $    $}} & {\footnotesize { $(0.170)$}}
  & {\footnotesize {$    $}} & {\footnotesize {$(0.144)$}}
  & {\footnotesize {$    $}} & {\footnotesize {$(0.144)$}}\\  \hline 
  \multirow{2}{*}{\footnotesize College or above}& $ $& $-0.085$ & $  $ & $-0.068$ & $  $ & $-0.498$ & $  $ & $-0.492$ \\ 
  & {\footnotesize { $ $}} & {\footnotesize { $ (  0.192) $}} 
  & {\footnotesize { $    $}} & {\footnotesize { $(0.192)$}}
  & {\footnotesize {$    $}} & {\footnotesize {$(0.135)$}}
  & {\footnotesize {$    $}} & {\footnotesize {$(0.135)$}}\\  \hline 
  \multirow{2}{*}{\footnotesize NE}& $ $& $-0.055$ & $  $ & $-0.041$ & $  $ & $0.043$ & $  $ & $0.048$ \\ 
  & {\footnotesize { $ $}} & {\footnotesize { $ (  0.149) $}} 
  & {\footnotesize { $    $}} & {\footnotesize { $(0.149)$}}
  & {\footnotesize {$    $}} & {\footnotesize {$(0.133)$}}
  & {\footnotesize {$    $}} & {\footnotesize {$(0.133)$}}\\  \hline 
  \multirow{2}{*}{\footnotesize SO}& $ $& $0.026$ & $  $ & $0.029$ & $  $ & $-0.240$ & $  $ & $-0.239$\\ 
  & {\footnotesize { $ $}} & {\footnotesize { $ (  0.115) $}} 
  & {\footnotesize { $    $}} & {\footnotesize { $(0.115)$}}
  & {\footnotesize {$    $}} & {\footnotesize {$(0.111)$}}
  & {\footnotesize {$    $}} & {\footnotesize {$(0.111)$}}\\  \hline 
  \multirow{2}{*}{\footnotesize WE}& $ $& $0.182$ & $  $ & $0.184$ &  $  $ & $0.067$ & $  $ & $0.058$\\ 
  & {\footnotesize { $ $}} & {\footnotesize { $ (  0.150) $}} 
  & {\footnotesize { $    $}} & {\footnotesize { $(0.150)$}}
  & {\footnotesize {$    $}} & {\footnotesize {$(0.131)$}}
  & {\footnotesize {$    $}} & {\footnotesize {$(0.131)$}}\\  \hline 
  \multirow{2}{*}{\footnotesize $\tau$} & $0.424$& $0.428$ & $0.405$ & $0.416$ & {\footnotesize { -}} & {\footnotesize { -}} & {\footnotesize { -}} & {\footnotesize { -}} \\ 
  & {\footnotesize { $ (  0.296) $}} & {\footnotesize { $ (  0.300) $}} 
  & {\footnotesize { $(0.296)$}} & {\footnotesize { $(0.300)$}}
  & {\footnotesize { -}} & {\footnotesize { -}} & {\footnotesize {-}} 
  & {\footnotesize {-}}\\   \hline
  {\footnotesize Starting Obj. Value} & $2.10^4$ & $5.10^4$ & $2.10^4$ & $5.10^4$ & {\footnotesize { -}} & {\footnotesize { -}} & {\footnotesize { -}} & {\footnotesize { -}} \\ 
  {\footnotesize Final Obj. Value } & $0.296$& $0.428$ & $0.360$ & $0.284$ & {\footnotesize { -}} & {\footnotesize { -}} & {\footnotesize { -}} & {\footnotesize { -}} \\    \hline
  {\footnotesize Number of Coef. } & $12$ & $30$ & $12$ & $30$ &  {\footnotesize { -}} & {\footnotesize { -}} & {\footnotesize { -}} & {\footnotesize { -}} \\ 
  {\footnotesize Number of Obs.} & $     1227 $ & $     1227 $ & $     1227 $ & $     1227 $ &  {\footnotesize { -}} & {\footnotesize { -}} & {\footnotesize { -}} & {\footnotesize { -}} \\ \hline
  {\footnotesize Computation Time }  & 12min & 14min & 12min & 14min & {\footnotesize { -}} & {\footnotesize { -}} & {\footnotesize { -}} & {\footnotesize { -}} \\ \hline \hline
\end{tabular}}\\
{\footnotesize Legend: s\textsc{gn}: $\gamma = 0.1$, $\alpha = 0.47$, $B = 500$ iterations in total. Husbands: - same as wives. Coefficients for wives and husbands are estimated jointly.}
\end{table}

\section{Comparisons with other methods} \label{apx:comp}
\subsection{Multi-start Algorithm} \label{apx:MultStart}
The following discusses the differences between the global convergence results in Lemma \ref{lem:cv_global} and global convergence using multiple starting values for the local optimizer in Lemma \ref{lem:cv_local}.  Recall from Lemma \ref{lem:cv_local} that for any $\|\theta_0 - \theta^\dagger\| \leq R$, we have $\|\theta_b - \theta^\dagger\| \leq (1-\overline{\gamma})^b R.$ Using covering arguments, running the local algorithm with $\tilde{k}\geq 1$ different starting values $(\theta^\ell)_{ 0 \leq \ell \leq \tilde{k}-1}$ results in convergence if $\|\theta^\ell - \theta^\dagger\|\leq R$ for some $\ell \in \{0,\dots,\tilde{k}-1\}$. Using covering arguments, this implies that $\tilde{k} \geq R^{-d_\theta} \text{vol}(\Theta)/\text{vol}(B)$. Because $R \geq \underline{r}_g$, using the lower bounds for $\tilde{k}$ and the $k$ required in Lemma \ref{lem:cv_global} we have $\tilde{k} \leq k$ for the same local step. 

As for Algorithm \ref{algo:sgn}, the $\tilde{k}$ needed in practice depends on the choice of covering sequence, the parameter space, and the moments. It is not known to the researcher so that it is often recommended to pick a large $\tilde{k}$, though some papers use only a handful of starting values in practice. This contrasts with Algorithm \ref{algo:sgn} where the user only needs to input a stopping criteria, it does not require to set $k$ explicitly as the Algorithm transitions from global to local convergence without any input from the user (cf. Lemma \ref{lem:cv_global}, Theorem \ref{th:sgn_global}). This implies that the $\tilde{k}$ used in practice could be greater than $k$, if the user chooses a very large $\tilde{k}$ to guarantee convergence. 
If the multi-start algorithm is run for $\tilde{b}$ iterations for each starting value. The cost of running the algorithm is at least $\tilde{k} \times \tilde{b}$. In comparison, the combined local and global steps are run for $b = k +j$ iterations, but each iteration involves both the local and global steps instead of just the local step. The combined local and global steps are less costly when the user would set $\tilde{k} > k$ to be very large to ensure convergence. Also note, that other choices of local minimizers could be associated with a different $R$, potentially smaller than $\underline{r}_g$, and a potentially slower rate of convergence. This could be the case when using a multi-start approach with the Nelder-Mead algorithm for the local search, for instance.

\subsection{Two-Step Algorithm} \label{apx:twoStep}
\citet{robinson1988} and \citet{andrews1997} suggest a two-step algorithm where a grid-search produces a consistent estimator $\tilde \theta_n$ and an additional k-step Newton-Raphson iterations yield an estimate which is asymptotically  first-order equivalent to $\hat\theta_n$, the required $k$ depends on the rate of convergence of $\tilde\theta_n$. Take $\alpha \in (0,1/2]$, the first-step requires $\|\tilde \theta_n - \hat\theta_n\| \leq O(n^{-\alpha})$, which implies $O(n^{\alpha d_\theta})$ iterations are required. For the second step, \citet{andrews1997} shows that $k > -\log(\alpha)/\log(2)$ iterations are required. Overall, the procedure requires $O(n^{d/\alpha}) -\log(\alpha)/\log(2)$ iterations. In comparision, Lemma \ref{lem:cv_global} implies that $k + O(|\log(n)|)$ iterations are required which is an order of magnitude smaller. Using the Sobol or Halton sequence, we have $k = O(\underline{r}_g^{-d_\theta})$, up to log terms, which is typically much smaller than $n^{\alpha d_\theta}$.

The key difference between the two approaches is that the local optimization step in \citet{robinson1988}, \citet{andrews1997} needs to be initialized in a vanishing neighborhood $\|\tilde\theta_n - \theta^\dagger\| = o_p(1)$ whereas Theorem \ref{th:sgn_local} shows that the local step is valid for $\|\theta_0 - \hat\theta_n\| \leq R_{n,\varepsilon} = R - o(1)$ with high probability, which is a non-vanishing neighborhood; a requirement to improve the global rate of convergence. Here, the local step does not require to first produce a consistent estimator, unlike existing methods.

\section{Illustration: choice of covering sequence} \label{apx:cover}
\begin{figure} \caption{Illustration: low-dispersion sequence vs. random uniform draws} \label{fig:cover}
  \begin{center}
    \includegraphics[scale=0.45]{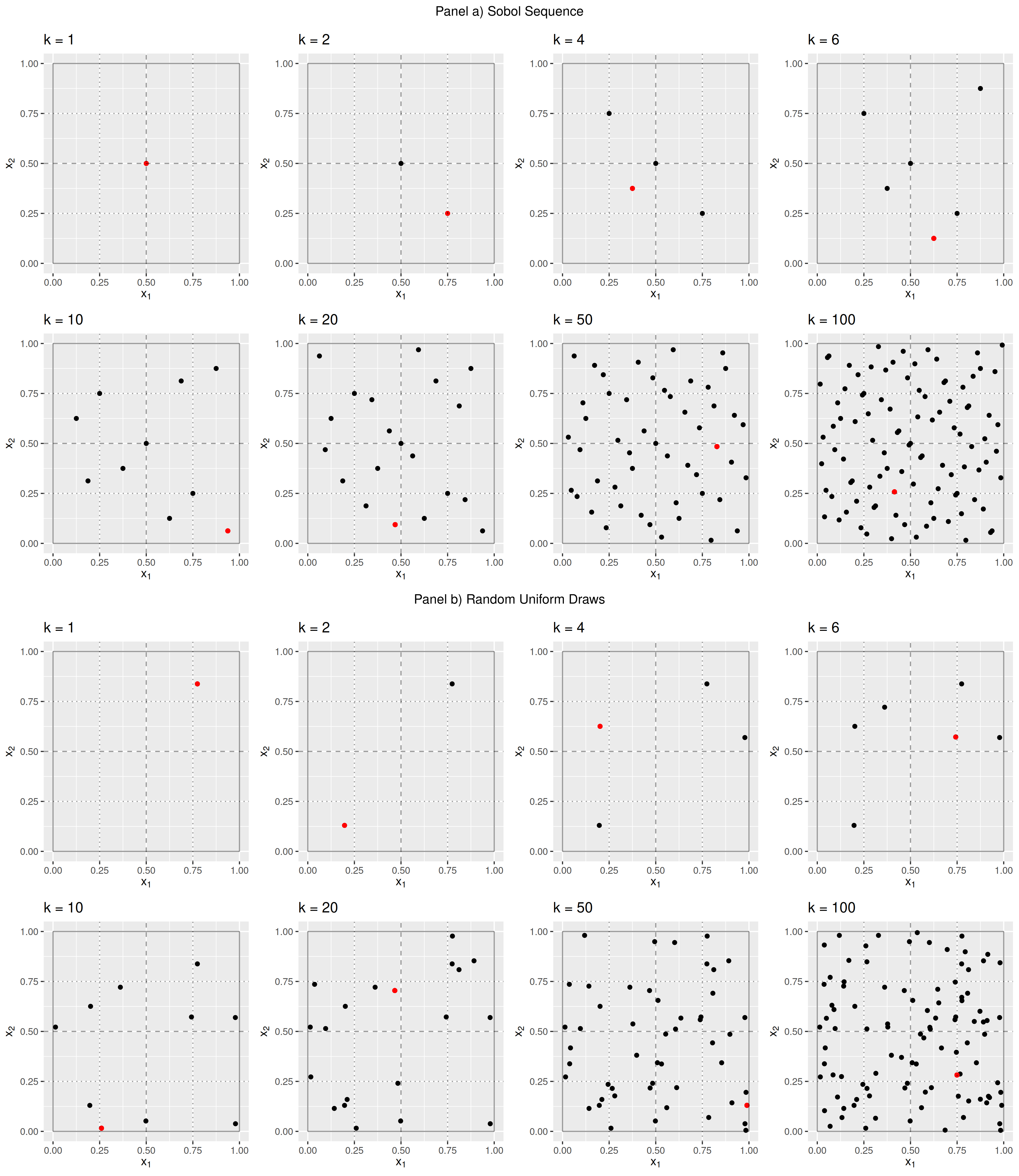}
  \end{center}
  {\footnotesize Legend: Each plot represents the set of points $\{\theta^j\}_{j=0,\dots,k-1}$, for the Sobol sequence (panel a) and a set of random uniform draws (panel b) on $[0,1]^2$. Red dot: last point $\theta^{k-1}$, back dots: previous points. }
\end{figure}

Figure \ref{fig:cover} illustrates the discussion on the choice of covering sequence in Section \ref{sec:popu}. The top panel a shows a Sobol sequence on $[0,1]^2$ for $k=1,2,\dots,100$ and the bottom panel are random uniform draws on the same set. Notice that the region $[0.5,1]^2$ in panel b has very few draws for $k=20$ compared to 5 points in panel a. Likewise, for $k=50,100$ the points are spaced more evenly in the top than the bottom panel. The following gives some numerical explanation for this phenomenon.

Because the draws are random uniform, the probability that a given point is not in the region $[0.5,1]^2$ is $3/4$. There are no points in $[0.5,1]^2$ with probability $5.6\%$ for $k=10$, and $0.3\%$ for $k=20$. 
The dimension of the parameter space is quite critical in this calculation. Suppose there are $d=5$ parameters and $\Theta = [0,1]^d$. The probability that $\theta^j \in [0.5,1]^d$ is $p = 1/2^5 \simeq 3\%$. On average $k = 32$ draws are needed before this occurs, and $k = 87$ ensures it occurs with a probability of a least $50\%$ probability and $k=108$ for $95\%$ probability. In comparison, the Sobol sequence requires $k=57$ to have a point in the same region, but with probability $1$. For $d=10$, on average $k=1024$ draws are needed, $k = 6385$ for $50\%$ and $k=7042$ for $95\%$ probability; compared with $k=1149$ with probability $1$ for Sobol. The curse of dimensionality affects both, 

\section{R code to implement Algorithm \ref{algo:sgn}} \label{sec:Code}

{
  \begin{lstlisting}[basicstyle=\linespread{0.4}\ttfamily\footnotesize]
  # The implementation uses quasi Gauss-Newton Monte Carlo and Acceleration
  # (Extensions 1 and 2 of Algorithm 1)
  # Disclaimer: There is some but not full error handling. If the moments return Inf, NA, etc., the code will likely crash
  
  library(randtoolbox) # Used to generate the covering sequence
  
  sGN <- function( mom, learn = 0.1, alpha = 0.47, maxit = 250, eps = 0.1,
                  LM = TRUE, seed = NULL, L = NULL, init = NULL, W = NULL,      
                  lb = NULL, ub = NULL, global = FALSE, verbose = FALSE ) {
      
      # mom: function which returns moments
      # learn = learning rate in (0,1)
      # alpha = momentum in [0,1)
      # maxit = maximum number of iterations (stopping criterion)
      # eps = smoothing parameter
      # LM = use Levenberg–Marquardt algorithm ? (True / False)
      # seed : used for the Monte Carlo draws used to compute G_{n,eps}
  
      # L = number of Monte Carlo draws used to compute G_{n,eps}, 
      #     defaults to ceil( max(25,p*1.5 ) ), p = no. of parameters
      #     for very small values of eps, it is recommended to use a larger L
      #     to compensate the increase in variance
      #     this code only updates one direction at each iteration, in some cases 
      #     updating several in parallel can speed-up convergence
  
      # init = starting value, optional when global = TRUE, required if global = FALSE
      # lb, ub = lower/upper bounds for parameters, optional if global = FALSE
      # global = FALSE: local step only
      # global = TRUE:  local and global step
      # verbose = TRUE: print a few coefficients and objective value
  
      if (!is.null(seed)) {
          set.seed(seed) # locally set the seed (within the function)
      }
      pb = txtProgressBar(min=0,max=maxit) # A progress bar
  
      # Check if required inputs were given
      if (global == FALSE) {
          if (is.null(init)) {
              print("Error: local step only, starting value required")
              return();
          } else {
              p = length(init)
          }
      } else {
          if ( is.null(lb) || is.null(ub) ) {
              print("Error: global step, parameter bounds required")
              return()
          } else { # Construct covering sequence using Sobol & bounds
              p = length(lb)
              grid = sobol(maxit,p,scrambling = 0)
  
              for (j in 1:p) { 
                  grid[,j] = lb[j] + (ub[j]-lb[j])*grid[,j] 
              }
              if (!is.null(init)) {
                  grid = rbind(init,grid)
              } else {
                  init = grid[1,]
              }
          }
      }
  
      if (is.null(L)) {
          L = max(25,ceiling(1.5*p) ) # Default choice for Algorithm 2
      } else if  (L < p) {
          print("Error: L is too small")
          return();
      }
  
      pens = exp( seq(log(1e-5),log(1),( log(1)-log(1e-5) )/10) )
      if (LM == FALSE) { pen = 0 }
  
      coefs = matrix(0,maxit,p) # matrix of coefficients
      objs  = rep(0,maxit)      # keep track of objective values
  
      coefs[1,] = init # starting value
      sb = rep(0,p)    # momentum initialized at 0
  
      par = coefs[1,]
      mom0 = mom(par) # moments at starting value
      
      k = length(mom0) # number of moments
  
      if (is.null(W)) {
          W = diag(k) # Identity weighting by default
      }
  
      objs[1] = t(mom0)%*%W%*%mom0
  
      # Initialize the Jacobian matrix G_{n,eps}:
      Zb = matrix(0,L,p)
      Yb = matrix(0,L,k)
  
      for (j in 1:L) {
              Zb[j,] = rnorm(p) # Z ~ N(0,1)
              Yb[j,] = (mom(par+eps*Zb[j,])-mom0)/eps
      }
  
      Gb = t(lm(Yb~Zb)$coef[2:(p+1),]) # Jacobian estimate
      I = diag(p)
  
      if (verbose == TRUE) {
          b = 1
          cat('=====================\n')
              cat(paste(
                " Iteration:", b,"\n",
                " // Objective value:", round(objs[b],4),"\n"))
      }
  
      # Main Loop:
      for (b in 2:maxit) {
          # Local Step
          Hb = t(Gb)%*%W%*%Gb
  
          if (LM == TRUE) { # If LM = TRUE
              res = Yb - cbind(1,Zb)%*%(lm(Yb~Zb)$coef)
              pen = 1e-3*sqrt(mean(res^2))*(norm(W%*%Gb,"F"))
          } else  { pen = 0 }
          
          if ( rcond( Hb + pen*norm(Hb,"F")*I ) < 1e-12 ) { # Safeguard against non-invertible matrix
              rc <- function(pen) { return( rcond( Hb + pen*I ) ) }
              pen = pens[ min( which( sapply(pens,rc) > 1e-9 ) ) ]
          }
  
          coefs[b,] = coefs[b-1,] - learn*solve( Hb + pen*norm(Hb,"F")*I, t(Gb)%*%W%*%mom0 ) + alpha*sb
  
          sb      = coefs[b,] - coefs[b-1,] # update momentum
          mom0    = mom(coefs[b,])          # update moments
          objs[b] = t(mom0)%*%W%*%mom0      # update objective
  
          update = TRUE # update G_{n,eps}
  
          if (global == TRUE) { # Global step
              mom1 = mom(grid[b,])
              obj1 = t(mom1)%*%W%*%mom1
  
              if (obj1 < objs[b]) {  # Global step binds
                  # The global step is binding
                  print('Switch: global step strictly improves on local step')
                  # Swap values:
                  coefs[b,] = grid[b,]
                  mom0 = mom1
                  objs[b] = obj1
                  sb = rep(0,p) # reset momentum to zero
  
                  # Reset the jacobian
                  par = grid[b,]
                  for (j in 1:L) {
                      Zb[j,] = rnorm(p) # Z ~ N(0,1)
                      Yb[j,] = (mom(par+eps*Zb[j,])-mom0)/eps
                  }
                  update = FALSE # G_{n,eps} has already been updated
              }
          }
  
          if (update == TRUE) {
              par = coefs[b,]
              Zb  = rbind( Zb[-1,], rnorm(p) )
              Yb  = rbind( Yb[-1,], (mom(par+eps*Zb[L,])-mom0)/eps )
          }
  
          Gb = t(lm(Yb~Zb)$coef[2:(p+1),]) # New Jacobian estimate
  
          if ( (verbose == TRUE) && (b %% 50 == 0) ) {
              cat('=====================\n')
              cat(paste(
                " Iteration:", b,"\n",
                " // Objective value:", round(objs[b],4),"\n"))
          }
      }
      close(pb) # Close the progress bar
      return( list( par = coefs[which(objs==min(objs))[1],], objs = objs, coefs = coefs  ) )
  
  }

\end{lstlisting}}
\end{appendices}
\end{document}